\documentclass[journal ]{new-aiaa}
\usepackage[utf8]{inputenc}
\usepackage{textcomp}
\usepackage{overpic}
\usepackage{booktabs}
\usepackage{subcaption}
\usepackage{bm}
\usepackage{amsmath}

\usepackage{amsthm}
\usepackage{booktabs}
\usepackage{footnpag}%
\usepackage{graphicx} %
\usepackage{url}
\usepackage[utf8]{inputenc}%
\usepackage{textcomp}%
\usepackage{cancel}
\usepackage{hhline}
\usepackage{amsfonts}
\usepackage{amssymb}
\usepackage{mathtools}
\usepackage{siunitx} %
\usepackage{longtable,tabularx}
\usepackage{tikz}
\usepackage{subcaption}
\usepackage{sidecap}

\usepackage{algorithm}
\usepackage{algpseudocode}
\usepackage{acronym}
\usepackage{ifthen}
\usepackage{caption}

\usetikzlibrary{decorations.pathreplacing, calligraphy, fit, patterns, patterns.meta, shapes.geometric}
\usepackage[dvipsnames, x11names]{xcolor}
\usepackage{xkcdcolors}

\usepackage[most]{tcolorbox}
\newtcbox{\inlinegreenbox}[1][]{enhanced,
    box align=base,
    nobeforeafter,
    colback=DarkSeaGreen1,
    colframe=Chartreuse4,
    size=small,
    left=0pt,
    right=0pt,
    boxsep=2pt,
    #1
}

\newtcbox{\inlineyellowbox}[1][]{enhanced,
    box align=base,
    nobeforeafter,
    colback=LemonChiffon1,
    colframe=DarkGoldenrod2,
    size=small,
    left=0pt,
    right=0pt,
    boxsep=2pt,
    #1
}

\newcommand{\norm}[1]{\left \lVert #1 \right \rVert}
\usepackage[version=4]{mhchem}
\acrodef{pdf}[pdf]{probability density function}
\acrodef{DRO}[DRO]{distant retrograde orbit}
\acrodef{ode}[ODE]{ordinary differential equation}
\acrodef{KF}[KF]{Kalman filter}
\acrodef{EKF}[EKF]{extended Kalman filter}
\acrodef{EOM}[EOM]{equations of motion}
\acrodef{UKF}[UKF]{unscented Kalman filter}
\acrodef{CKF}[CKF]{cubature Kalman filter}
\acrodef{GMF}[GMF]{Gaussian mixture filter}
\acrodef{GMEKF}[GM-EKF]{Gaussian mixture extended Kalman filter}
\acrodef{RTS}[RTS]{Rauch-Tung-Striebel}
\acrodef{TFS}[TFS]{two-filter smoother}
\acrodef{FBS}[FBS]{forward-backward smoother}
\acrodef{BC}[BC]{backward corrector}
\acrodef{LS}[LS]{least squares}
\acrodef{cdf}[cdf]{cumulative distribution function}
\acrodef{GM}[GM]{Gaussian mixture}
\acrodef{OC-IMM}[OC-IMM]{optimal control interactive multiple model}
\acrodef{AGM}[AGM]{adaptive Gaussian mixture}
\acrodef{STM}[STM]{state transition matrix}
\acrodef{MD}[MD]{Malanobis distance}
\acrodef{TSE}[TSE]{Taylor series expansion}
\acrodef{PDT}[PDT]{partial derivative tensor}
\acrodef{PF}[PF]{particle filter}
\acrodef{KL}[KL]{Kullback-Liebler}
\acrodef{SVD}[SVD]{singular value decomposition}
\acrodef{IEKF}[IEKF]{iterated extended Kalman filter}
\acrodef{MAP}[MAP]{maximum a posteriori}
\acrodef{NEES}[NEES]{normalized estimation error squared}
\acrodef{ANEES}[ANEES]{average normalized estimation error squared}
\acrodef{NRHO}[NRHO]{near-rectilinear halo orbit}

\acrodef{STT}[STT]{state transition tensor}
\acrodef{CR3BP}[CR3BP]{circular restricted 3-body problem}
\acrodef{PSD}[PSD]{positive semi-definite}

\acrodef{MM}[MM]{multiple model}
\acrodef{IMM}[IMM]{interacting multiple model}
\acrodef{AGMIMM}[AGMIMM]{adaptive Gaussian mixture interacting multiple model}
\acrodef{GMIMM}[GMIMM]{Gaussian mixture interacting multiple model}
\acrodef{ode}[ODE]{ordinary differential equations}
\acrodef{usfos}[US-FOS]{uncertainty-scaled first-order stretching}
\acrodef{wussolc}[W-US-SOLC]{whitened uncertainty-scaled second-order linearization change}

\newcommand{\PD}[2]{\displaystyle\frac{\partial {#1}}{\partial {#2}}}
\DeclareMathOperator{\EX}{\mathrm{E}}%

\title{Two-Filter Adaptive Gaussian Mixture Smoothing for Nonlinear Systems}

\author{Benjamin Schneiderheinze\footnote{Ph.D. Student, School of Aeronautics and Astronautics}, Andrea De Vittori\footnote{Postdoctoral Researcher, School of Aeronautics and Astronautics}, and Keith A. LeGrand\footnote{Assistant Professor, School of Aeronautics and Astronautics. Corresponding Author. klegrand@purdue.edu}}
\affil{Purdue University, West Lafayette, IN, 47907}
\author{Jill Bruer\footnote{Senior Research Aerospace Engineer}}
\affil{Air Force Research Laboratory, Space Vehicles Directorate, Albuquerque, NM}

\begin{document}

\maketitle

\begin{abstract}
Space object tracking poses challenging estimation problems due to the significantly non-Gaussian distributions that can arise, particularly under highly nonlinear dynamics or during periods of measurement unavailability.
Adaptive Gaussian mixture filters can dynamically adjust their mixture resolution to systematically approximate these non-Gaussian distributions, but challenging estimation problems can still produce highly uncertain or inaccurate estimates, especially during prolonged observation gaps.
Smoothing algorithms can significantly improve filtered estimates by incorporating future measurement information for applications where immediacy is not required, but smoothing in nonlinear, non-Gaussian settings poses additional theoretical and computational challenges.
This work develops a new recursive Bayesian smoothing algorithm for nonlinear systems that refines Gaussian mixture posteriors produced by a forward adaptive Gaussian mixture filter.
A two-filter smoothing approach approximates the future measurement information by an information-form Gaussian mixture in the state variable.
Techniques from nonlinear Gaussian mixture filtering including splitting, merging, and recursive measurement updating are also incorporated to improve the accuracy and computational efficiency of this approximation.
The proposed smoother's estimation capabilities are demonstrated on space object tracking problems for Molniya and Earth-Moon halo orbits and shown to significantly reduce estimation error and uncertainty compared to the forward filter.

\end{abstract}

\newpage

\section{Introduction}\label{sec:introduction}

\lettrine{A}{ccurate}
 orbit determination and trajectory estimation remain critical functions of space domain awareness and find uses in both online and forensic capacities.
However, this estimation problem presents significant challenges in certain orbital regimes characterized by strongly nonlinear dynamics and sparse observation opportunities.
For example, cislunar spacecraft can be subjected to chaotic dynamics, which when evolved over extended measurement gaps, produce highly non-Gaussian uncertainty structures.\cite{legrand2023BayesianAnglesOnlyCislunar,Iannamorelli_AGMIMM_2025}.
Thus, achieving accurate and uncertainty-realistic state estimates requires advanced nonlinear and non-Gaussian estimation approaches.

State estimation generally relies on filtering techniques that combine knowledge of the system dynamics with partial and noisy measurements to recursively produce real-time state estimates.
The Kalman filter \cite{kalman_1960, kalman_bucy_1961} and its variants \cite{Julier_EKF_1997, VanDerMerwe_UKF_2000, Arasaratnam_Haykin_CKF_2009} are the most standard filtering techniques, but these methods only estimate the mean and covariance of the state over time, and as a result their performance can degrade when the system models or state uncertainty are sufficiently non-Gaussian.
Alternatives to combat this issue include~\acp{GMF} \cite{Sorenson_GMF_1972, Sorenson_GM_1971}, which approximate non-Gaussian posteriors as weighted sums of Gaussian mixands.
Adaptive variants dynamically adjust the mixture complexity through splitting and reduction operations, forming the basis of \ac{AGM} filtering \cite{Hanebeck_progBayes_2003}, with numerous strategies proposed in the literature \cite{Huber_split_2008,demars_entropy-based_2013,tuggle_zanetti_split_2018,zanetti_novel_2018,Legrand_SplitHappens_2023}.
Still other work has leveraged \acp{GM} to track maneuvering cislunar space objects~\cite{Iannamorelli_AGMIMM_2025, Lin_Oguri_OCIMM_2025}.

\begin{figure}[htbp]
    \centering
    \begin{overpic}[width=0.49\textwidth]{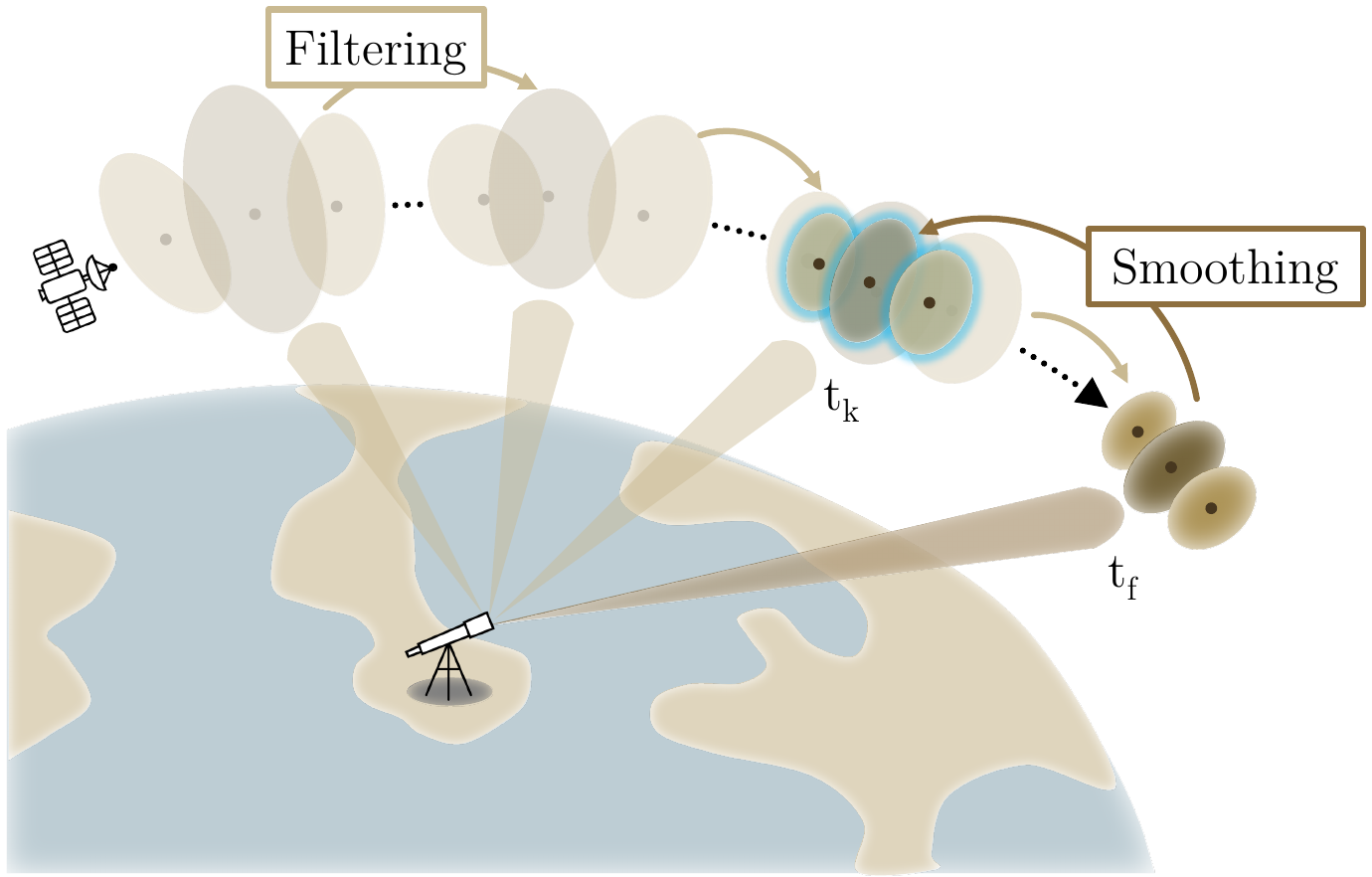}
        \put(2,60){\textbf{(a)}}
    \end{overpic}
    \hfill
    \begin{overpic}[width=0.49\textwidth]{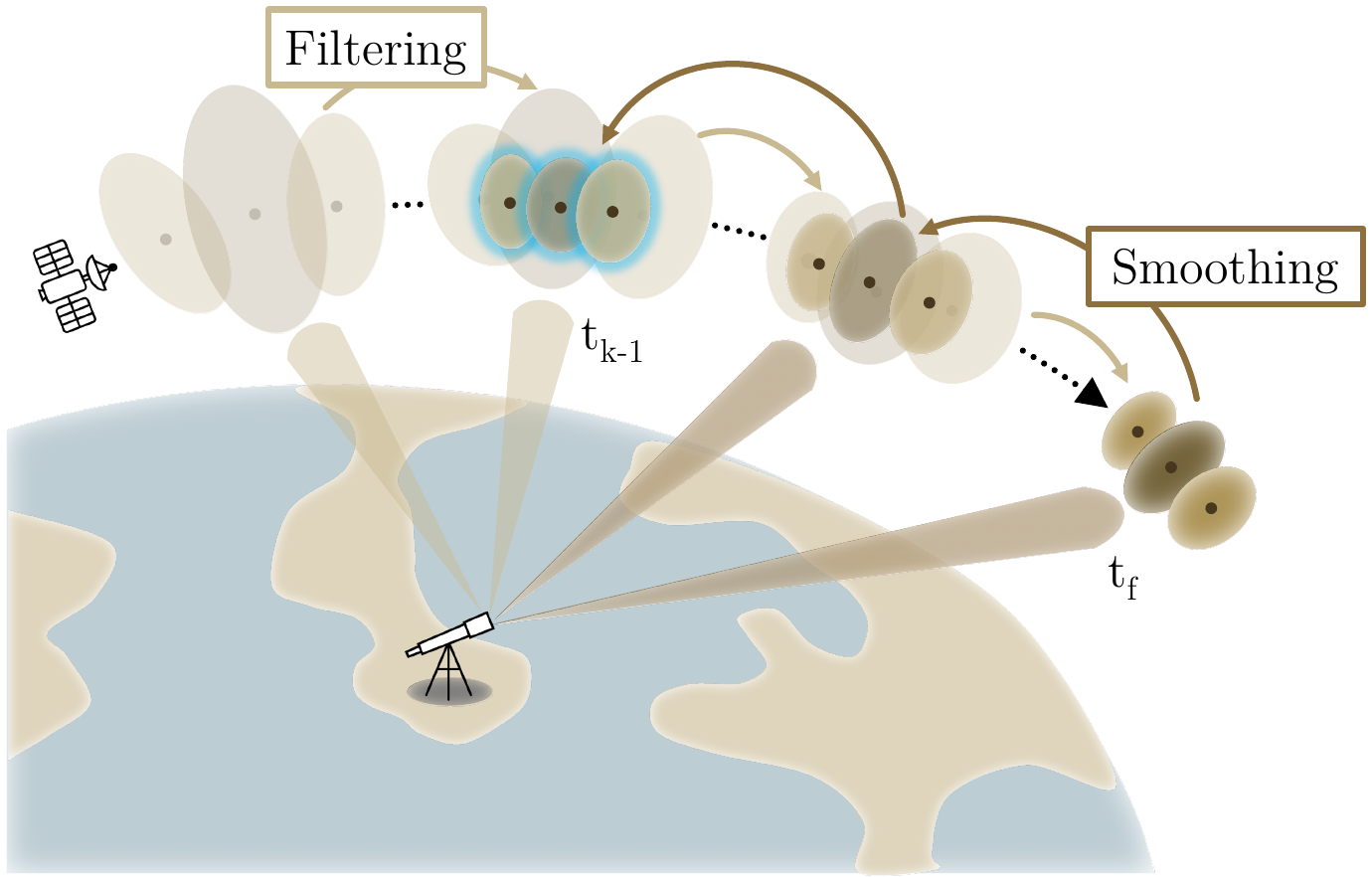}
        \put(2,60){\textbf{(b)}}
    \end{overpic}
    \caption{Smoothing visualization for space object tracking.}
    \label{fig:SmoothingVisualization}
\end{figure}

While filtering enables sequential inference, estimates computed during long observation gaps based on outdated measurement information can become highly uncertain and inaccurate.
In contrast, smoothing exploits both past and future observations to retrospectively estimate system states, yielding more accurate and uniformly informed estimates throughout the time interval of interest.
The smoothing process is illustrated in Fig.~\ref{fig:SmoothingVisualization} in the context of a space object tracking application.
Smoothing techniques exist in the literature using analogous techniques to the Kalman filter and its variants for nonlinear systems \cite{RTS_1965,cox_estimation_1964,melsa_estimation_1981,sarkka_unscented_2008}, as well as the particle filter \cite{kitagawa_monte_1996,godsill_monte_2004,lindsten_backward_2013}.
For \ac{GM} estimates, existing approaches to smoothing generally fall into two categories: forward-backward and two-filter.
Forward-backward smoothers directly compute a \ac{BC} term whose product with the filtering posterior yields the smoothed posterior.
The \ac{BC} term can be computed recursively in a manner that circumvents a troublesome quotient of \acp{GM} \cite{vo_vo_2012}, and can be adapted to nonlinear systems through linearization about mixand means \cite{lee_campbell_2015,Desai_agmsmooth_2026}.
However, both the memory requirement and linearization error of this \ac{BC} term grows over time, which can make standard fixed-interval smoothing impractical.
In contrast, two-filter methods filter the future measurement density backward in time and then combine it with the prior state density to form the smoothed state density \cite{sarkka_bayesian_2023}.
These methods previously required approximations or delayed initialization when the measurement model is not invertible \cite{Kitagawa_TwoFilterGaussianSum_1994}, but this problem has recently been circumvented by applying information-form representations of the future measurement density to the \ac{GM} smoothing problem \cite{balenzuela_new_2022,kitagawa_revisiting_2023}.
Importantly, these two-filter methods have been formulated only for linear systems, and the generalization to nonlinear systems is a non-trivial matter because of limitations of existing representations of the future measurement density.
A nonlinear batch processing approach for \ac{GM} estimates has also been demonstrated on initial orbit determination problems \cite{craft_gaussian_2025}, although the computational complexity of this approach scales exponentially with the number of measurements in the batch, making it best suited for problems involving very few measurements.

The contributions of this paper are as follows. Inspired by the approach of \cite{balenzuela_new_2022} for linear systems, this paper introduces the first \ac{AGM} smoother suitable for highly nonlinear systems. In addition to adaptive splitting steps performed in the forward filter, this two-filter algorithm performs splitting in the backward filter to obtain high accuracy approximations of the future measurement likelihood function. In the development of this smoother, other minor contributions are made with potential applications beyond smoothing. These include new precision-matrix forms of Gaussian merging and splitting equations suitable for degenerate mixtures. A recursive update technique recently proposed for nonlinear filtering problems is also employed in the smoother initialization phase to improve stability in cases of high-accuracy measurement updates of diffuse mixtures.

The paper is organized as follows. Section~\ref{sec:problem_formulation} introduces the problem formulation and the underlying assumptions. Section~\ref{sec:background} reviews relevant background material on Bayesian filtering and smoothing. Section~\ref{sec:methodology} presents the proposed smoothing methodology, and Section~\ref{sec:results} evaluates its estimation performance on representative nonlinear systems, including Keplerian and \ac{CR3BP} orbital estimation problems. Finally, Section~\ref{sec:conclusion} summarizes the main findings and outlines directions for future research.

\section{Problem Formulation}\label{sec:problem_formulation}
This work considers the estimation of states governed by nonlinear discrete-time systems or by discrete-time approximations of nonlinear continuous-time dynamical systems. For completeness, the remainder of this paper assumes the latter, as its development encompasses the simplified former.
The system state~$\mathbf{x}\in\mathbb{R}^{n_x}$ evolves over some time span~$[t_0,\,t_T]$ according to the system of stochastic \acp{ode} given by
\begin{equation}
    \displaystyle\frac{\mathrm{d}\mathbf{x}}{\mathrm{d}t} = \mathbf{f}(\mathbf{x},t) + \bm{\Gamma}(t)\mathbf{w}(t)
\end{equation}
where ~$\mathbf{f}:\mathbb{R}^{n_x},\, \mathbb{R}\rightarrow\mathbb{R}^{n_x}$ is the dynamics model,~$\boldsymbol{\Gamma}(t) \in \mathbb{R}^{n_x \times n_w}$ is the noise input matrix, and~$\mathbf{w}\sim\mathcal{N}\!(\mathbf{0}, \mathbf{Q}_c)$ is white, Gaussian, continuous process noise representing unknown disturbances in the system with power spectral density~$\mathbf{Q}_c$.
The corresponding continuous-time process noise covariance is
\begin{equation}
\mathbf{Q}_c \,\delta(t-\tau) = \mathrm{E}[\mathbf{w}(t)\mathbf{w}^\top(\tau)]
\end{equation}
where~$\delta(\cdot)$ denotes the Dirac delta function. The solution flow of the deterministic portion of the system is defined as
\begin{equation}
    \bm{\varphi}_{\Delta t}(\mathbf{x}(t_0), t_0) = \mathbf{x}(t_0 + \Delta t) = \mathbf{x}(t_0) + \int_{t_0}^{t_0+\Delta t}{\mathbf{f}(\mathbf{x}(\tau),\tau)\ \mathrm{d}\tau}
\end{equation}

For nonlinear stochastic \acp{ode}, the resulting moment equations give rise to an infinite dimensional system of fully-coupled \acp{ode}, and thus solutions are generally unavailable \cite{Kuehn_MomentClosure_2015}.
Thus, an approximate discrete-time stochastic process is used instead, expressed as
\begin{equation}
    \mathbf{x}_{k} = \bm{\varphi}_{t_k - t_{k-1}}(\mathbf{x}_{k-1}, t_{k-1}) + \mathbf{w}_{k-1}
\end{equation}
where the discrete-time process noise~$\mathbf{w}_{k-1} \sim \mathcal{N}\!(\mathbf{0},\mathbf{Q}_{k-1})$ is white and Gaussian characterized by discrete-time process noise covariance of
\begin{equation}
    \mathbf{Q}_k \delta_{k}^{k'} = \mathrm{E}[\mathbf{w}_{k}\mathbf{w}_{k'}^\top]
\end{equation}
where~$\mathbf{Q}_k$ denotes the discrete-time process noise covariance and~$\delta_k^{k'}$ denotes the Kronecker delta.
The stochastic portion of the system must be approximated both in the conversion from a continuous-time to a discrete-time system and in the propagation for a given~$\Delta t$.
For this work, the discrete process noise integral is approximated for nonlinear systems as
\begin{equation}
    \mathbf{Q}_k|_{\mathbf{x}_k} \approx
    \int_{t_k}^{t_{k+1}}
    \bm{\Phi}(t_{k+1}, \tau)\,
    \bm{\Gamma}(\tau) \mathbf{Q}_c \bm{\Gamma}(\tau)^\top
    \bm{\Phi}(t_{k+1}, \tau)^\top
    \,\mathrm{d}\tau \label{eqn:Q_integral}
\end{equation}
where~$\bm{\Phi}$ is the \ac{STM}, which describes the first-order sensitivity of deviations in the output state to perturbations to the input state and is defined as
\begin{equation}
  \bm{\Phi}(t_{k+1},t_k) = \PD{\bm\varphi_{t_{k+1}-t_{k}}(\mathbf{x}_k,t_k)}{\mathbf{x}_k}
\end{equation}
Note that the discrete-time process noise covariance is a function of the state~$\mathbf{x}_k$ due to the \ac{STM} in~\eqref{eqn:Q_integral}.

This system is also characterized by measurements~$\mathbf{z}\in\mathbb{R}^{n_z}$, which are related to the state by the nonlinear function
\begin{equation}
    \mathbf{z}_k = \mathbf{h}_k(\mathbf{x}_k) + \mathbf{v}_k \label{eqn:meas_model}
\end{equation}
where~$\mathbf{h}_k:\mathbb{R}^{n_x}\rightarrow\mathbb{R}^{n_z}$ is the measurement model and~$\mathbf{v}_k\sim\mathcal{N}\!(\mathbf{0}, \mathbf{R}_k)$ is additive, white, Gaussian measurement noise.
The corresponding measurement noise covariance is defined as
\begin{equation}
    \mathbf{R}_k = \mathrm{E}\left[\mathbf{v}_k\mathbf{v}_k^\top\right]
\end{equation}

\section{Background}\label{sec:background}

\subsection{Bayesian Filtering}
Bayesian filtering consists of estimating the \ac{pdf} of the state~$\mathbf{x}_k$ as it evolves over time and new measurement information becomes available.
The complete set of measurements~$\mathbf{z}_k$ up to time~\(t_k\) is denoted by
\begin{equation}
    \mathbf{z}_{1:k} = \{\mathbf{z}_1, \mathbf{z}_2, \dots, \mathbf{z}_k \}
\end{equation}

The estimate of the state at time~\(t_k\), conditioned on all measurements up to and including that time, is given by the posterior filtering density~$p(\mathbf{x}_k | \mathbf{z}_{1:k})$. The state evolution is assumed to satisfy the Markov property, meaning that the future state~\(\mathbf{x}_{k+1}\) only depends on the current state~\(\mathbf{x}_k\) and the process model, not on the full history of past states \cite{crassidis_book_2011}. Additionally, each measurement~$\mathbf{z}_k$ is assumed to depend only on the current state~$\mathbf{x}_k$, being conditionally independent of the measurements and states at all other times.
Whenever a new measurement~\(\mathbf{z}_{k+1}\) becomes available, the posterior~\(p(\mathbf{x}_k|\mathbf{z}_{1:k})\) is obtained through a \emph{prediction step}, based on the Chapman--Kolmogorov equation, followed by an \emph{update step}, based on Bayes' rule \cite{Jazwinski_StochasticProcesses_1970,anderson_optimalfiltering_2012} as
\begin{align}
    p(\mathbf{x}_{k+1} | \mathbf{z}_{1:k})
        &= \int p(\mathbf{x}_k | \mathbf{z}_{1:k})\,
               p(\mathbf{x}_{k+1} | \mathbf{x}_k) \, d\mathbf{x}_k
        \label{eqn:bayes_pred_k1} \\
    p(\mathbf{x}_{k+1} | \mathbf{z}_{1:k+1})
        &= p(\mathbf{x}_{k+1} | \mathbf{z}_{1:k})\,
           \displaystyle\frac{p(\mathbf{z}_{k+1} | \mathbf{x}_{k+1})}
                {\int p(\mathbf{z}_{k+1} | \mathbf{x}_{k+1})\,
                      p(\mathbf{x}_{k+1} | \mathbf{z}_{1:k}) \, d\mathbf{x}_{k+1}}
        \label{eqn:bayes_upd_k1}
\end{align}
where~\(p(\mathbf{x}_{k+1}|\mathbf{x}_k)\) is the state transition density and~\(p(\mathbf{z}_{k+1} |\mathbf{x}_{k+1})\) denotes the measurement likelihood for a measurement~\(\mathbf{z}_{k+1}\) at time~\(t_{k+1}\) given state~\(\mathbf{x}_{k+1}\). The prediction and update steps together constitute the standard Bayesian filtering recursion.

\subsection{Bayesian Smoothing}
While Bayesian filtering produces a state estimate based on all measurements available up to the current time~$t_k$, it does not exploit future measurements beyond~$t_k$ that may provide additional information about the present state.
In applications where the complete measurement sequence is available \emph{a posteriori}, a Bayesian smoothing framework can be formulated, in which the entire set of measurements~$\mathbf{z}_{1:T}$ is used to inform the state estimate at time~$t_k$, with~$T \ge k$ \cite{sarkka_bayesian_2023}.
Whereas filtering proceeds forward in time from~$t_0$ to~$t_T$, Bayesian smoothing begins with the final state estimate at~$t_T$ and recursively propagates information backward to obtain improved estimates at earlier times.
This backward recursion is initialized at the final time with a smoothed posterior equal to the filtered posterior since there is no future measurement information to incorporate. For any other time index~$k < T$, the smoothed distribution~$p(\mathbf{x}_k \mid \mathbf{z}_{1:T})$ can be obtained through either the forward-backward or two-filter formulations.

The forward-backward formulation computes the smoothed density as
\begin{align}
    p(\mathbf{x}_k|\mathbf{z}_{1:T}) = p(\mathbf{x}_k|\mathbf{z}_{1:k}) \int \frac{p(\mathbf{x}_{k+1}|\mathbf{z}_{1:T})}{p(\mathbf{x}_{k+1}|\mathbf{z}_{1:k})}p(\mathbf{x}_{k+1}|\mathbf{x}_k)\textrm{d}\mathbf{x}_{k+1}
\end{align}
Since this formulation operates only on state densities, it is generally the most straightforward for deriving smoothers for filters that assume Gaussian state densities. This includes the \ac{RTS} smoother for linear Gaussian systems \cite{RTS_1965}, as well as its nonlinear variants based on the \ac{EKF} \cite{cox_estimation_1964,melsa_estimation_1981} and \ac{UKF} \cite{sarkka_unscented_2008}. However, the quotient of densities can limit the applicability of this method to more general \ac{pdf} representations such as \acp{GM} \cite{Kitagawa_TwoFilterGaussianSum_1994}.

In contrast, the two-filter framework is based in the factorization of the smoothed density using Bayes' rule as
\begin{align}\label{Eq:two_filter_update}
    p(\mathbf{x}_k|\mathbf{z}_{1:T}) = p(\mathbf{x}_k|\mathbf{z}_{1:k})\displaystyle\frac{p(\mathbf{z}_{k+1:T}|\mathbf{x}_k)}{p(\mathbf{z}_{k+1:T}|\mathbf{z}_{1:k})}
\end{align}
which requires the backward filter likelihood~$p(\mathbf{z}_{k+1:T}|\mathbf{x}_k)$ and the normalization term~$p(\mathbf{z}_{k+1:T}|\mathbf{z}_{1:k})$. These terms can be calculated recursively as
\begin{subequations}
    \begin{align}
        p(\mathbf{z}_{k:T}|\mathbf{x}_{k-1}) &= \int p(\mathbf{z}_{k+1:T}|\mathbf{x}_{k})p(\mathbf{z}_{k}|\mathbf{x}_{k})p(\mathbf{x}_{k}|\mathbf{x}_{k-1})\textrm{d}\mathbf{x}_{k} \label{eqn:twofilter_recurse_bfl} \\
        p(\mathbf{z}_{k:T}|\mathbf{z}_{1:k-1}) &= p(\mathbf{z}_{k+1:T}|\mathbf{z}_{1:k})\int p(\mathbf{z}_{k}|\mathbf{x}_{k})p(\mathbf{x}_{k}|\mathbf{z}_{1:k-1})\textrm{d}\mathbf{x}_{k} \label{eqn:twofilter_recurse_norm}
    \end{align}
\end{subequations}
This method is referred to as a two-filter smoother because the recursion for the backward filter likelihood~$p(\mathbf{z}_{k:T}|\mathbf{x}_{k-1})$ resembles a filtering pass performed backwards through the measurement history, complete with associated update and prediction steps. By nature of it being a density over the future measurement space rather than the state-space, the backward filter likelihood can be difficult to approximate for nonlinear systems. However, the two-filter formulation does not contain a quotient of densities, which can make it preferable over the forward-backward formulation for general \ac{pdf} representations.

\subsection{Gaussian Mixtures}
While Gaussianity is tacitly assumed in many estimation algorithms, this assumption is generally only valid over short time horizons for nonlinear dynamical systems.
As the system evolves and nonlinear effects accumulate, a Gaussian approximation of the underlying \ac{pdf} can rapidly deteriorate, leading to degraded estimation performance.
This necessitates a \ac{pdf} representation capable of capturing significant non-Gaussianities.
One approach is to approximate each \ac{pdf} with a weighted sum of Gaussian mixands, forming a \ac{GM} \cite{Sorenson_GMF_1972, Sorenson_GM_1971}.
This representation is able to capture the multimodal and non-Gaussian uncertainties that arise in nonlinear systems while retaining the analytical tractability of Gaussian densities.
Mathematically, the \ac{GM} approximation yields an estimated \ac{pdf} of the form
\begin{equation}
    p(\mathbf{x}_k|\mathbf{z}_{1:k}) \approx \sum_{n=1}^{N_k} w^{(n)}_{k} \, \mathcal{N}\!\left(\mathbf{x}_{k}; \mathbf{m}^{(n)}_{k}, \mathbf{P}^{(n)}_{k} \right)
\end{equation}
where~$w_k^{(n)} \in [0,1]$,~$\mathbf{m}^{(n)}_k \in \mathbb{R}^{n_x}$, and~$\mathbf{P}^{(n)}_k \in \mathbb{S}_{++}^{n_x}$ are the weight, mean, and positive-definite covariance matrix of the~\(n\)-th mixand (\(n=1,\dots,N_k\)) of the \ac{GM}, where the set of positive-definite matrices is defined as
\begin{equation}
\mathbb{S}_{++}^{n} = \left\{ \mathbf{A}\in\mathbb{R}^{n\times n} \;\middle|\; \mathbf{A}=\mathbf{A}^\top,\; \mathbf{x}^\top\mathbf{A}\mathbf{x}>0, \ \forall\,\mathbf{x}\in\mathbb{R}^n\setminus\{\mathbf{0}\} \right\}.
\end{equation}
The weights satisfy the normalization condition
\begin{equation}
    \sum_{n=1}^{N_k} w^{(n)}_{k} = 1
\end{equation}
such that the mixture is a valid probability density.

\subsection{Gaussian Splitting}\label{sec:background_split}
Although \ac{GM}-based estimation frameworks generally outperform linear estimators in nonlinear regimes, their efficacy depends heavily on the local approximations used to propagate and update the individual mixands. When local linearization or low-order expansions are applied, significant approximation errors can still arise if the dynamics or measurement models are highly nonlinear over the support of an individual mixand. This limitation can be effectively addressed through Gaussian splitting, whereby a single mixand with a large support is partitioned into a sum of multiple, tightly constrained mixands. By decreasing the spatial support of each individual mixand, the accuracy of local linearizations is significantly improved, thereby reducing truncation errors during nonlinear transformations.

Gaussian splitting is typically accomplished through the use of a univariate splitting library \cite{Hanebeck_progBayes_2003}, which stores approximate solutions~$\{\tilde{w}^{(i)},\tilde{m}^{(i)},\tilde{\sigma}^{(i)}\}$ to the mixture representation of a standard univariate Gaussian
\begin{align}
    \mathcal{N}\!\left(x;0,1\right) \approx \sum_{i=1}^{L} \tilde{w}^{(i)}\mathcal{N}\!\left(x;\tilde{m}^{(i)},\left(\tilde{\sigma}^{(i)}\right)^2\right)
\end{align}
where~$L$ is the resulting number of mixands in the split approximation.
For multivariate Gaussian distributions, early  works relied solely on minimizing the $L_2$ distance between the \ac{GM} and the original single-mixand density \cite{demars_entropy-based_2013}.
Other approaches leverage library-based solutions that strictly preserve the mean and variance of a standard univariate Gaussian \cite{Tuggle_thesis_2020}.
To simplify the underlying optimization problem, many of these methods additionally enforce homoscedastic mixtures with equally-spaced means, although such constraints are not strictly necessary for multivariate generalizations \cite{kulik2025NonlinearityUncertaintyInformed}.

To split a multivariate Gaussian $\mathcal{N}(\mathbf{x}, \mathbf{m}, \mathbf{P})$ along a chosen direction $\hat{\bm\delta}^*$, the univariate library solution $\{\tilde{w}^{(i)}, \tilde{m}^{(i)}, \tilde{\sigma}^{(i)}\}$ is lifted to the full state space by placing mixand means along $\hat{\bm\delta}^*$,
\begin{equation}
    \mathbf{m}^{(i)} = \mathbf{m} + m^{(i)} \hat{\bm\delta}^*,\qquad m^{(i)} = \frac{\tilde{m}^{(i)}}{\sqrt{(\hat{\bm\delta}^*)^\top \mathbf{P}^{-1} \hat{\bm\delta}^*}}
\end{equation}
and assigning each mixand the covariance $\left(\tilde{\sigma}^{(i)}\right)^2 \bar{\mathbf{P}}$, where $\bar{\mathbf{P}}$ is the rank-1 downdate of the parent covariance
\begin{equation}
    \bar{\mathbf{P}} = \frac{\mathbf{P} - \sum_{i=1}^L w^{(i)} (\mathbf{m}^{(i)} - \mathbf{m})(\mathbf{m}^{(i)} - \mathbf{m})^\top}{\sum_{i=1}^L w^{(i)} \left(\tilde{\sigma}^{(i)}\right)^2}
\end{equation}
which guarantees that the resulting mixture exactly preserves the mean and covariance of the original mixand \cite{kulik2025NonlinearityUncertaintyInformed}.

Within this framework, the optimal splitting direction is selected by maximizing a functional $\mathcal{F}$ that measures the nonlinearity and/or uncertainty evolution associated with a transform~$\mathbf{g}$ as
\begin{equation}
    \hat{\bm\delta}^* \propto \underset{\norm{\bm\delta}_{\mathbf{A}}=1}{\arg\max} \;
    \mathcal{F}(\mathbf{g}, \bm\delta)
    \label{eq:opt_splitting}
\end{equation}
where~$\norm{\bm\delta}_\mathbf{A} \triangleq \sqrt{\bm\delta^\top \mathbf{A} \bm\delta}$ denotes the norm induced by~$\mathbf{A} \in \mathbb{S}_{++}^{n_x}$. Let
\begin{equation}
    \mathbf{G} \triangleq \left.\frac{\partial \mathbf{g}}{\partial \mathbf{x}} \right|_{\mathbf{m}}, \qquad
    \mathbf{G}^{(2)} \triangleq \left.\frac{\partial^2 \mathbf{g}}{\partial \mathbf{x}^2} \right|_{\mathbf{m}}
\end{equation}
be the first- and second-order derivatives of $\mathbf{g}$, defined as either the solution flow of the dynamics or the measurement function evaluated at the mean state. Furthermore, let $\mathbf{P}_z = \mathbf{G}\mathbf{P}_x\mathbf{G}^\top$ represent the linearly transformed covariance, and let $\mathbf{v}_{\max}(\cdot)$ and $\lambda_{\max}(\cdot)$ be the maximal right singular vector and eigenvalue respectively of the target matrix operator. While several formulations for the splitting functional $\mathcal{F}$ have been proposed \cite{tuggle_zanetti_split_2018, kulik2025NonlinearityUncertaintyInformed}, a representative subset is adopted in this work  and summarized in Table~\ref{tab:splitting_heuristics}.

\begin{table}[h]
\footnotesize
\caption{\label{tab:splitting_heuristics} Summary of splitting heuristics considered under the constraint $\norm{\bm\delta}_{\mathbf{P}_x^{-1}} = 1$.}
\centering
\renewcommand{\arraystretch}{1.3} %
\begin{tabular}{cccc}
\hline\hline
\textbf{Heuristic} & \textbf{Objective, \(\mathcal{F}(\mathbf{g},\bm\delta)\)} & \textbf{Solution, \(\hat{\bm\delta}^*\)} & \textbf{Criterion} \\
\hline
\textit{maxvar} &
$\| \bm\delta \|_2$ &
\(\mathbf{P}_x^{1/2}\mathbf{v}_{\max}(\mathbf{P}_x^{1/2})\) &
$\lambda_{\max}(\mathbf{P}_x^{1/2})$ \\

US-FOS &
\(\|\mathbf{G}\bm\delta\|_2\) &
\(\mathbf{P}_x^{1/2}\mathbf{v}_{\max}(\mathbf{G}\mathbf{P}_x^{1/2})\) &
$\lambda_{\max}(\mathbf{G}\mathbf{P}_x^{1/2})$ \\

W-US-SOLC &
\(\left\|
\mathbf{P}_z^{-1/2}
\left(\mathbf{G}^{(2)}\bm\delta\right)
\mathbf{P}_x^{1/2}
\right\|_F^2\) &
\(\mathbf{P}_x^{1/2}\mathbf{v}_{\max}(\bar{\mathbf{G}}_w^{(2)})\) &
$\lambda_{\max}(\bar{\mathbf{G}}_w^{(2)})$ \\
\hline\hline
\end{tabular}
\end{table}
The solution for \ac{wussolc} relies on the reshaped second-order derivative tensor defined in Einstein notation as
\begin{subequations}
\begin{align}
    (G_w^{(2)})_{j,k}^i &=
    (P_z^{-1/2})^i_l \,
    (G^{(2)})_{p,q}^l \,
    (P_x^{1/2})^p_j \,
    (P_x^{1/2})^q_k, \\
    (\bar{G}_w^{(2)})_k^{ni+j} &= (G_w^{(2)})_{j,k}^i.
\end{align}
\end{subequations}
Optimization of the ``\textit{maxvar}'' heuristic yields the direction of maximum prior uncertainty, corresponding to the dominant eigenvector of $\mathbf{P}_x$, independently of the nonlinear mapping $\mathbf{g}$.
\Ac{usfos} identifies directions of maximum local linear stretching induced by $\mathbf{g}$, weighted by the input uncertainty, while \ac{wussolc} targets directions exhibiting the strongest variation of the Jacobian through a whitened formulation that ensures invariance to coordinate scaling in both input and output spaces.

Among the three heuristics, \ac{wussolc} generally produces the best performing split directions, as it leverages second-order derivative information.
\Ac{usfos} offers a lower-cost alternative that is often well aligned with \ac{wussolc} in astrodynamics applications, although its performance may degrade depending on the nonlinear dynamics.
In contrast, \textit{maxvar} is the simplest and least computationally intensive method, but it neglects the nonlinear mapping entirely and therefore performs poorly under strongly nonlinear propagation.

Once a univariate split solution is obtained, a multivariate split can be performed using the univariate solution and a given unit split direction~$\hat{\bm\delta}$. The multivariate Gaussian can then be approximated by a mixture matching its mean and covariance as \cite{kulik2025NonlinearityUncertaintyInformed}
\begin{subequations}\label{eqn:multivariate_split}
    \begin{align}
        \mathcal{N}\!\left(\mathbf{x};\mathbf{m},\mathbf{P}\right) &\approx \sum_{i=1}^{L}w^{(i)}\mathcal{N}\!\left(\mathbf{x},\mathbf{m}^{(i)},\mathbf{P}^{(i)}\right) \\
        w^{(i)} &= \tilde{w}^{(i)} \\
        \mathbf{m}^{(i)} &= \mathbf{m} + \displaystyle\frac{\tilde{m}^{(i)}}{\sqrt{\hat{\bm\delta}^\top \mathbf{P}^{-1} \hat{\bm\delta}}}\hat{\bm\delta} \\
        \mathbf{P}^{(i)} &= \frac{\left(\tilde{\sigma}^{(i)}\right)^2}{\sum_{j=1}^L \tilde{w}^{(j)}\left(\tilde{\sigma}^{(j)}\right)^2}\left(\mathbf{P} - \sum_{j=1}^L \tilde{w}^{(j)}(\mathbf{m}^{(j)}-\mathbf{m})(\mathbf{m}^{(j)}-\mathbf{m})^\top\right) \label{eqn:multivariate_split_cov}
    \end{align}
\end{subequations}

To split a mixture, the direction and criterion for the selected heuristic from Table~\ref{tab:splitting_heuristics} is computed for each mixand. If the criterion for a mixand exceeds a specified split tolerance, then that mixand is split. This criterion can also be scaled by the mixand weight so that mixands comprising negligible probability mass are less likely to be split \cite{siciliano2025HigherOrderTensorBasedDeferral}. Once this is done for all mixands, the process can be repeated until a maximum recursion depth has been reached or no mixands exceed the split tolerance.

\subsection{Gaussian Merging}\label{sec:background_merge}
Without intervention, repeated splitting operations will produce increasingly large mixtures, which increases the overall computation cost of the filter.
In order to maintain computational tractability, the size of the mixture must occasionally be reduced.
This can be accomplished through Gaussian merging, whereby similar Gaussian mixands are merged into a single Gaussian mixand through moment-matching.
For a pair of mixands, the resulting merged mixand can be computed as
\begin{subequations}\label{eqn:multivariate_merge}
    \begin{align}
        w^{(i)}\mathcal{N}\!\left(\mathbf{x};\mathbf{m}^{(i)},\mathbf{P}^{(i)}\right) + w^{(j)}\mathcal{N}\!\left(\mathbf{x};\mathbf{m}^{(j)},\mathbf{P}^{(j)}\right) &\approx w^{(ij)}\mathcal{N}\!\left(\mathbf{x};\mathbf{m}^{(ij)},\mathbf{P}^{(ij)}\right) \\
       w^{(ij)} &= w^{(i)} + w^{(j)} \\
        \alpha^{(i)} &= \displaystyle\frac{w^{(i)}}{w^{(i)} + w^{(j)}},\: \alpha^{(j)} = \displaystyle\frac{w^{(j)}}{w^{(i)} + w^{(j)}} \label{eqn:multivariate_merge_alpha} \\
        \mathbf{m}^{(ij)} &= \alpha^{(i)}\mathbf{m}^{(i)} + \alpha^{(j)}\mathbf{m}^{(j)} \\
        \mathbf{P}^{(ij)} &= \alpha^{(i)}\mathbf{P}^{(i)} + \alpha^{(j)}\mathbf{P}^{(j)} + \alpha^{(i)}\alpha^{(j)}\left(\mathbf{m}^{(i)} - \mathbf{m}^{(j)}\right)\left(\mathbf{m}^{(i)} - \mathbf{m}^{(j)}\right)^\top \label{eqn:multivariate_merge_cov}
    \end{align}
\end{subequations}

The selection of mixands to merge can be achieved using the Runnalls' merging algorithm \cite{Runnalls_gmreduction_2007}, whereby mixands are merged pairwise based on an upper bound of the \ac{KL} divergence between them, given by
\begin{align}
    D_{\textrm{KL}}[i||j] \leq B(i,j) = \displaystyle\frac{1}{2}\left(\left(w^{(i)} + w^{(j)}\right)\log\left|\mathbf{P}^{(ij)}\right| - w^{(i)}\log\left|\mathbf{P}^{(i)}\right| - w^{(j)}\log\left|\mathbf{P}^{(j)}\right|\right)
\end{align}

This metric is computed for each pair of mixands in the mixture, and the pair with the smallest value is merged into a single mixand. This process is repeated until a specified maximum number of mixands~$N_{\textrm{max}}$ has been reached, and further merging can be performed between mixand pairs with a \ac{KL} divergence upper bound lower than a specified maximum~$B_{\textrm{max}}$. This ensures that the number of mixands does not become computationally intractable, while also ensuring that the approximation error of each merge is not significant.

\section{Methodology}\label{sec:methodology}
Before smoothing can be performed, the filtered posteriors~$p(\mathbf{x}_k|\mathbf{z}_{1:k})$ must be computed in the form of a \ac{GM} for all time steps~$0\leq k \leq T$. Although the following proposed smoothing algorithm makes no assumptions as to how these filtered estimates are obtained, an adaptive Gaussian mixture extended Kalman filter is employed in this work, and is outlined in Appendix~\ref{sec:appendix_GMfilter}. Once the filtered posteriors are obtained, then the backward filter recursion can be performed to incorporate future measurement information and produce the smoothed estimates.

\subsection{Backward Information Filter for Nonlinear Systems}
\subsubsection{Likelihood Component Mixture Representation}
In order to perform two-filter smoothing, the backward filter likelihood~$p(\mathbf{z}_{k+1:T}|\mathbf{x}_k)$ must be recursively computed for all~$0\leq k \leq T-1$ using~\eqref{eqn:twofilter_recurse_bfl}.
Crucially, this backward filter likelihood is a function over~$\mathbf{x}_k$ rather than a density over~$\mathbf{z}_{k+1:T}$, since the set of future measurements is realized and known.
This distinction motivates a representation of the backward filter likelihood over the state-space rather than the future measurement space.
A state-space representation makes the information provided by the future measurements to the current state more explicit, and ensures that the backward filter likelihood does not grow in dimension as the number of future measurements increases when recursing backwards.

Drawing from \cite{balenzuela_new_2022}, this work represents the backward filter likelihood as a mixture of likelihood components, written explicitly as
\begin{align}
    p(\mathbf{z}_{k+1:T}|\mathbf{x}_k)= \sum_{m=1}^{M_{k|k+1}} \nu_{k|k+1}^{(m)}\mathcal{L}\!\left(\mathbf{x}_{k};\bm\mu_{k|k+1}^{(m)},\bm\Lambda_{k|k+1}^{(m)}\right)
\end{align}
where each likelihood component is given by
\begin{align}
    \mathcal{L}\!(\mathbf{x};\bm\mu,\bm\Lambda) = \exp\left\{-\displaystyle\frac{1}{2}(\mathbf{x}-\bm\mu)^\top \bm\Lambda(\mathbf{x}-\bm\mu)\right\}
\end{align}
Here,~$M_{k|k+1}$ denotes the total number of likelihood components, and~$\nu^{(m)}_{k|k+1} \in \mathbb{R}$,~$\bm\mu^{(m)}_{k|k+1} \in \mathbb{R}^{n_x}$, and~$\bm\Lambda^{(m)}_{k|k+1} \in \mathbb{S}_{+}^{n_x}$ denote the weight, center, and positive semi-definite precision matrix of each likelihood component respectively. The~$k|k+1$ subscript indicates that the mixture serves as a prior for the backward filter, as it is conditioned on the state at time step~$k$ and incorporates all measurement information from time step~$k+1$ onward.

This representation of the backward filter likelihood is conceptually equivalent to a Gaussian mixture representation, with the distinction that the precision matrix~$\bm\Lambda$ of each likelihood is permitted to be singular. This allows the backward filter likelihood to capture measurement information that cannot be inverted to obtain a unique state estimate, which has been a difficulty of previous two-filter approaches \cite{Kitagawa_TwoFilterGaussianSum_1994}. However, a center~$\bm\mu$ is still used to parameterize each term rather than an information vector, as it provides a point in state-space around which to perform any approximations of the true nonlinear dynamics and measurements. When the precision matrix~$\bm\Lambda$ is nonsingular,~$\bm\mu$ corresponds to the mean of the component. This is not required for the linear case, which is why prior approaches have utilized an information vector parameterization \cite{balenzuela_new_2022}. The weights~$\bm\nu_{k|k+1}^{(m)}$ also do not generally sum to one, since the mixture is not a valid density over~$\mathbf{x}_k$ and each likelihood term is unnormalized.

With the assumed mixture form, the backward filter recursion of~\eqref{eqn:twofilter_recurse_bfl} separates naturally into
an update and a prediction step:
\begin{subequations}\label{eqn:BIF_recursion}
    \begin{align}
        p(\mathbf{z}_{k:T}|\mathbf{x}_k)
        &= p(\mathbf{z}_{k+1:T}|\mathbf{x}_{k}) p(\mathbf{z}_k|\mathbf{x}_k)
        \label{eqn:BIF_update_step} \\
        p(\mathbf{z}_{k:T}|\mathbf{x}_{k-1})
        &= \int p(\mathbf{z}_{k:T}|\mathbf{x}_{k})
           p(\mathbf{x}_k|\mathbf{x}_{k-1})
           \textrm{d}\mathbf{x}_k
        \label{eqn:BIF_prediction_step}
    \end{align}
\end{subequations}
The update and prediction steps are discussed separately in Sections~\ref{sec:filter Update} and~\ref{sec:filter Predict} respectively.

\subsubsection{Backward Information Filter Update}\label{sec:filter Update}
With the assumed mixture form of the backward filter likelihood and nonlinear measurement model of~\eqref{eqn:meas_model}, the measurement update for the backward filter likelihood from~\eqref{eqn:BIF_update_step} is given by
\begin{equation}
    p(\mathbf{z}_{k:T}|\mathbf{x}_k) = \sum_{m=1}^{M_{k|k+1}}
    \nu_{k|k+1}^{(m)}
    \mathcal{L}\!\left(\mathbf{x}_k;\bm\mu_{k|k+1}^{(m)},\bm\Lambda_{k|k+1}^{(m)}\right)
    \mathcal{N}\!\left(\mathbf{z}_k;\mathbf{h}_k(\mathbf{x}_k),\mathbf{R}_k\right)
\end{equation}
The nonlinear measurement update can be approximated for each component of the backward filter likelihood in a number of ways, including linearization, sigma point approximation, and cubature methods. This work employs a linearization-based approach analogous to the \ac{EKF}, whereby the nonlinear measurement function is approximated via a first-order Taylor series expansion as
\begin{align}
    \mathbf{h}_k\left(\mathbf{x}_k\right) \approx \mathbf{h}_k\left(\bm\mu_{k|k+1}^{(m)}\right) + \mathbf{H}_k|_{\bm\mu_{k|k+1}^{(m)}}\left(\mathbf{x}_k - \bm\mu_{k|k+1}^{(m)}\right) \label{eqn:first_order_meas_func}
\end{align}

This expansion highlights the utility of using a state-space center to parameterize the likelihood term rather than an information vector, as otherwise the choice of a suitable linearization point would be a non-trivial matter. With this, the measurement update for the backward filter likelihood is computed as
\begin{subequations}\label{eqn:BIF_update}
    \begin{align}
        p(\mathbf{z}_{k:T}|\mathbf{x}_k) &\approx \sum_{m=1}^{M_{k|k+1}}\nu_{k|k+1}^{(m)}\mathcal{L}\!\left(\mathbf{x}_k;\bm\mu_{k|k+1}^{(m)},\bm\Lambda_{k|k+1}^{(m)}\right)\mathcal{N}\!\left(\mathbf{z}_k;\mathbf{h}_k\left(\bm\mu_{k|k+1}^{(m)}\right) + \mathbf{H}_k|_{\bm\mu_{k|k+1}^{(m)}}\left(\mathbf{x}_k - \bm\mu_{k|k+1}^{(m)}\right),\mathbf{R}_k\right) \\
        &= \sum_{m=1}^{M_{k|k}} \nu_{k|k}^{(m)}\mathcal{L}\!\left(\mathbf{x}_k;\bm\mu_{k|k}^{(m)},\bm\Lambda_{k|k}^{(m)}\right) \label{eqn:BIF_update_2}
    \end{align}
\end{subequations}
where
\begin{subequations}\label{eqn:BIF_update_params}
    \begin{align}
        M_{k|k} &= M_{k|k-1} \\
        \bm\Lambda_{k|k}^{(m)} &= \bm\Lambda_{k|k+1}^{(m)} + \left(\mathbf{H}_k|_{\bm\mu_{k|k+1}^{(m)}}\right)^\top \mathbf{R}_k^{-1}\mathbf{H}_k|_{\bm\mu_{k|k+1}^{(m)}}\\
        \bm\mu_{k|k}^{(m)} &= \bm\mu_{k|k+1}^{(m)} + \left(\bm\Lambda_{k|k}^{(m)}\right)^\dagger\left(\mathbf{H}_k|_{\bm\mu_{k|k+1}^{(m)}}\right)^\top \mathbf{R}_k^{-1}\left(\mathbf{z}_k - \mathbf{h}_k\left(\bm\mu_{k|k+1}^{(m)}\right)\right)\\
        \nu_{k|k}^{(m)} &= \nu_{k|k+1}^{(m)}\mathcal{N}\!\left(\mathbf{z}_k;\mathbf{h}_k\left(\bm\mu_{k|k+1}^{(m)}\right),\mathbf{R}_k\right)\exp\left\{-\displaystyle\frac{1}{2}\left(\mathbf{z}_k-\mathbf{h}_k\left(\bm\mu_{k|k+1}^{(m)}\right)\right)^\top\mathbf{R}_k^{-1}\mathbf{H}_k|_{\bm\mu_{k|k+1}^{(m)}}\left(\bm\mu_{k|k+1}^{(m)} - \bm\mu_{k|k}^{(m)}\right)\right\}
    \end{align}
\end{subequations}
and~$(\cdot)^\dagger$ denotes the Moore-Penrose inverse.
A proof of this measurement update is given in Appendix~\ref{sec:appendix_upd}.
The subscript~$k|k$ in~\eqref{eqn:BIF_update_2} indicates that the mixture serves as a posterior for the backward filter, as it is conditioned on the state at time step~$k$ and incorporates all measurement information from time step~$k$ onward.
In the case where~$\bm\Lambda_{k|k}^{(m)}$ is singular, an infinite number of solutions exist for the posterior center~$\bm\mu_{k|k}^{(m)}$ due to the ambiguity in nullspace directions. The use of the Moore-Penrose inverse provides the unique solution with the minimum-norm adjustment to the center, which ensures further linearization points remain close to the filtered distribution.
Importantly, the backward filter likelihood remains a likelihood mixture under the linearized measurement update.

Although the measurement update based on a direct first-order approximation of the measurement function is generally valid, it can perform poorly when the measurement function is highly nonlinear over the support of a component. This is especially problematic for singular likelihood components, as an inaccurate measurement update can shift the center outside of the support of the filtered distribution in directions of zero information, resulting in unreasonable linearization points at subsequent smoothing time steps. Several techniques from standard nonlinear Kalman filtering can be applied to improve the accuracy of the measurement update, including the \ac{IEKF} update \cite{gelb_Applied_1974,barshalom_estimation_2002,sarkka_bayesian_2023}, the Bayesian recursive update \cite{michaelson2023bayesianrecursiveupdateensemble}, and second-order update \cite{gelb_Applied_1974,sarkka_bayesian_2023}. For this work, the Bayesian recursive update of \cite{michaelson2023bayesianrecursiveupdateensemble} is employed due to its adaptability and its straightforward implementation into the existing methodology.

The Bayesian recursive update utilizes the following representation of the component-wise update:
\begin{subequations}
    \begin{align}
        \mathcal{L}\!\left(\mathbf{x}_k;\bm\mu_{k|k+1}^{(m)},\bm\Lambda_{k|k+1}^{(m)}\right) \mathcal{N}\!\left(\mathbf{z}_k;\mathbf{h}_k(\mathbf{x}_k),\mathbf{R}_k\right) &= \mathcal{L}\!\left(\mathbf{x}_k;\bm\mu_{k|k+1}^{(m)},\bm\Lambda_{k|k+1}^{(m)}\right) \prod_{l=1}^L \mathcal{N}\!\left(\mathbf{z}_k;\mathbf{h}_k\left(\mathbf{x}_k\right),\mathbf{R}_k\right)^{c_l} \\
        &= \mathcal{L}\!\left(\mathbf{x}_k;\bm\mu_{k|k+1}^{(m)},\bm\Lambda_{k|k+1}^{(m)}\right) \prod_{l=1}^L \mathcal{N}\!\left(\mathbf{z}_k;\mathbf{h}_k\left(\mathbf{x}_k\right),\mathbf{R}_k/c_l\right)
    \end{align}
\end{subequations}
where~$c_l > 0$, and~$\sum_l c_l = 1$.
This allows a single measurement update to be replaced by~$L$ gradual updates using an inflated measurement noise covariance~$\mathbf{R}_k/c_l$.
This series of measurement updates is given explicitly by
\begin{subequations}
    \begin{align}
        \bm\Lambda_{k|k}^{(m)(0)} &= \bm\Lambda_{k|k+1}^{(m)},\quad \bm\mu_{k|k}^{(m)(0)} = \bm\mu_{k|k+1}^{(m)(0)},\quad \nu_{k|k}^{(m)(0)} = \nu_{k|k+1}^{(m)} \\
        \bm\Lambda_{k|k}^{(m)(l)} &= \bm\Lambda_{k|k}^{(m)(l-1)} + \left(\mathbf{H}_k|_{\bm\mu_{k|k}^{(m)(l-1)}}\right)^\top \left(\mathbf{R}_k/c_l\right)^{-1}\mathbf{H}_k|_{\bm\mu_{k|k}^{(m)(l-1)}} \\
        \bm\mu_{k|k}^{(m)(l)} &= \bm\mu_{k|k}^{(m)(l-1)} + \left(\bm\Lambda_{k|k}^{(m)(l)}\right)^\dagger \left(\mathbf{H}_k|_{\bm\mu_{k|k}^{(m)(l-1)}}\right)^\top \left(\mathbf{R}_k/c_l\right)^{-1}\left(\mathbf{z}_k - \mathbf{h}_k\left(\bm\mu_{k|k}^{(m)(l-1)}\right)\right) \\
        \nu_{k|k}^{(m)(l)} &= \nu_{k|k}^{(m)(l-1)}\mathcal{N}\!\left(\mathbf{z}_k;\mathbf{h}_k\left(\bm\mu_{k|k}^{(m)(l-1)}\right),\mathbf{R}_k/c_l\right) \\
        &\quad\cdot \exp\left\{-\displaystyle\frac{1}{2}\left(\mathbf{z}_k-\mathbf{h}_k\left(\bm\mu_{k|k}^{(m)(l-1)}\right)\right)^\top\left(\mathbf{R}_k/c_l\right)^{-1}\mathbf{H}_k|_{\bm\mu_{k|k}^{(m)(l-1)}}\left(\bm\mu_{k|k+1}^{(m)} - \bm\mu_{k|k}^{(m)}\right)\right\} \nonumber \\
        \bm\Lambda_{k|k}^{(m)} &= \bm\Lambda_{k|k}^{(m)(L)},\quad \bm\mu_{k|k}^{(m)} = \bm\mu_{k|k}^{(m)(L)},\quad \nu_{k|k}^{(m)} = \nu_{k|k}^{(m)(L)}
    \end{align}
\end{subequations}
Importantly, the previous step's posterior center~$\bm\mu_{k|k}^{(m)(l-1)}$ is used as the linearization point for the next step, and this repeated relinearization results in a more accurate representation of the true nonlinear measurement update. A variable step size of $c_l = \displaystyle\frac{l^3 - (l-1)^3}{L^3}$ is used to ensure the weight of the update increases as the linearization point gains agreement with the measurement. This work uses~$L=20$ total steps for all measurement updates. Adaptive step size selection based on the size of the measurement update can also be employed to improve performance and computational efficiency \cite{fife_discrete_2025}.

The initialization of the backward filter likelihood is a special case of a measurement update, wherein the final measurement likelihood~$p(\mathbf{z}_T|\mathbf{x}_T)$ must be cast as a likelihood mixture. Although this cannot be accomplished exactly for nonlinear measurement models, the final filtered posterior provides a natural set of points in state-space to center a linearized approximation, as accurate knowledge of $p(\mathbf{z}_T | \mathbf{x}_T)$ far from the filtered state is largely irrelevant to the smoothing solution. Multiplying $p(\mathbf{z}_T | \mathbf{x}_T)$ by unity, expressed as a sum of zero-information components centered at and weighted by the final filtered posterior, yields the exact pre-update representation
\begin{equation}\label{eqn:BIF_init_pre}
    p(\mathbf{z}_T|\mathbf{x}_T) =
    \left(\sum_{n=1}^{N_T} w_{T}^{(n)}
    \mathcal{L}\!\left(\mathbf{x}_T;\,\mathbf{m}_T^{(n)},\mathbf{0}_{n_x\times n_x}\right)
    \right)
    \mathcal{N}\!\left(\mathbf{z}_T;\,\mathbf{h}_T(\mathbf{x}_T),\mathbf{R}_T\right)
\end{equation}
where~$\mathcal{L}(\mathbf{x}_T;\mathbf{m}_T^{(n)},\mathbf{0}_{n_x\times n_x})=1$ so the equality is
exact.
The centers and weights used in this unity term only matter insofar as they provide and weight the initial linearization points for the backward filter likelihood.
Using the means and weights of the final filtered posterior ensures that the final measurement likelihood is accurately approximated over the posterior's support.
The measurement update methodology of~\eqref{eqn:BIF_update}-\eqref{eqn:BIF_update_params} is then applied, yielding the initialized backward filter likelihood in standard mixture form
\begin{equation}\label{eqn:BIF_init}
    p(\mathbf{z}_T|\mathbf{x}_T) \approx
    \sum_{n=1}^{N_T} \nu_{T|T}^{(n)}\,
    \mathcal{L}\!\left(\mathbf{x}_T;\,\bm\mu_{T|T}^{(n)},\,\bm\Lambda_{T|T}^{(n)}\right).
\end{equation}
If more initial mixands are required to ensure an accurate approximation of the final measurement likelihood, the mixands of the final filtered posterior can be
split into additional mixands according to their uncertainty and the nonlinearity of
$\mathbf{h}_T$ \cite{kulik2025NonlinearityUncertaintyInformed}.

Figure~\ref{fig:BIF_init_example} demonstrates this approximation of the final measurement likelihood on a two-dimensional example, where a range measurement is taken with respect to the origin at the final time. The red dots denote the means and centers of the components of the filtered distribution and final measurement likelihood respectively. The approximation of the measurement likelihood as a likelihood component mixture is clearly only locally valid, but it accurately represents the true measurement likelihood over the support of the filtered posterior. The centers of the approximation are also within the support of the posterior, meaning that further linearization approximations will remain valid.

\newlength{\figrowwidth}
\setlength{\figrowwidth}{0.95\linewidth}
\begin{figure}[bht!]
    \par\medskip
    \centering
    \begin{subfigure}{0.33\figrowwidth}
        \begin{overpic}[width=\linewidth]{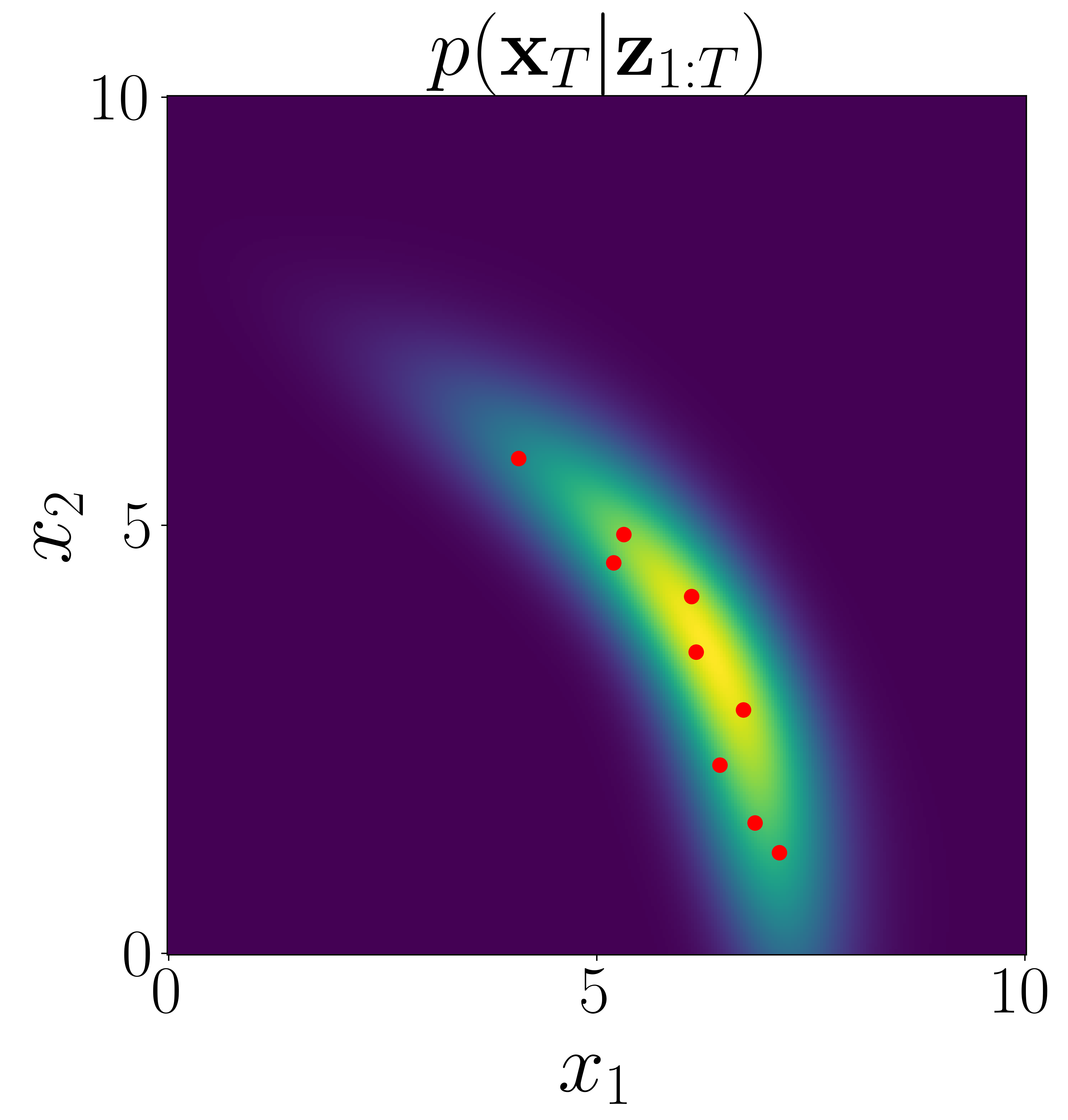}
            \put(3,95){\large (a)}
        \end{overpic}
        \label{fig:BIF_init_example_filter}
    \end{subfigure} %
    \begin{subfigure}{0.33\figrowwidth}
        \begin{overpic}[width=\linewidth]{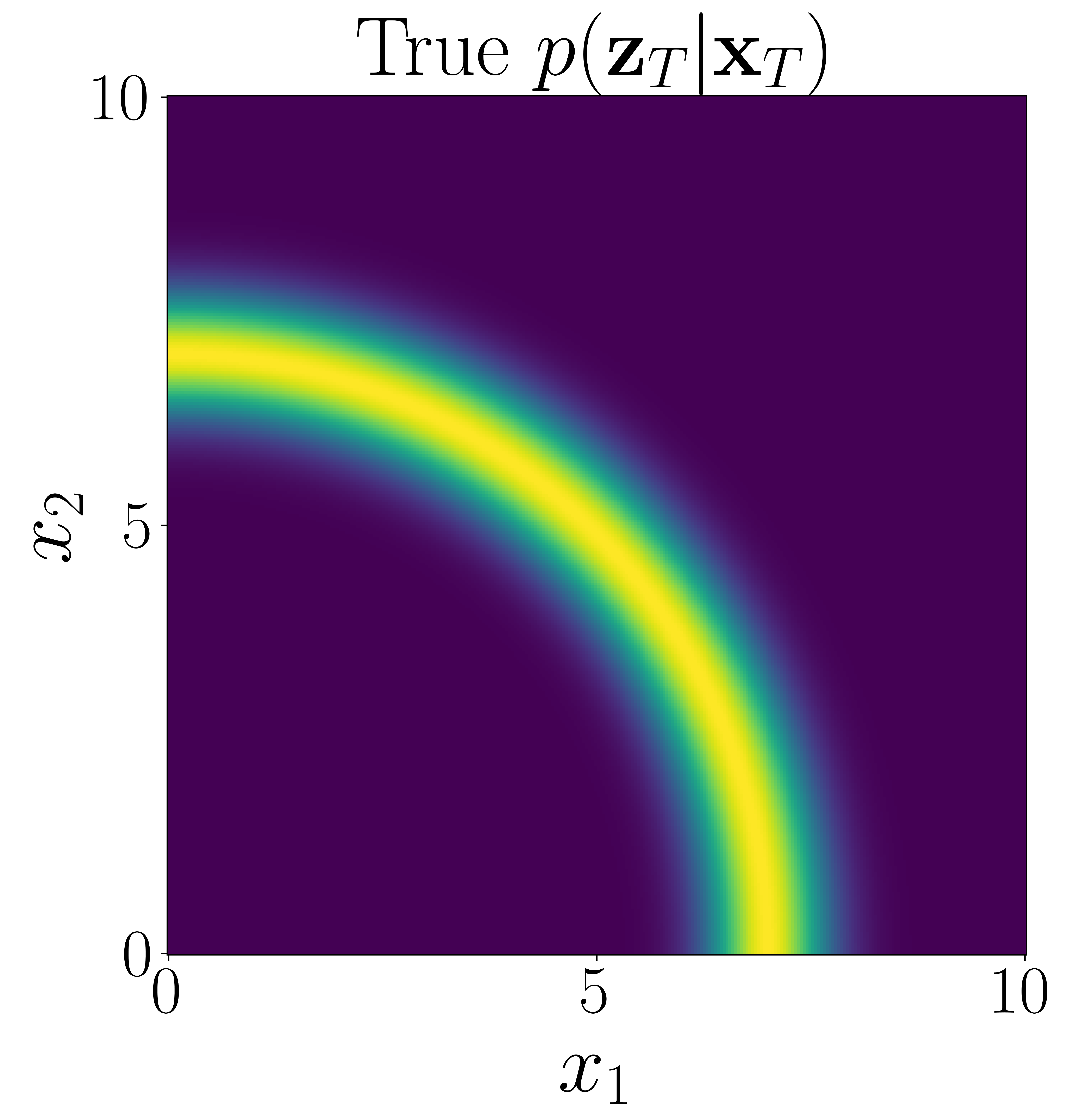}
            \put(3,95){\large (b)}
        \end{overpic}
        \label{fig:BIF_init_example_trueMeasLikelihood}
    \end{subfigure} %
    \begin{subfigure}{0.33\figrowwidth}
        \begin{overpic}[width=\linewidth]{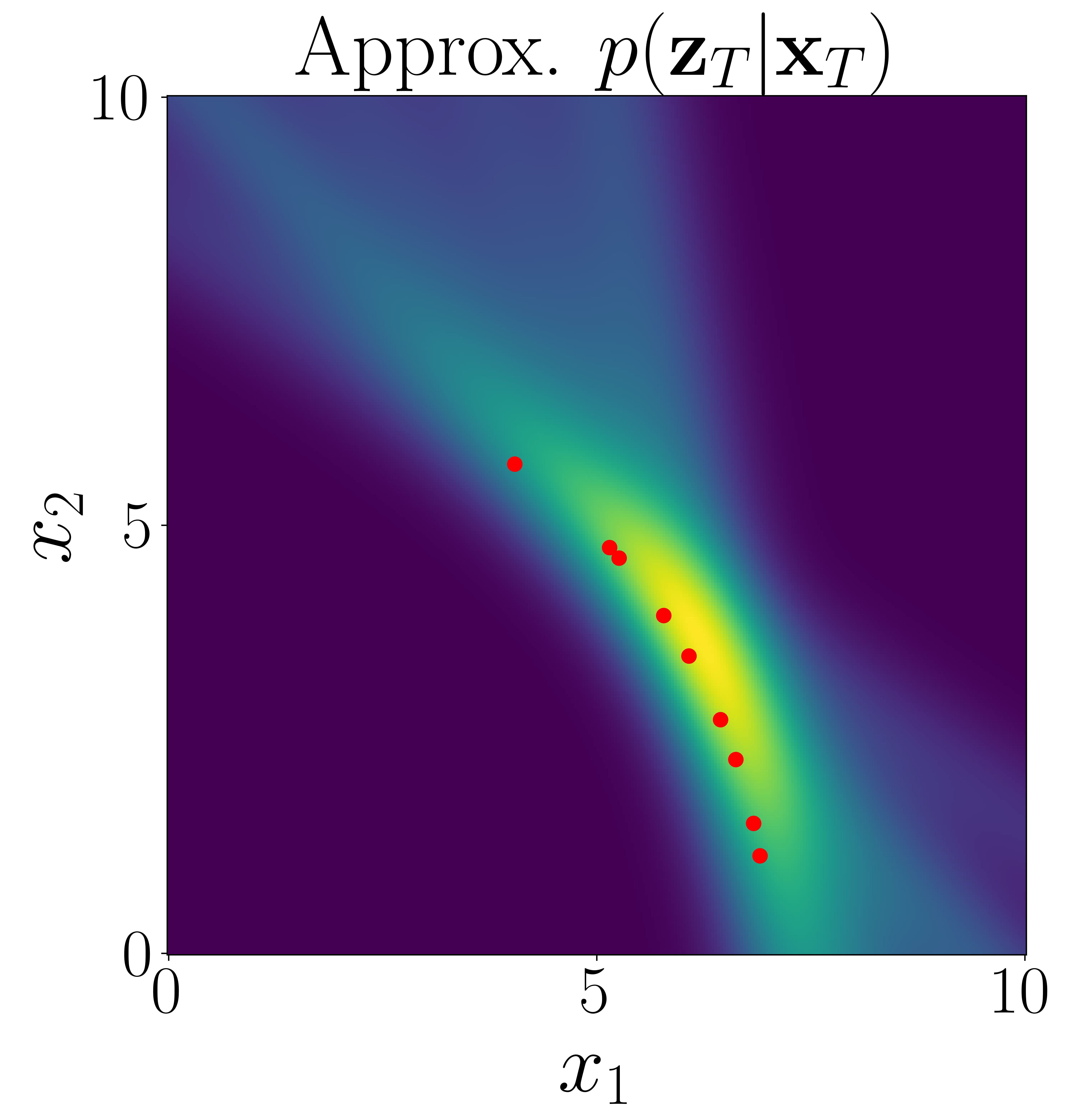}
            \put(3,95){\large (c)}
        \end{overpic}
        \label{fig:BIF_init_example_approxMeasLikelihood}
    \end{subfigure}

    \caption{Backward filter initialization on range measurement example.}
    \label{fig:BIF_init_example}
\end{figure}

\subsubsection{Backward Information Filter Prediction}\label{sec:filter Predict}
Following the update step, the backward filter likelihood must be propagated from time step~$k$ to~$k-1$ to allow smoothing to be performed at time step~$k-1$.
The prediction step from \eqref{eqn:BIF_prediction_step}  for the backward filter likelihood is given by
\begin{equation}
        p(\mathbf{z}_{k:T}|\mathbf{x}_{k-1})
        = \sum_{m=1}^{M_{k|k}} \nu_{k|k}^{(m)} \int \mathcal{L}\!\left(\mathbf{x}_k;\bm\mu_{k|k}^{(m)},\bm\Lambda_{k|k}^{(m)}\right)\mathcal{N}\!\left(\mathbf{x}_{k};\bm\varphi_{k-1}^k(\mathbf{x}_{k-1}),\mathbf{Q}_{k-1}|_{\mathbf{x}_{k-1}}\right)\textrm{d}\mathbf{x}_{k}
\end{equation}
where the notation~$\bm{\varphi}_{k-1}^k(\mathbf{x}_{k-1}) \triangleq \bm{\varphi}_{t_k-t_{k-1}}(\mathbf{x}_{k-1},t_{k-1})$ is used for compactness.

The nonlinear propagation function can be approximated using a first-order Taylor series expansion as
\begin{align}
    \bm\varphi_{k-1}^k(\mathbf{x}_{k-1}) \approx \bm\mu_{k|k}^{(m)} + \left.\bm\Phi_{k-1}^{k}\right|_{\bm\varphi_{k}^{k-1}\left(\bm\mu_{k|k}^{(m)}\right)}\left(\mathbf{x}_{k-1} - \bm\varphi_{k}^{k-1}\left(\bm\mu_{k|k}^{(m)}\right)\right)
\end{align}
where the \ac{STM} is denoted by~$\bm\Phi_{k-1}^k \triangleq \bm\Phi(t_k,t_{k-1})$ for compactness and~$\bm\varphi_k^{k-1}(\mathbf{x}_k)$ denotes the inverse of the dynamics solution flow from~$t_k$ to~$t_{k-1}$.

With this, the prediction step for the backward filter likelihood can be performed as
\begin{subequations}\label{eqn:BIF_prop}
    \begin{align}
        p(\mathbf{z}_{k:T}|\mathbf{x}_{k-1}) &\approx \sum_{m=1}^{M_{k|k}} \nu_{k|k}^{(m)} \int \mathcal{L}\!\left(\mathbf{x}_k;\bm\mu_{k|k}^{(m)},\bm\Lambda_{k|k}^{(m)}\right) \\
        &\qquad\qquad\qquad \cdot\mathcal{N}\!\left(\mathbf{x}_{k};\bm\mu_{k|k}^{(m)} + \left.\bm\Phi_{k-1}^{k}\right|_{\bm\varphi_{k}^{k-1}\left(\bm\mu_{k|k}^{(m)}\right)}\left(\mathbf{x}_{k-1} - \bm\varphi_{k}^{k-1}\left(\bm\mu_{k|k}^{(m)}\right)\right),\left.\mathbf{Q}_{k-1}\right|_{\bm\varphi_{k}^{k-1}\left(\bm\mu_{k|k}^{(m)}\right)}\right) \textrm{d}\mathbf{x}_{k} \nonumber \\
        &= \sum_{m=1}^{M_{k-1|k}} \nu_{k-1|k}^{(m)}\mathcal{L}\!\left(\mathbf{x}_{k-1};\bm\mu_{k-1|k}^{(m)},\bm\Lambda_{k-1|k}^{(m)}\right)
    \end{align}
\end{subequations}
where
\begin{subequations}\label{eqn:BIF_prop_params}
    \begin{align}
        M_{k-1|k} &= M_{k|k} \\
        \bm\Lambda_{k-1|k}^{(m)} &= \left(\left.\bm\Phi_{k-1}^{k}\right|_{\bm\varphi_{k}^{k-1}\left(\bm\mu_{k|k}^{(m)}\right)}\right)^\top \bm\Lambda_{k|k}^{(m)}\left(\mathbf{I} + \left.\mathbf{Q}_{k-1}\right|_{\bm\varphi_{k}^{k-1}\left(\bm\mu_{k|k}^{(m)}\right)}\bm\Lambda_{k|k}^{(m)}\right)^{-1} \left.\bm\Phi_{k-1}^{k}\right|_{\bm\varphi_{k}^{k-1}\left(\bm\mu_{k|k}^{(m)}\right)} \\
        \bm\mu_{k-1|k}^{(m)} &= \bm\varphi_{k}^{k-1}\left(\bm\mu_{k|k}^{(m)}\right) \\
        \nu_{k-1|k}^{(m)} &= \nu_{k|k}^{(m)}\left|\mathbf{I}+\left.\mathbf{Q}_{k-1}\right|_{\bm\varphi_{k}^{k-1}\left(\bm\mu_{k|k}^{(m)}\right)}\bm\Lambda_{k|k}^{(m)}\right|^{-1/2}
    \end{align}
\end{subequations}
A proof of this prediction step is given in Appendix~\ref{sec:appendix_prop}.

\subsubsection{Process Noise Discretization}\label{sec noise dicscretization}

In the evaluation of the predicted backwards filter likelihood $p(\mathbf{z}_{k:T}|\mathbf{x}_{k-1})$, the discrete-time process noise matrix $\left.\mathbf{Q}_{k-1}\right|_{\bm\varphi_{k}^{k-1}\left(\bm\mu_{k|k}^{(m)}\right)}$ appears in the product with the transition density~$p(\mathbf{x}_k|\mathbf{x}_{k-1})$.
  To clarify its derivation, it is convenient to first simplify the notation and consider the standard continuous-to-discrete integration of the process noise covariance $\mathbf{Q}_k$ for a single Gaussian component over the generic time interval $[t_k, t_{k+1}]$.
The matrix  $\mathbf{Q}_k$  is computed by solving the continuous-time stochastic integral via Van Loan's matrix exponential method~\cite{VanLoan_processNoise_1978}, which yields
\begin{equation}
    \mathbf{Q}_k = \bm{\Psi}_{k,22}^\top \bm{\Psi}_{k,12}, \qquad
    \bm{\Psi}_k
    = e^{\bm{\Theta}_k (t_{k+1}-t_k)}
    = \begin{bmatrix}
        \bm{\Psi}_{k,11} & \bm{\Psi}_{k,12} \\
        \mathbf{0}       & \bm{\Psi}_{k,22}
    \end{bmatrix},
    \qquad
    \bm{\Theta}_k =
    \begin{bmatrix}
        -\mathbf{A}_k^\top & \bm{\Gamma}(t_k) \mathbf{Q}_c \bm{\Gamma}(t_k)^\top \\
        \mathbf{0}         & \mathbf{A}_k
    \end{bmatrix},
\end{equation}
where $\mathbf{A}_k$ is the Jacobian matrix of the dynamics $\mathbf{f}(\cdot)$ evaluated at the beginning of the integration interval~$t_k$ and held constant via a zeroth-order-hold approximation. To mitigate the linearization and integration errors introduced by a single-step approximation over long propagation intervals, the time span $[t_k, t_{k+1}]$ is subdivided into smaller sub-steps, allowing the discrete-time process noise to be recursively accumulated. Expanding $\mathbf{Q}_{k,n}$ over the time partition $t_k = \tau_0 \leq \tau_1 \leq \cdots \leq \tau_N = t_{k+1}$ yields
\begin{align}
    \mathbf{Q}_{k,n}
    &= \int_{\tau_0}^{\tau_n}
        \bm{\Phi}(\tau_n,\sigma)\,\bm{\Gamma}(\tau)\,\mathbf{Q}_c\,
        \bm{\Gamma}(\tau)^\top\, \bm{\Phi}^\top(\tau_n,\sigma)\,
       \mathrm{d}\sigma \notag\\
    &= \int_{\tau_0}^{\tau_{n-1}}
        \bm{\Phi}(\tau_n,\sigma)\,\bm{\Gamma}(\tau)\,\mathbf{Q}_c\,
        \bm{\Gamma}(\tau)^\top\, \bm{\Phi}^\top(\tau_n,\sigma)\,
       \mathrm{d}\sigma
     + \int_{\tau_{n-1}}^{\tau_n}
        \bm{\Phi}(\tau_n,\sigma)\,\bm{\Gamma}(\tau)\,\mathbf{Q}_c\,
        \bm{\Gamma}(\tau)^\top\, \bm{\Phi}^\top(\tau_n,\sigma)\,
       \mathrm{d}\sigma \notag\\
    &= \bm{\Phi}(\tau_n,\tau_{n-1})
       \underbrace{\left(
         \int_{\tau_0}^{\tau_{n-1}}
           \bm{\Phi}(\tau_{n-1},\sigma)\,\bm{\Gamma}(\tau)\,\mathbf{Q}_c\,
           \bm{\Gamma}(\tau)^\top\, \bm{\Phi}^\top(\tau_{n-1},\sigma)\,
         \mathrm{d}\sigma
       \right)}_{\mathbf{Q}_{k,n-1}}
       \bm{\Phi}^\top(\tau_n,\tau_{n-1}) \notag\\
    &\quad
     + \int_{\tau_{n-1}}^{\tau_n}
        \bm{\Phi}(\tau_n,\sigma)\,\bm{\Gamma}(\tau)\,\mathbf{Q}_c\,
        \bm{\Gamma}(\tau)^\top\, \bm{\Phi}^\top(\tau_n,\sigma)\,
       \mathrm{d}\sigma,
\end{align}
which leads to the following recursive accumulation scheme:
\begin{subequations}\label{eqn:process_noise_accumulation}
    \begin{align}
        \mathbf{Q}_{k,0} &= \mathbf{0}_{n_x\times n_x}, \\
        \mathbf{Q}_{k,n} &= \bm{\Phi}(\tau_{n},\tau_{n-1})\,
                             \mathbf{Q}_{k,n-1}\,
                             \bm{\Phi}^\top(\tau_{n},\tau_{n-1})
                           + \int_{\tau_{n-1}}^{\tau_n}
                               \bm{\Phi}(\tau_n,\sigma)\,\bm{\Gamma}(\tau)\,
                               \mathbf{Q}_c\,\bm{\Gamma}(\tau)^\top\,
                               \bm{\Phi}^\top(\tau_n,\sigma)\,
                             \mathrm{d}\sigma, \\
        \mathbf{Q}_{k}   &= \mathbf{Q}_{k,N},
    \end{align}
\end{subequations}
where  the inner integral at each sub-step is evaluated via Van Loan's method applied to the sub-interval $[\tau_{m-1}, \tau_m]$. This process noise discretization applies to both the filtering and smoothing operations.

\subsection{Smoothing Step}
Once the backward filter likelihood is computed recursively for all time steps~$0\leq k\leq T-1$, the smoothed density is a simple product of mixtures. Given the backward filter likelihood and the filtered posterior, then by~\eqref{Eq:two_filter_update}, the smoothed density becomes
\begin{subequations}\label{eqn:smoothed_dist}
    \begin{align}
        p(\mathbf{x}_{k-1}|\mathbf{z}_{1:T}) &= p(\mathbf{x}_{k-1}|\mathbf{z}_{1:k-1})\frac{p(\mathbf{z}_{k:T}|\mathbf{x}_{k-1})}{p(\mathbf{z}_{k:T}|\mathbf{z}_{1:k-1})} \\
        &= \frac{1}{p(\mathbf{z}_{k:T}|\mathbf{z}_{1:k-1})}\left(\sum_{n=1}^{N_{k-1}} w_{k-1}^{(n)}\mathcal{N}\!\left(\mathbf{x}_{k-1};\mathbf{m}_{k-1}^{(n)},\mathbf{P}_{k-1}^{(n)}\right)\right)\left(\sum_{m=1}^{M_{k-1|k}}\nu_{k-1|k}^{(m)}\mathcal{L}\!\left(\mathbf{x}_{k-1};\bm\mu_{k-1|k}^{(m)},\bm\Lambda_{k-1|k}^{(m)}\right)\right) \\
        &= \sum_{n=1}^{N_{k-1}}\sum_{m=1}^{M_{k-1|k}} w_{k-1|T}^{(nm)}\mathcal{N}\!\left(\mathbf{x}_{k-1};\mathbf{m}_{k-1|T}^{(nm)},\mathbf{P}_{k-1|T}^{(nm)}\right)
    \end{align}
\end{subequations}
where
\begin{subequations}\label{eqn:smoothed_params}
    \begin{align}
        \mathbf{P}_{k-1|T}^{(nm)} &= \left(\left(\mathbf{P}_{k-1}^{(n)}\right)^{-1} + \bm\Lambda_{k-1|k}\right)^{-1} \\
        \mathbf{m}_{k-1|T}^{(nm)} &= \mathbf{P}_{k-1|T}^{(nm)}\left(\left(\mathbf{P}_{k-1}^{(n)}\right)^{-1}\mathbf{m}_{k-1}^{(n)} + \bm\Lambda_{k-1|k}^{(m)}\bm\mu_{k-1|k}^{(m)}\right) \\
        \bar{w}_{k-1|T}^{(nm)} &= w_{k-1}^{(n)}\nu_{k-1|k}^{(m)}\left|\mathbf{I}+\mathbf{P}_{k-1}^{(n)}\bm\Lambda_{k-1|k}^{(m)}\right|^{-1/2}\\
        &\cdot\exp\left\{\displaystyle\frac{1}{2}\left[\left(\mathbf{m}_{k-1|T}^{(nm)}\right)^\top\left(\mathbf{P}_{k-1|T}^{(nm)}\right)^{-1}\mathbf{m}_{k-1|T}^{(nm)} - \left(\mathbf{m}_{k-1}^{(n)}\right)^\top \left(\mathbf{P}_{k-1}^{(n)}\right)^{-1}\mathbf{m}_{k-1}^{(n)} - \left(\bm\mu_{k-1|k}^{(m)}\right)^\top \bm\Lambda_{k-1|k}^{(m)}\bm\mu_{k-1|k}^{(m)}\right]\right\} \nonumber \\
        w_{k-1|T}^{(nm)} &= \frac{\bar{w}_{k-1|T}^{(nm)}}{\sum_{i=1}^{N_{k-1}} \sum_{j=1}^{M_{k-1|k}} \bar{w}_{k-1|T}^{(ij)}}
    \end{align}
\end{subequations}
A proof of this smoothing operation is given in Appendix~\ref{sec:appendix_smooth}. The smoothed weights are normalized to sum to one in this smoothing operation to ensure that the smoothed distribution is a valid \ac{pdf}. Although this smoothing operation results in a large number of mixands if both the filtered distribution and backward filter likelihood have many components, most of the resulting mixands will have negligible weight and can be safely pruned from the smoothed distribution. Mixands can either be pruned if their normalized weight is below a specified threshold, or the lowest weight mixands can be pruned until their total weight reaches a specified threshold.

As a result, the backward filter likelihood is initialized with~\eqref{eqn:BIF_init}, recursively updated and propagated with~\eqref{eqn:BIF_update}-\eqref{eqn:BIF_update_params} and~\eqref{eqn:BIF_prop}-\eqref{eqn:BIF_prop_params} respectively, and the smoothed distribution is obtained with~\eqref{eqn:smoothed_dist}-\eqref{eqn:smoothed_params}.
Together, these operations constitute a complete \ac{GM} smoothing algorithm for nonlinear systems.
However, the accuracy of the approximations made in the update and prediction equations depend largely on the nonlinearity of the corresponding functions over the effective support of each component.
Gaussian mixture splitting improves the approximation accuracy by adaptively increasing the mixture resolution in regions of high nonlinearity and/or uncertainty. To combat the accompanying growth in mixture size, merging algorithms combine similar mixands to maintain computational tractability. In the following section, the translation of these techniques to the backward filter gives rise to \emph{adaptive} \ac{GM} smoothing, which is shown to enable accurate Bayesian estimation even in strongly nonlinear and non-Gaussian settings.

\subsection{Adaptive Mixture Techniques for Backward Information Filter}

\subsubsection{Likelihood Component Splitting}
Splitting is performed in the backward filter prior to both the prediction and update steps in order to improve the accuracy of their approximations. As with the splitting procedure used in the forward filter and described in Section~\ref{sec:background_split}, likelihood components are selected for splitting based on their uncertainty, the nonlinearity of the associated function for the prediction or update, and their contribution to the total distribution's probability mass. If a component is selected for splitting, then it is split along the direction of highest nonlinearity and/or uncertainty as determined by a split heuristic. Splitting is also performed recursively to ensure sufficient mixture resolution in cases where a single split does not provide adequate fidelity.

Because the likelihood components being used are structurally very similar to Gaussians, their splitting and merging are accomplished in much the same manner. The main caveat to this is that the precision matrix~$\bm\Lambda$ is allowed to be singular following initialization when there is not yet enough future measurement information to fully constrain the state. In these cases, a scaled Gaussian representation of the likelihood component cannot be directly recovered. However, splitting and merging can still be formulated on the range of a likelihood component's precision matrix, which can be obtained using a \ac{SVD} as
\begin{align}
    \bm\Lambda = \begin{bmatrix}
        \mathbf{U}_r & \mathbf{U}_0
    \end{bmatrix} \begin{bmatrix}
        \bm\Sigma & \mathbf{0} \\
        \mathbf{0} & \mathbf{0}
    \end{bmatrix} \begin{bmatrix}
        \mathbf{U}_r^\top \\
        \mathbf{U}_0^\top
    \end{bmatrix} \label{eqn:SVD}
\end{align}
where the columns of~$\mathbf{U}_r$ form an orthonormal basis for the range of~$\bm\Lambda$, the columns of~$\mathbf{U}_0$ form an orthonormal basis for the nullspace of~$\bm\Lambda$, and~$\bm\Sigma$ is a diagonal matrix of the non-zero singular values of~$\bm\Lambda$. By this decomposition, a likelihood component with a singular precision matrix can be rewritten as a lower-dimensional likelihood component with a full-rank precision matrix as
\begin{subequations}\label{eqn:likelihood_lowdim}
    \begin{align}
        \mathcal{L}\!\left(\mathbf{x};\bm\mu,\bm\Lambda\right) &=\mathcal{L}\!\left(\tilde{\mathbf{x}};\bm\eta,\bm\Sigma\right) \\
        \tilde{\mathbf{x}} &= \mathbf{U}_r^\top \mathbf{x} \\
        \bm\eta &= \mathbf{U}_r^\top \bm\mu \\
        \bm\Sigma &= \mathbf{U}_r^\top \bm\Lambda \mathbf{U}_r
    \end{align}
\end{subequations}

Once a likelihood component is written in this full-rank form, it can be equivalently represented by a scaled Gaussian as
\begin{align}
    \mathcal{L}\!\left(\tilde{\mathbf{x}};\bm\eta,\bm\Sigma\right) &= \left|\displaystyle\frac{1}{2\pi}\bm\Sigma\right|^{-1/2}\mathcal{N}\!\left(\tilde{\mathbf{x}};\bm\eta,\bm\Sigma^{-1}\right) \label{eqn:likelihood2Gaussian}
\end{align}
With this representation, standard directional Gaussian splitting can now be employed according to~\eqref{eqn:multivariate_split}, resulting in
\begin{align}
    \mathcal{N}\!\left(\tilde{\mathbf{x}};\bm\eta,\bm\Sigma^{-1}\right) &\approx \sum_{i=1}^L w^{(i)} \mathcal{N}\!\left(\tilde{\mathbf{x}},\bm\eta^{(i)},\left(\bm\Sigma^{(i)}\right)^{-1}\right)
\end{align}
The Sherman-Morrison formula can be applied to the split covariance formula~\eqref{eqn:multivariate_split_cov} so that no covariance matrices need to be explicitly computed, which is given in terms of the split direction~$\hat{\bm\delta}^*$ as
\begin{align}
    \bm\Sigma^{(i)} &= \displaystyle\frac{\sum_{j=1}^L \tilde{w}^{(j)}\left(\tilde{\sigma}^{(j)}\right)^2}{\left(\tilde{\sigma}^{(i)}\right)^2}\left(\bm\Sigma + \displaystyle\frac{\bm\Sigma\hat{\bm\delta}^* (\hat{\bm\delta}^*)^\top\bm\Sigma}{\left(\sum_{j=1}^L \tilde{w}^{(j)} \left(m^{(j)}\right)^2\right)^{-1} - (\hat{\bm\delta}^*)^\top \bm\Sigma \hat{\bm\delta}^*}\right)
\end{align}
With this, each Gaussian mixand can be equivalently expressed as a likelihood component as
\begin{align}
    \mathcal{N}\!\left(\tilde{\mathbf{x}};\bm\eta^{(i)},\left(\bm\Sigma^{(i)}\right)^{-1}\right) = \left|\displaystyle\frac{1}{2\pi}\bm\Sigma^{(i)}\right|^{1/2}\mathcal{L}\!\left(\tilde{\mathbf{x}};\bm\eta^{(i)},\bm\Sigma^{(i)}\right)
\end{align}
and the resulting likelihood component is embedded in the higher-dimensional space as
\begin{subequations}\label{eqn:raised_likelihood_split}
    \begin{align}
        \mathcal{L}\!\left(\tilde{\mathbf{x}};\bm\eta^{(i)},\bm\Sigma^{(i)}\right) &= \mathcal{L}\!\left(\mathbf{x}; \bm\mu^{(i)},\bm\Lambda^{(i)}\right) \\
        \bm\mu^{(i)} &= \mathbf{U}_r\bm\eta^{(i)} + \mathbf{U}_0\mathbf{U}_0^\top \bm\mu\\
        \bm\Lambda^{(i)} &= \mathbf{U}_r \bm\Sigma^{(i)}\mathbf{U}_r^\top
    \end{align}
\end{subequations}
When recovering this higher-dimensional likelihood component, the value of each recovered center~$\bm\mu^{(i)}$ is unconstrained in the nullspace directions of the original precision matrix~$\bm\Lambda$ dictated by the columns of~$\mathbf{U}_0$. Mathematically, each split center can take on any value~$\bm\mu^{(i)} = \mathbf{U}_r\bm\eta^{(i)} + \mathbf{u}_0$ , where~$\mathbf{u}_0$ is in the nullspace of~$\bm\Lambda$. Choosing~$\mathbf{u}_0$ as the projection of the original component center~$\bm\mu$ into the nullspace of~$\bm\Lambda$ as is done in~\eqref{eqn:raised_likelihood_split} provides the minimum-norm solution with respect to~$\bm\mu$, which is desirable to ensure the split centers remain reasonable linearization points.

Figure~\ref{fig:split_example} demonstrates this singular likelihood component splitting on a two-dimensional example. The colored lines in Fig.~\ref{fig:split_example}a and 3c denote 3-$\sigma$ contours, and the dots denote the component centers. Figure~\ref{fig:split_example}a shows the original component with the range and nullspace directions identified. The component is then projected down to the one-dimensional subspace of its precision matrix as shown in Fig.~\ref{fig:split_example}b, where standard Gaussian splitting is performed. Finally, the resulting split components are lifted back to the state-space as shown in Fig.~\ref{fig:split_example}c.

\setlength{\figrowwidth}{0.95\linewidth}
\begin{figure}[bht!]
    \par\medskip
    \centering
    \begin{subfigure}{0.33\figrowwidth}
        \begin{overpic}[width=\linewidth]{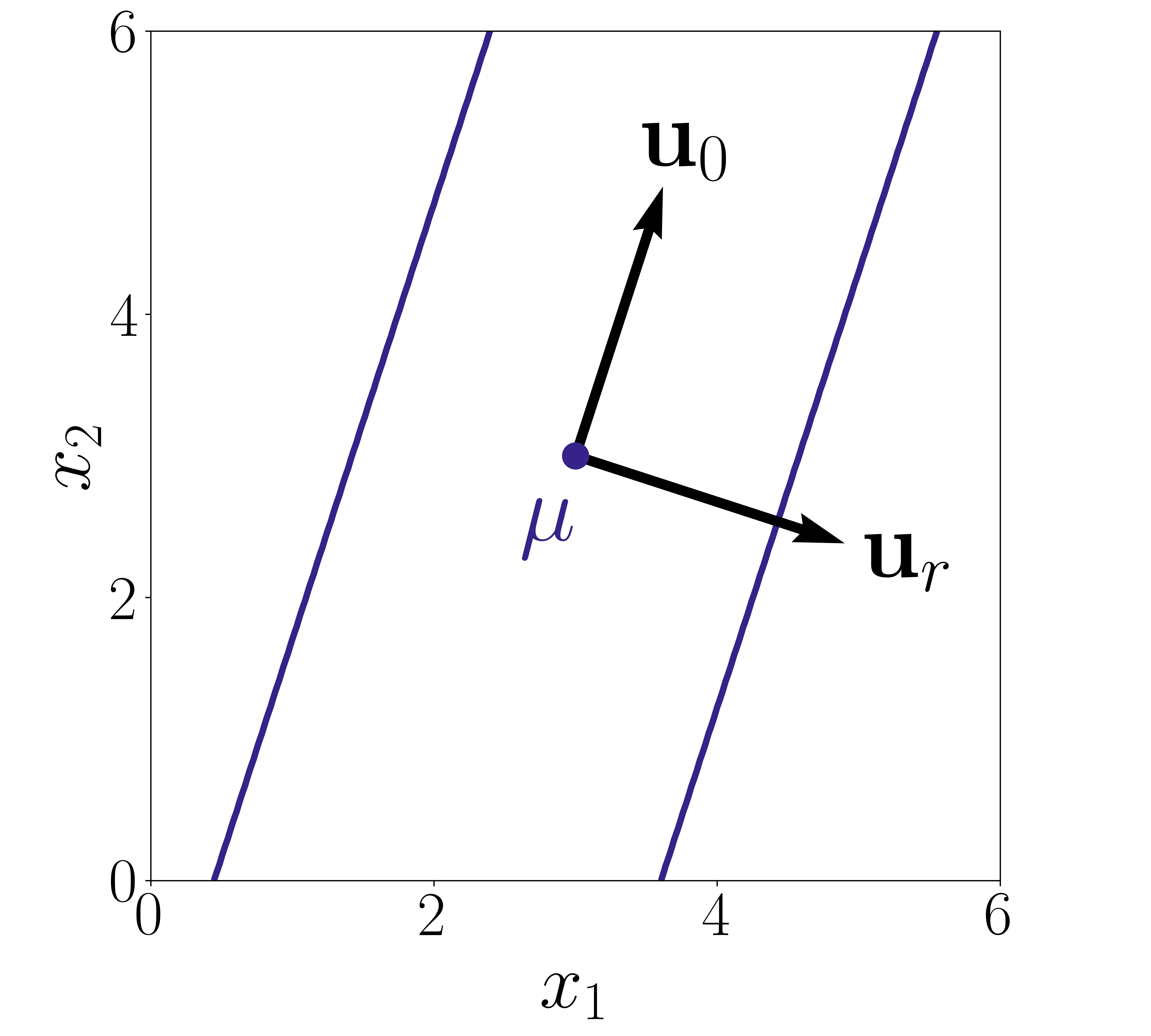}
            \put(0,85){\large (a)}
        \end{overpic}
        \label{fig:split_example_preSplit}
    \end{subfigure} %
    \begin{subfigure}{0.33\figrowwidth}
        \begin{overpic}[width=\linewidth]{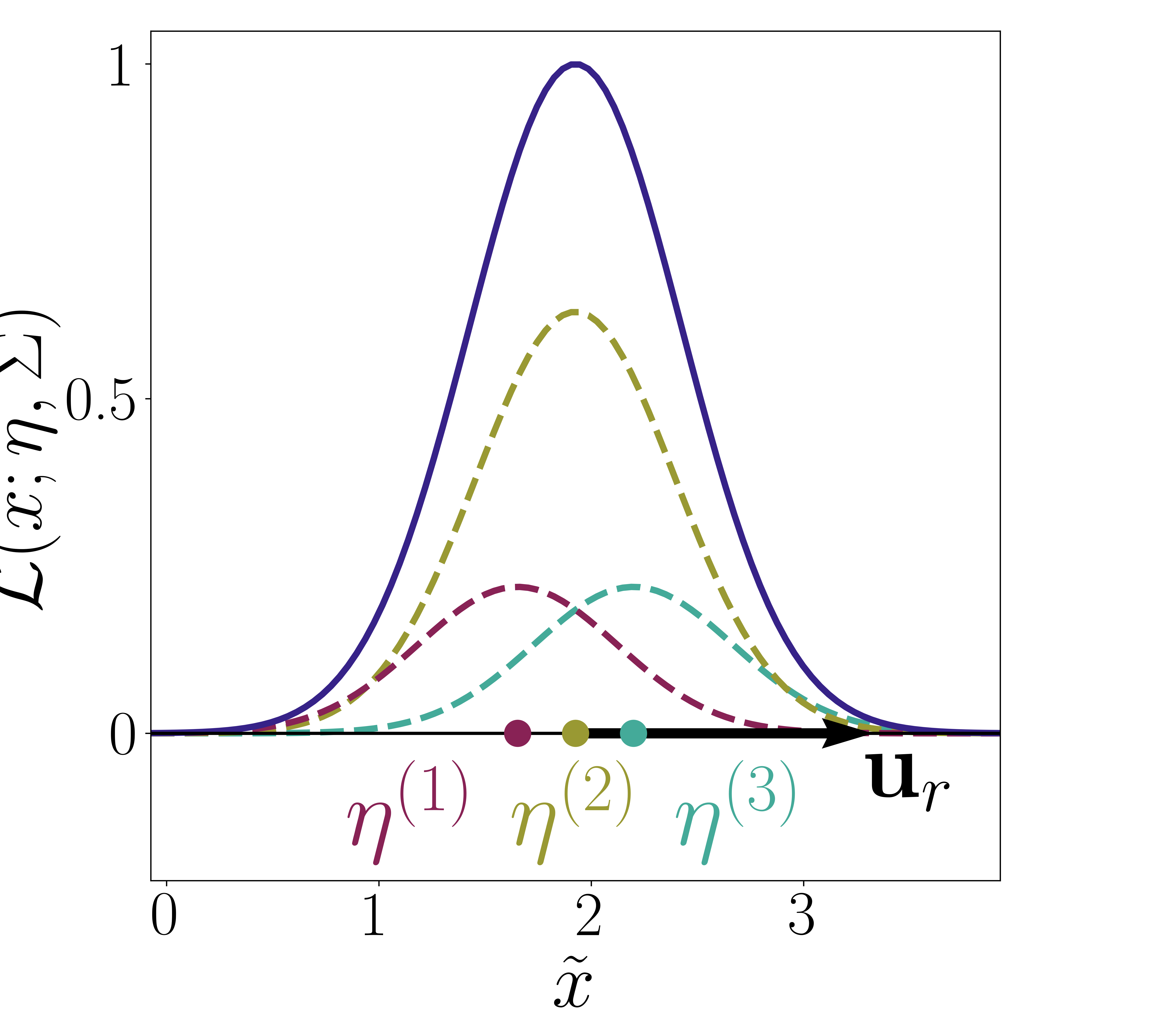}
            \put(0,85){\large (b)}
        \end{overpic}
        \label{fig:split_example_subspace}
    \end{subfigure} %
    \begin{subfigure}{0.33\figrowwidth}
        \begin{overpic}[width=\linewidth]{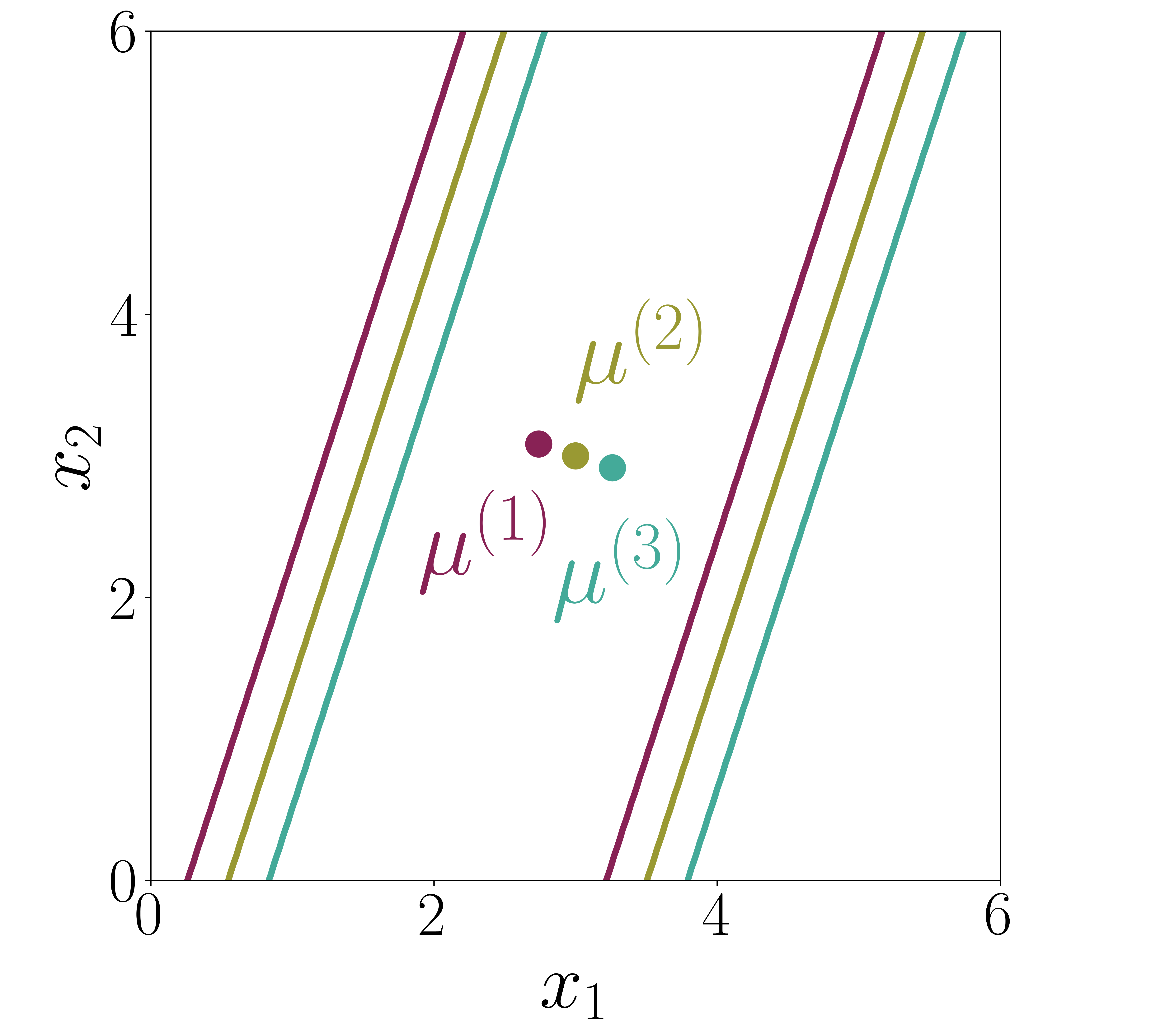}
            \put(0,85){\large (c)}
        \end{overpic}
        \label{fig:split_example_postSplit}
    \end{subfigure}

    \caption{Likelihood component split with singular precision matrix example.}
    \label{fig:split_example}
\end{figure}

In order to determine an ideal split direction, a constrained optimization problem is cast which aims to maximize some measure of uncertainty, nonlinearity, and/or non-Gaussianity. In this framework, the split-direction heuristics introduced in \cite{kulik2025NonlinearityUncertaintyInformed} and summarized in Section~\ref{sec:background_split}, namely \textit{maxvar}, \ac{usfos}, and \ac{wussolc}, are extended to the splitting of likelihood components. Since any direction outside the range of a likelihood component's precision matrix is fully uninformative, the full-space optimization over $\|\bm\delta\|_{\boldsymbol{\Lambda}}=1$ is ill-posed. Instead, restricting to the range via $\tilde{\mathbf{x}} = \mathbf{U}_r^\top \mathbf{x}$ maps the singular constraint $\|\bm\delta\|_{\boldsymbol{\Lambda}}=1$ to the well-posed $\|\tilde{\bm\delta}\|_{\bm\Sigma}=1$. In order to isolate the nonlinearities within this range, the function
\begin{align}
    \tilde{\mathbf{g}}(\tilde{\mathbf{x}}) = \mathbf{g}(\mathbf{U}_r\tilde{\mathbf{x}} + \mathbf{U}_0\mathbf{U}_0^\top\bm\mu)
\end{align}
is defined, where~$\bm\mu$ is the evaluation point of interest in the state-space, with Jacobian and \ac{PDT}
\begin{equation} \label{eqn:aug_G_PDT}
    \tilde{G}^i_j = G^i_k (U_r)^k_j, \qquad
    (\tilde{G}^{(2)})^i_{j,k} = (G^{(2)})^i_{m,n}(U_r)^m_j(U_r)^n_k,
\end{equation}
expressed using Einstein notation. The composition of the original nonlinear function~$\mathbf{g}(\mathbf{x})$ with~$\mathbf{U}_r\tilde{\mathbf{x}} + \mathbf{U}_0\mathbf{U}_0^\top\bm\mu$ ensures that only nonlinearities in potential split directions are used to inform splitting.

When splitting for the measurement update, $\mathbf{g}(\mathbf{x})$ is the nonlinear measurement function~$\mathbf{h}_k(\mathbf{x})$, and the associated \acp{PDT} are
\begin{equation}
    \mathbf{G} = \mathbf{H}_k,\qquad \left(G^{(2)}\right)^{i}_{j,m} = \PD{\left(H_k\right)^i_j}{x^m}
\end{equation}

In the backward time propagation step, the nonlinear function in question is the inverse dynamics flow $\mathbf{g}(\mathbf{x}) = \bm\varphi_{k}^{k-1}(\mathbf{x}_k)$, and the associated \acp{PDT} are the \ac{STM} and \ac{STT},
\begin{equation}
    \mathbf{G} = \bm\Phi_{k}^{k-1},\qquad \left(G^{(2)}\right)^{i}_{j,m} = \PD{\left(\Phi_k^{k-1}\right)^i_j}{x^m}
\end{equation}

All three heuristics are summarized in Table~\ref{tab:split_heuristics_range}, together with their solutions in terms of the maximal right singular vector $\mathbf{v}_{\max}(\cdot)$. The \ac{wussolc} solution requires the reshaped tensor in Einstein notation
\begin{subequations}
\begin{align}
    (\tilde{G}_w^{(2)})_{j,k}^i &=
    (\tilde{S}^{-1/2})^i_l \,
    (\tilde{G}^{(2)})_{p,q}^l \,
    (\Sigma^{-1/2})^p_j \,
    (\Sigma^{-1/2})^q_k, \\
    (\bar{G}_w^{(2)})_k^{ni+j} &= (\tilde{G}_w^{(2)})_{j,k}^i,
\end{align}
\end{subequations}
where $\tilde{\mathbf{S}} = \tilde{\mathbf{G}}\bm\Sigma^{-1}\tilde{\mathbf{G}}^\top$ represents the linearly predicted output covariance in the range space, which serves as a whitening transformation. Through this whitening transformation, the \ac{wussolc} measure is interpretable as a Mahalanobis distance and is invariant to linear changes of basis, including changes in units, as discussed in~\cite{kulik2025NonlinearityUncertaintyInformed}.

\begin{table}[h]
\centering
\caption{Split direction heuristics in range formulation ($\norm{\tilde{\bm\delta}}_{\bm\Sigma} =1$).}
\label{tab:split_heuristics_range}
\renewcommand{\arraystretch}{1.5} %
\begin{tabular}{cccc}
\toprule
\toprule
Heuristic & Objective $\mathcal{F}(\mathbf{g},\tilde{\bm\delta})$ & Solution $\hat{\bm\delta}^*$ & Criterion \\
\midrule
\textit{maxvar} &
  $\|\tilde{\bm\delta}\|_2$ &
  $\bm\Sigma^{-1/2}\mathbf{v}_{\max}(\bm\Sigma^{-1/2})$&
  $\lambda_{\max}(\bm\Sigma^{-1/2})$\\
[0.5ex] %
\ac{usfos} &
  $\|\tilde{\mathbf{G}}\tilde{\bm\delta}\|_2$ &
  $\bm\Sigma^{-1/2}\mathbf{v}_{\max}(\tilde{\mathbf{G}}\bm\Sigma^{-1/2})$&
  $\lambda_{\max}(\tilde{\mathbf{G}}\bm\Sigma^{-1/2})$\\
[0.5ex]
\ac{wussolc} &
  $\displaystyle\left\|\tilde{\mathbf{S}}^{-1/2} \left(\tilde{\mathbf{G}}^{(2)}\tilde{\bm\delta}\right)\bm\Sigma^{-1/2}\right\|_F^2$ &
  $\bm\Sigma^{-1/2}\mathbf{v}_{\max}(\bar{\mathbf{G}}_w^{(2)})$&
  $\lambda_{\max}(\bar{\mathbf{G}}_w^{(2)})$\\
\bottomrule
\bottomrule
\end{tabular}
\end{table}

If no likelihood components with singular precision matrices exist in the mixture, then the split criteria of Table~\ref{tab:split_heuristics_range} is scaled by the component's proportion of the total distribution mass~$w^{(m)}/\sum_{j=1}^M w^{(j)}$, where~$w^{(m)} = \nu^{(m)}\left|\frac{1}{2\pi}\Sigma^{(m)}\right|^{1/2}$ is the volume of each component. This ensures that components with negligible mass are split less often, and reduces the computational burden without significantly sacrificing accuracy. This scaling is not done when components with singular precision matrices exist, since the total distribution mass is unbounded.

\subsubsection{Likelihood Component Merging}
Merging is performed in the backward filter following both the prediction and update steps to ensure the number of components in the backward filter likelihood remains computationally tractable. Likelihood component merging is accomplished in a very similar manner to likelihood component splitting. Yet, care must be taken when merging components with singular precision matrices, as this could lead to poor approximations and false observability if the components' precision matrices do not share a common range. Because of this, only components whose precision matrices share a common range are considered for merging. The methodology employed in~\eqref{eqn:SVD}-\eqref{eqn:likelihood2Gaussian} can be used to represent a mixture of two likelihood components as lower-dimensional Gaussian mixands, written explicitly as
\begin{align}
    \nu^{(i)}\mathcal{L}\!\left(\mathbf{x};\bm\mu^{(i)},\bm\Lambda^{(i)}\right) + \nu^{(j)}\mathcal{L}\!\left(\mathbf{x};\bm\mu^{(j)},\bm\Lambda^{(j)}\right) = w^{(i)}\mathcal{N}\!\left(\tilde{\mathbf{x}};\bm\eta^{(i)},\left(\bm\Sigma^{(i)}\right)^{-1}\right) + w^{(j)}\mathcal{N}\!\left(\tilde{\mathbf{x}};\bm\eta^{(j)},\left(\bm\Sigma^{(j)}\right)^{-1}\right)
\end{align}

The two mixands can be approximated by a single Gaussian mixand matching the mean and covariance of the mixture using~\eqref{eqn:multivariate_merge}, resulting in
\begin{align}
    w^{(i)}\mathcal{N}\!\left(\tilde{\mathbf{x}};\bm\eta^{(i)},\left(\bm\Sigma^{(i)}\right)^{-1}\right) + w^{(j)}\mathcal{N}\!\left(\tilde{\mathbf{x}};\bm\eta^{(j)},\left(\bm\Sigma^{(j)}\right)^{-1}\right) &\approx w^{(ij)}\mathcal{N}\!\left(\tilde{\mathbf{x}};\bm\eta^{(ij)},\left(\bm\Sigma^{(ij)}\right)^{-1}\right)
\end{align}
The matrix inversion lemma can be employed on the merged covariance formula~\eqref{eqn:multivariate_merge_cov} so that only one matrix inversion is present in the formula. This rearrangement is given by
\begin{align}
    \bm\Sigma^{(ij)} &= \mathbf{A} - \displaystyle\frac{\alpha^{(i)}\alpha^{(j)}\mathbf{A}(\bm\eta^{(i)} - \bm\eta^{(j)})(\bm\eta^{(i)} - \bm\eta^{(j)})^\top \mathbf{A}}{1 + \alpha^{(i)}\alpha^{(j)}(\bm\eta^{(i)} - \bm\eta^{(j)})^\top \mathbf{A} (\bm\eta^{(i)} - \bm\eta^{(j)})} \\
    \mathbf{A} &= \displaystyle\frac{1}{\alpha^{(i)}}\bm\Sigma^{(i)} - \displaystyle\frac{1}{\alpha^{(i)}}\bm\Sigma^{(i)}\left(\displaystyle\frac{1}{\alpha^{(i)}}\bm\Sigma^{(i)} + \displaystyle\frac{1}{\alpha^{(j)}}\bm\Sigma^{(j)}\right)^{-1}\displaystyle\frac{1}{\alpha^{(i)}}\bm\Sigma^{(i)}
\end{align}
where~$\alpha^{(i)}$ and~$\alpha^{(j)}$ are defined in~\eqref{eqn:multivariate_merge_alpha}. The selection of components to merge can then be achieved using the Runnalls' merging algorithm outlined in Section~\ref{sec:background_merge}. Alternative Gaussian mixture reduction algorithms exist \cite{Crouse_gmreduction_2011}, but Runnalls' algorithm provides a good balance between computation and accuracy, and generalizes more easily to the merging of components with singular precision matrices than other methods. The upper bound on the \ac{KL} divergence required for the Runnalls' algorithm can be reformulated in terms of precision matrices as
\begin{align}
    D_{\textrm{KL}}[i||j] \leq B(i,j) = \displaystyle\frac{1}{2}\left(w^{(i)}\log\left|\bm\Sigma^{(i)}\right| + w^{(j)}\log\left|\bm\Sigma^{(j)}\right| - \left(w^{(i)} + w^{(j)}\right)\log\left|\bm\Sigma^{(ij)}\right|\right)
\end{align}
The component pair with the lowest value of~$B(i,j)$ is repeatedly merged until the number of components is less than a specified maximum~$M_{\textrm{max}}$. To further reduce the mixture size, component pairs can continue to be merged until the \ac{KL} divergence upper bound for all pairs is greater than a specified maximum~$B_{\textrm{max}}$. Since the backward filter likelihood does not generally integrate to unity and~$B(i,j)$ scales with the component masses, the metric must be scaled by the total distribution mass for a predefined maximum to be generally applicable. This yields the condition~$B(i,j) /\sum_{m=1}^M w^{(m)} > B_{\textrm{max}}$ to dictate whether a component pair is too dissimilar to merge, where~$w^{(m)} = \nu^{(m)}\left|\frac{1}{2\pi}\Sigma^{(m)}\right|^{1/2}$ is the mass of each component. Since the total mass of the backward filter likelihood is unbounded when components have singular precision matrices, this additional merging based on~$B_{\textrm{max}}$ is simply omitted in these cases.

Once a merge is performed, the resulting Gaussian can be equivalently represented by a likelihood component as
\begin{subequations}
    \begin{align}
        w^{(ij)}\mathcal{N}\!\left(\tilde{\mathbf{x}};\bm\eta^{(ij)},\left(\bm\Sigma^{(ij)}\right)^{-1}\right) = \left|\displaystyle\frac{1}{2\pi}\bm\Sigma^{(ij)}\right|^{1/2}w^{(ij)}\mathcal{L}\!\left(\tilde{\mathbf{x}};\bm\eta^{(ij)},\bm\Sigma^{(ij)}\right) = \nu^{(ij)}\mathcal{L}\!\left(\tilde{\mathbf{x}};\bm\eta^{(ij)},\bm\Sigma^{(ij)}\right)
    \end{align}
\end{subequations}
and the higher-dimensional likelihood is recovered as
\begin{subequations}\label{eqn:raised_likelihood_merge}
    \begin{align}
        \nu^{(ij)}\mathcal{L}\!\left(\tilde{\mathbf{x}};\bm\eta^{(ij)},\bm\Sigma^{(ij)}\right) &= \nu^{(ij)}\mathcal{L}\!\left(\mathbf{x};\bm\mu^{(ij)},\bm\Lambda^{(ij)}\right) \\
        \bm\mu^{(ij)} &= \mathbf{U}_r \bm\eta^{(ij)} + \mathbf{U}_0\mathbf{U}_0^\top \left(\alpha^{(i)}\bm\mu^{(i)} + \alpha^{(j)}\bm\mu^{(j)}\right) \label{eqn:raised_likelihood_merge_center} \\
        \bm\Lambda^{(ij)} &= \mathbf{U}_r \bm\Sigma^{(ij)}\mathbf{U}_r^\top
    \end{align}
\end{subequations}

As is the case with splitting, the center~$\bm\mu^{(ij)}$ of this recovered higher-dimensional likelihood component is unconstrained in the nullspace directions. However, a unique solution is obtained when the center is chosen to lie on the line in state-space between the centers of the original two components~$\bm\mu^{(i)}$ and~$\bm\mu^{(j)}$, which is the value given by~\eqref{eqn:raised_likelihood_merge_center}.

Although the determination of ranges and nullspaces of likelihood components via~\eqref{eqn:SVD} is simple mathematically, it can be less trivial in practice when numerical error results in components having near-zero but still positive singular values. For this work, a singular value is treated as zero if it is smaller than a specified relative tolerance times the largest singular value of the matrix, and two components are treated to share a nullspace if the sum of their precision matrices has the same number of singular values as each matrix individually as measured by the same criterion. This tolerance is set at~$10^{-10}$ during the initialization phase, which is large enough to ensure that numerical error does not result in false observability. Once the precision matrices of all likelihood components in the mixture reach full rank, this singularity check is no longer required and therefore omitted.

A summary of the \ac{AGM} smoother for nonlinear systems is given as Algorithm \ref{alg:BIF_smoother}. The function~$\texttt{split}(p,\mathbf{g})$ splits a \ac{GM}~$p$ according to the nonlinear function~$\mathbf{g}$, where additional parameters including the recursion depth, split tolerance, and split heuristic are constant across the algorithm and thus omitted. The function~$\texttt{merge}(p)$ merges a \ac{GM}~$p$, where additional parameters including the maximum allowed number of components and maximum \ac{KL} divergence upper bound are also constant across the algorithm and thus omitted.

\begin{algorithm}[bht!]
\caption{\texttt{\ac{AGM} Smoother}}
\label{alg:BIF_smoother}

\begin{algorithmic}[1]
    \Require Filtered distributions
    $p(\mathbf{x}_k|\mathbf{z}_{1:k})$
    for all $0 \leq k \leq T$

    \State Initialize
    $p(\mathbf{z}_T|\mathbf{x}_T)$
    using
    $p(\mathbf{x}_T|\mathbf{z}_{1:T})$
    via~\eqref{eqn:BIF_init}

    \State $p(\mathbf{z}_{T}|\mathbf{x}_{k-1}) \gets \texttt{BIFPredict}(p(\mathbf{z}_T|\mathbf{x}_T))$

    \State Obtain
    $p(\mathbf{x}_{T-1}|\mathbf{z}_{1:T})$
    using
    $p(\mathbf{x}_{T-1}|\mathbf{z}_{1:T-1})$
    and
    $p(\mathbf{z}_{T}|\mathbf{x}_{T-1})$
    via~\eqref{eqn:smoothed_dist}--\eqref{eqn:smoothed_params}

    \For{$k = T-1,\ldots,1$}
        \State $p(\mathbf{z}_{k:T}|\mathbf{x}_{k}) \gets \texttt{BIFUpdate}(p(\mathbf{z}_{k+1:T}|\mathbf{x}_k))$

        \State $p(\mathbf{z}_{k:T}|\mathbf{x}_{k-1}) \gets \texttt{BIFPredict}(p(\mathbf{z}_{k:T}|\mathbf{x}_k))$

        \State Obtain
        $p(\mathbf{x}_{k-1}|\mathbf{z}_{1:T})$
        using
        $p(\mathbf{x}_{k-1}|\mathbf{z}_{1:k})$
        and
        $p(\mathbf{z}_{k:T}|\mathbf{x}_{k-1})$
        via~\eqref{eqn:smoothed_dist}--\eqref{eqn:smoothed_params}

    \EndFor

    \Return Smoothed distributions
    $p(\mathbf{x}_k|\mathbf{z}_{1:T})$
    for all $0 \leq k \leq T$

\end{algorithmic}

\vspace{0.3em}
\end{algorithm}

\begin{algorithm}[bht!]
    \caption{\texttt{BIFUpdate}}
    \label{alg:BIF_smoother_update}
    \begin{algorithmic}[1]
        \Require Prior backward filter likelihood~$p(\mathbf{z}_{k+1:T}|\mathbf{x}_k)$

        \State $p(\mathbf{z}_{k+1:T}|\mathbf{x}_k)
        \gets
        \texttt{split}
        (p(\mathbf{z}_{k+1:T}|\mathbf{x}_k),
        \mathbf{h}_k(\mathbf{x}_k))$

        \State Obtain
        $p(\mathbf{z}_{k:T}|\mathbf{x}_k)$
        via~\eqref{eqn:BIF_update}--\eqref{eqn:BIF_update_params}

        \State $p(\mathbf{z}_{k:T}|\mathbf{x}_k)
        \gets
        \texttt{merge}
        (p(\mathbf{z}_{k:T}|\mathbf{x}_k))$

        \Return $p(\mathbf{z}_{k:T}|\mathbf{x}_k)$
    \end{algorithmic}
\end{algorithm}

\begin{algorithm}[bht!]
    \caption{\texttt{BIFPredict}}
    \label{alg:BIF_smoother_prop}
    \begin{algorithmic}[1]
        \Require Posterior backward filter likelihood~$p(\mathbf{z}_{k:T}|\mathbf{x}_k)$

        \State $p(\mathbf{z}_{k:T}|\mathbf{x}_k)
        \gets
        \texttt{split}
        (p(\mathbf{z}_{k:T}|\mathbf{x}_k),
        \bm{\varphi}_{T}^{T-1}(\mathbf{x}_T))$

        \State Obtain
        $p(\mathbf{z}_{k:T}|\mathbf{x}_{k-1})$
        via~\eqref{eqn:BIF_prop}--\eqref{eqn:BIF_prop_params}

        \State $p(\mathbf{z}_{k:T}|\mathbf{x}_{k-1})
        \gets
        \texttt{merge}
        (p(\mathbf{z}_{k:T}|\mathbf{x}_{k-1}))$

        \Return $p(\mathbf{z}_{k:T}|\mathbf{x}_{k-1})$
    \end{algorithmic}
\end{algorithm}

\section{Results}\label{sec:results}
\subsection{Scenario Setup}
The proposed \ac{AGM} smoother is evaluated on two space object tracking scenarios:
a Molniya orbit and an Earth-Moon southern $\text{L}_2$ halo orbit. In both cases,
the state is represented by Cartesian position and velocity in the respective
coordinate frame,
\begin{equation}
    \mathbf{x}_k = \begin{bmatrix}
        \mathbf{r}_k^\top & \mathbf{v}_k^\top
    \end{bmatrix}^\top = \begin{bmatrix}
        x_k & y_k & z_k & \dot{x}_k & \dot{y}_k & \dot{z}_k
    \end{bmatrix}^\top,
\end{equation}
and the two-filter approach is evaluated over 250 Monte Carlo trials, with the truth being deterministic and the initial mean and measurement noise sequence being randomly varied in each trial. The target's initial state is Gaussian distributed with mean drawn from $\mathcal{N}(\mathbf{x}_0, \mathbf{P}_0)$,
where $\mathbf{x}_0 = [\mathbf{r}_0^\top\ \mathbf{v}_0^\top]^\top$ is the true initial
state and
\begin{equation}
    \mathbf{P}_0 = \mathrm{blockdiag}\!\left(\sigma_r^2\mathbf{I}_3,\,\sigma_v^2\mathbf{I}_3\right),
\end{equation}
with coordinates standard deviation  $\sigma_r$ and $\sigma_v$. The operator~$\mathrm{blockdiag}(\mathbf{A}_1,\ldots,\mathbf{A}_n)$ returns the block-diagonal matrix whose diagonal blocks are~$\mathbf{A}_1,\ldots,\mathbf{A}_n$.

The dynamics model for the filter and smoother is formulated in continuous time with additive process noise
$\mathbf{w}(t) \sim \mathcal{N}(\mathbf{0}, \mathbf{Q}_c)$, acting on the velocity
 through the mapping matrix $\bm\Gamma(t)$:
\begin{equation}
    \mathbf{Q}_c = \mathrm{diag}\!\left(q_{w_x}^2,\, q_{w_y}^2,\, q_{w_z}^2\right), \qquad
    \bm\Gamma(t) = \begin{bmatrix}
        \mathbf{0}_{3\times 3} \\ \mathbf{I}_{3\times 3}
    \end{bmatrix},
\end{equation}
where $q_{w_x}^2, q_{w_y}^2, q_{w_z}^2$ are the acceleration noise intensities along
each axis, whose effect is discretized in time as detailed in Section~\ref{sec noise dicscretization}. The operator~$\mathrm{diag}(a_1,\ldots,a_n)$ returns the diagonal matrix whose diagonal elements are~$a_1,\ldots,a_n$.

The filtered estimate is obtained using an adaptive \ac{GM}-\ac{EKF}
(Appendix~\ref{sec:appendix_GMfilter}), with nonlinearity- and uncertainty-informed
Gaussian splitting \cite{kulik2025NonlinearityUncertaintyInformed,
siciliano2025HigherOrderTensorBasedDeferral} applied before both the prediction and
update steps. Following each Bayesian update, the mixture
is reduced via Runnalls' algorithm \cite{Runnalls_gmreduction_2007} with the mixand
count capped at $N_{\max}$. The splitting and merging parameters for the forward and backward filters are given in Table~\ref{tab:split_merge_results}, with all values applying to both filters. The forward filter also employs the Bayesian recursive update \cite{michaelson2023bayesianrecursiveupdateensemble} with the same step sizes as the backward filter.
The \acp{STM} and \acp{STT} are computed via differential algebra using a custom, optimized build of DACEyPy \cite{massari_differential_2018}\footnote{\url{https://github.com/scope-lab/daceypy}}.This version enables the extraction of maps at desired time instants, just as a traditional \ac{ode} solver outputs states, and it is internally refactored to significantly reduce computational time. Regarding the splitting heuristics and handling of mixtures, an open-source Python
implementation is available in the PyEst library\footnote{\url{https://github.com/scope-lab/pyest}},

\begin{table}[h]
\centering
\caption{Splitting and merging parameters for forward and backward filter.}
\label{tab:split_merge_results}
\renewcommand{\arraystretch}{1.5} %
\begin{tabular}{ccc}
\toprule
\toprule
Parameter & Value \\
\midrule
Split heuristic & \ac{wussolc} \\
Prediction split tolerance & $10^{-1}$ \\
Update split tolerance & $10^{-3}$ \\
Components per split ($L$) & 3 \\
Split recursion depth & 4 \\
Minimum weight for splitting ($w_{\textrm{min}}$) & $0.01$ \\
Max. number of components ($N_{\textrm{max}}, M_{\textrm{max}}$) & 100 \\
Max. \ac{KL} div. for merge ($B_{\textrm{max}}$) & $0.01$ \\
\bottomrule
\bottomrule
\end{tabular}
\end{table}

\subsection{Evaluation Metrics}
To evaluate the performance of the filter and smoother relative to the ground truth, the state estimate is taken as the mean of the filtered and smoothed distributions, respectively, so that the estimation error is defined as
\begin{align}
    \mathbf{e}_k = \mathbf{x}_k - \EX[\mathbf{x}_k]
\end{align}
The primary performance metrics utilized to compare the filter and the smoother at each time step are the average position and velocity estimation errors. For each Monte Carlo trial, these errors are computed as the norms of the position and velocity components of the state estimation error vector, $\mathbf{e}_k$. The final metrics are then obtained by averaging these individual errors across all Monte Carlo trials.

To compare the credibility of the filter and the smoother, the~\ac{NEES} \cite{li2002estimator, Oliver1998, BARSHALOM1983431, barshalom2001estimation} is computed at each timestep and averaged across the Monte Carlo trials. This results in the \ac{ANEES}, written explicitly as
\begin{align}
    \textrm{ANEES} = \displaystyle\frac{1}{n_x}\frac{1}{N_{\text{MC}}}\sum_{i=1}^{N_{\text{MC}}} \mathbf{e}_{k,i}^\top \mathbf{P}_{k,i}^{-1}\mathbf{e}_{k,i}
\end{align}
where $n_x$ is the state dimension,  $N_{\text{MC}}$ denotes the number of Monte Carlo trials, and $\mathbf{e}_{k,j}$ and $\mathbf{P}_{k,j}$ are the filter-reported estimation error and covariance associated with the $j$-th trial at time step $k$. Perfectly credible estimators produce an \ac{ANEES} equal to one.  Higher and lower value of \ac{ANEES} indicate optimistic and pessimistic behaviour, respectively.

\subsection{Molniya Orbit}

For the Molniya orbit case, a Keplerian two-body \ac{EOM} with no perturbations is adopted in the simulation for simplicity. A ground-based observer collects topocentric right ascension and declination measurements. The noisy measurements are modeled according to~\eqref{eqn:meas_model}, where
\begin{equation}
    \mathbf{h}(\mathbf{x}) = \begin{bmatrix}
    \arctan\left(\displaystyle\displaystyle\frac{\hat{\mathbf{u}}\cdot\hat{\mathbf{y}}}{\hat{\mathbf{u}}\cdot\hat{\mathbf{x}}}\right) \\
    \arctan\left(\displaystyle\frac{\hat{\mathbf{u}}\cdot\hat{\mathbf{z}}}{\sqrt{(\hat{\mathbf{u}}\cdot\hat{\mathbf{x}})^2+(\hat{\mathbf{u}}\cdot\hat{\mathbf{y}})^2}}\right)
    \end{bmatrix},\quad \hat{\mathbf{u}} = \displaystyle\frac{\mathbf{r}-\mathbf{r_{obs}}}{||\mathbf{r}-\mathbf{r_{obs}}||}
\end{equation}

\begin{figure}[bht!]
    \centering
    \includegraphics[width=0.5\linewidth]{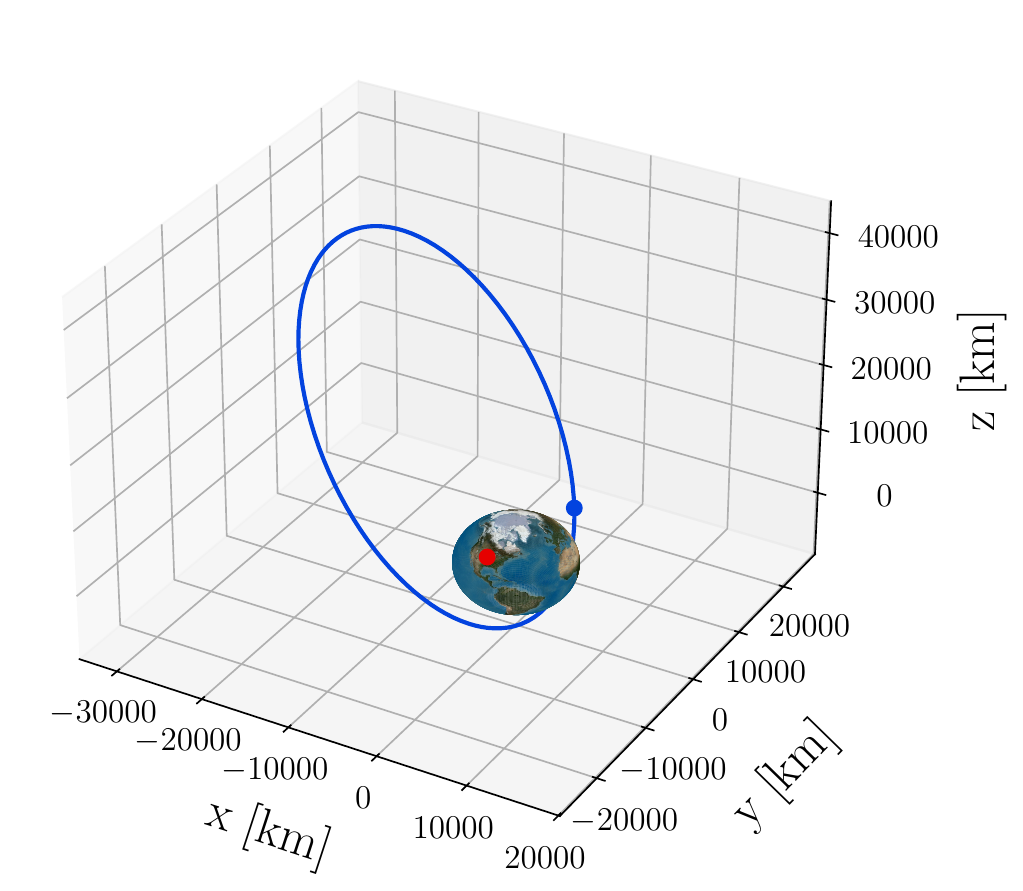}
    \caption{Molniya orbit scenario, where the target's Molniya orbit is blue, and the observer's initial position is red.}
    \label{fig:MolniyaOrbits}
\end{figure}

The tracking simulation spans 72 hours, during which the observer samples a single measurement at most once every hour when lighting and elevation-based visibility conditions permit. This amounts to five observations per target orbit on average. Table~\ref{tab:MolniyaScenario} summarizes the parameters for the Molniya orbit case, including the target's orbital elements, the ground-based observer location, the measurement noise, the initial target state uncertainty, and the continuous-time process noise statistics. The dynamics normalization constants are also reported, which ensure the covariance and precision matrices remain numerically well-conditioned. The orbit of the target and initial position of the observer are shown in Figure~\ref{fig:MolniyaOrbits}.
\begin{table}[hbt!]
    \centering
    \caption{Parameters for Molniya orbit scenario.}
    \label{tab:MolniyaScenario}

    \begin{subtable}{\textwidth}
        \centering
        \caption{Target Orbit Parameters}
        \label{tab:SubOrbit}
        \vspace{2pt}
        \begin{tabular}{lcc}
            \toprule \toprule
            \textbf{Parameter Description} & \textbf{Symbol} & \textbf{Value} \\
            \midrule
            Semi-major axis & $a$ & $26{,}553\,[\text{km}]$ \\
            Eccentricity & $e$ & $0.737$ \\
            Inclination & $i$ & $63.4^\circ$ \\
            Right ascension of the ascending node (RAAN) & $\Omega$ & $90^\circ$ \\
            Argument of periapsis (AoP) & $\omega$ & $270^\circ$ \\
            True anomaly & $\nu_0$ & $90^\circ$ \\
            Characteristic length & LU & $6{,}371\,[\text{km}]$\\
            Characteristic time & TU & $805.45739 \ [s]$\\
            \bottomrule \bottomrule
        \end{tabular}
    \end{subtable}

    \vspace{15pt} %

    \begin{subtable}[t]{0.48\textwidth}
        \centering
        \caption{Ground Observer Location}
        \label{tab:SubObserver}
        \vspace{2pt}
        \begin{tabular}{lcc}
            \toprule \toprule
            \textbf{Parameter} & \textbf{Symbol} & \textbf{Value} \\
            \midrule
            Latitude & $\phi$ & $40.4237^\circ$ \\
            Longitude & $\lambda$ & $-86.9212^\circ$ \\
            Altitude & $h$ & $0\,[\text{m}]$ \\

            \bottomrule \bottomrule
        \end{tabular}
    \end{subtable}
    \hfill
    \begin{subtable}[t]{0.48\textwidth}
        \centering
        \caption{Filter Parameters \& Noise}
        \label{tab:SubFilter}
        \vspace{2pt}
        \begin{tabular}{lcc}
            \toprule \toprule
            \textbf{Parameter} & \textbf{Symbol} & \textbf{Value} \\
            \midrule
            Pos. uncertainty & $\sigma_r$ & $1\,[\text{km}]$ \\
            Vel. uncertainty & $\sigma_v$ & $0.1\,[\text{m/s}]$ \\
            Process noise & $q_{w_{x,y,z}}$ & $10^{-6}\,[\text{m}/\text{s}^{5/2}]$ \\
           Measurement noise & v & $10\,[\text{arcsec}]$\\
            \bottomrule \bottomrule
        \end{tabular}
    \end{subtable}
\end{table}

\begin{figure}[bht!]
    \centering
    \includegraphics[width=0.5\linewidth]{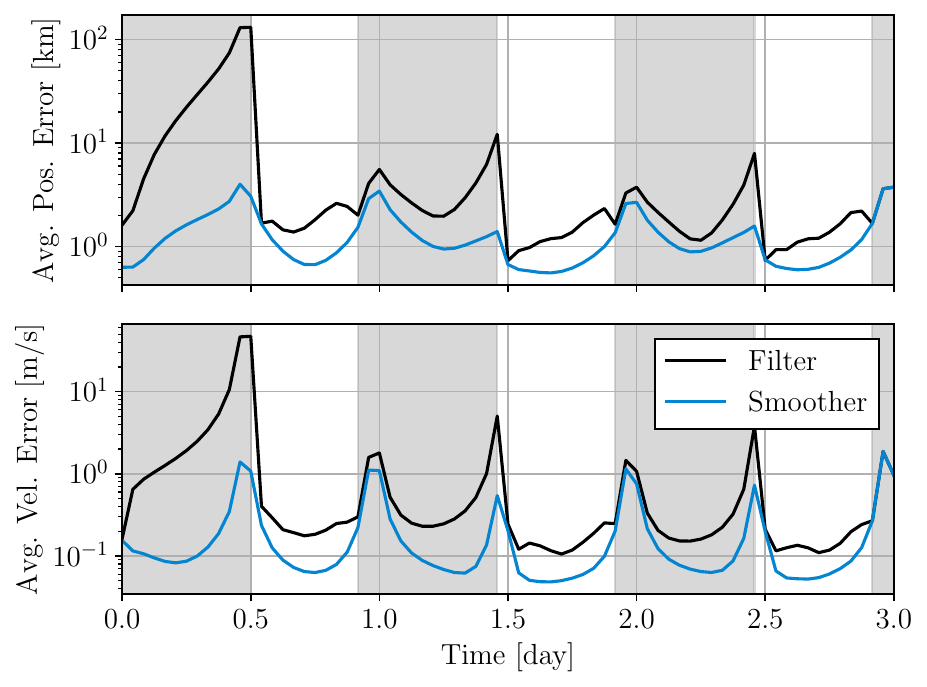}
    \caption{Average position and velocity error over 250 Monte Carlo trials for Molniya orbit scenario.}
    \label{fig:MolniyaError}
\end{figure}

Average position and velocity errors across all Monte Carlo trials are shown in
Fig.~\ref{fig:MolniyaError}, where gray bars denote periods of measurement
unavailability. The smoother consistently improves the filter estimate  across the entire
72-hour window, with the most pronounced improvements occurring during measurement gaps. Since the smoother incorporates measurement information from both sides of a measurement gap, it is able to significantly improve estimates during periods of measurement unavailability, which is generally when the filter estimation error is the largest. The total computation time of the smoother is comparable to that of the forward filter, with computation time generally increasing during periods of high non-Gaussianity in the backward filter likelihood. This increase in computation time is due to the per-component nonlinearity and uncertainty-informed splitting criteria being met more often, resulting in more total components following splitting.

Figure~\ref{fig:MolniyaXZMarginalComparison} shows the~$x$-$z$ marginals for the filter and smoother during a perigee passage near the end of the initial measurement gap. The white and blue text boxes denote the marginals of the filter and smoother respectively, while the red and white diamond denotes the true target state.
The filter marginal shows non-Gaussianity in the form of a heavy tail during the peak of perigee, with significant uncertainty and stretching occurring at the time steps before and after this peak.
The smoother, thanks to the future information, is able to reach a much more certain estimate. Most of this uncertainty reduction is along the track of the orbit, as the smoother reduces the in-track standard deviation at the peak of perigee at 0.4799~[day] from 239.9~[km] to 8.4~[km].

\setlength{\figrowwidth}{0.95\linewidth}
\begin{figure}[bht!]
    \par\medskip
    \centering
    \begin{subfigure}{0.33\figrowwidth}
        \begin{overpic}[width=\linewidth]{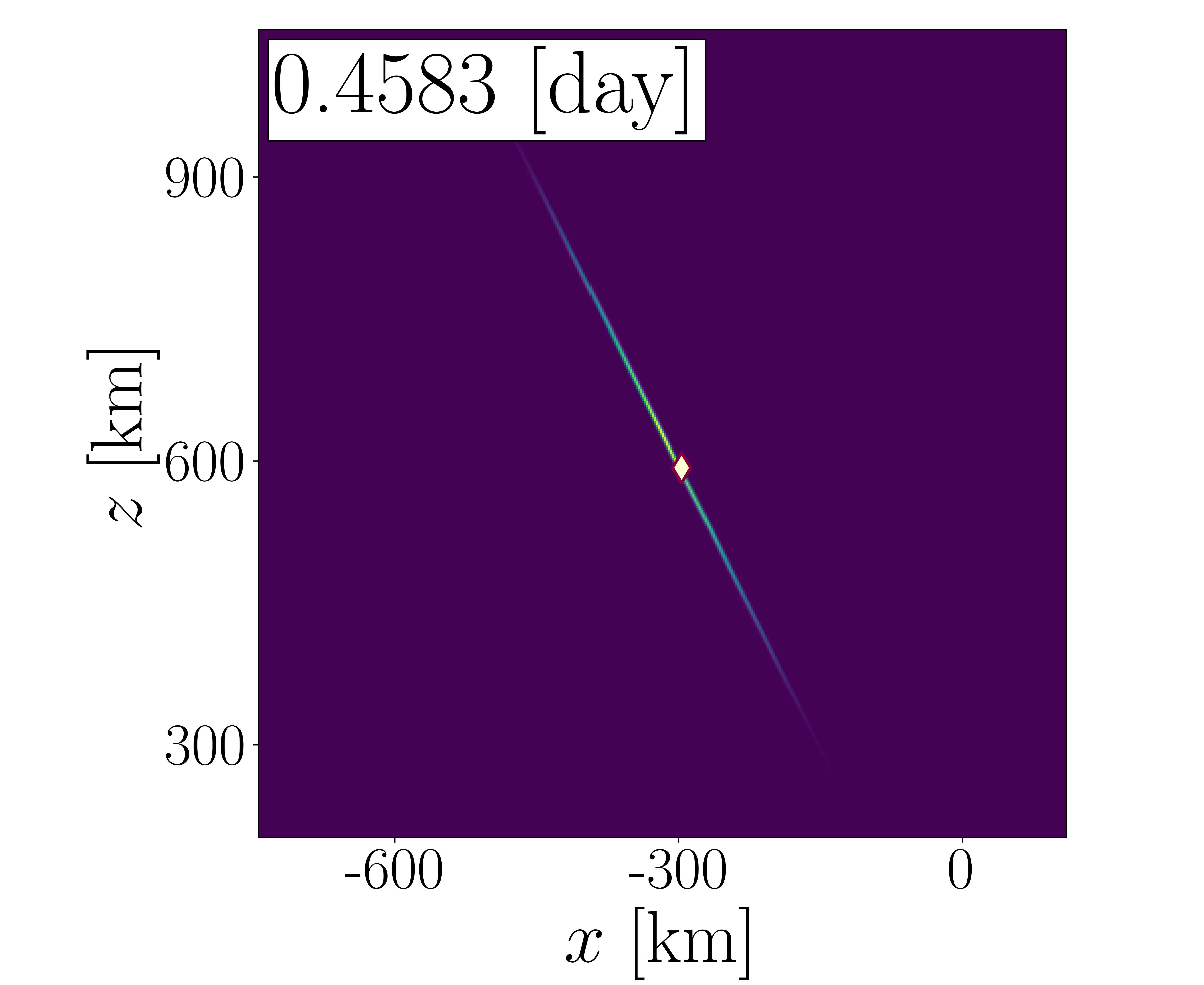}
            \put(8,80){\large (a)}
        \end{overpic}
        \label{fig:MolniyaFilterXZMarginal11}
    \end{subfigure} %
    \begin{subfigure}{0.33\figrowwidth}
        \begin{overpic}[width=\linewidth]{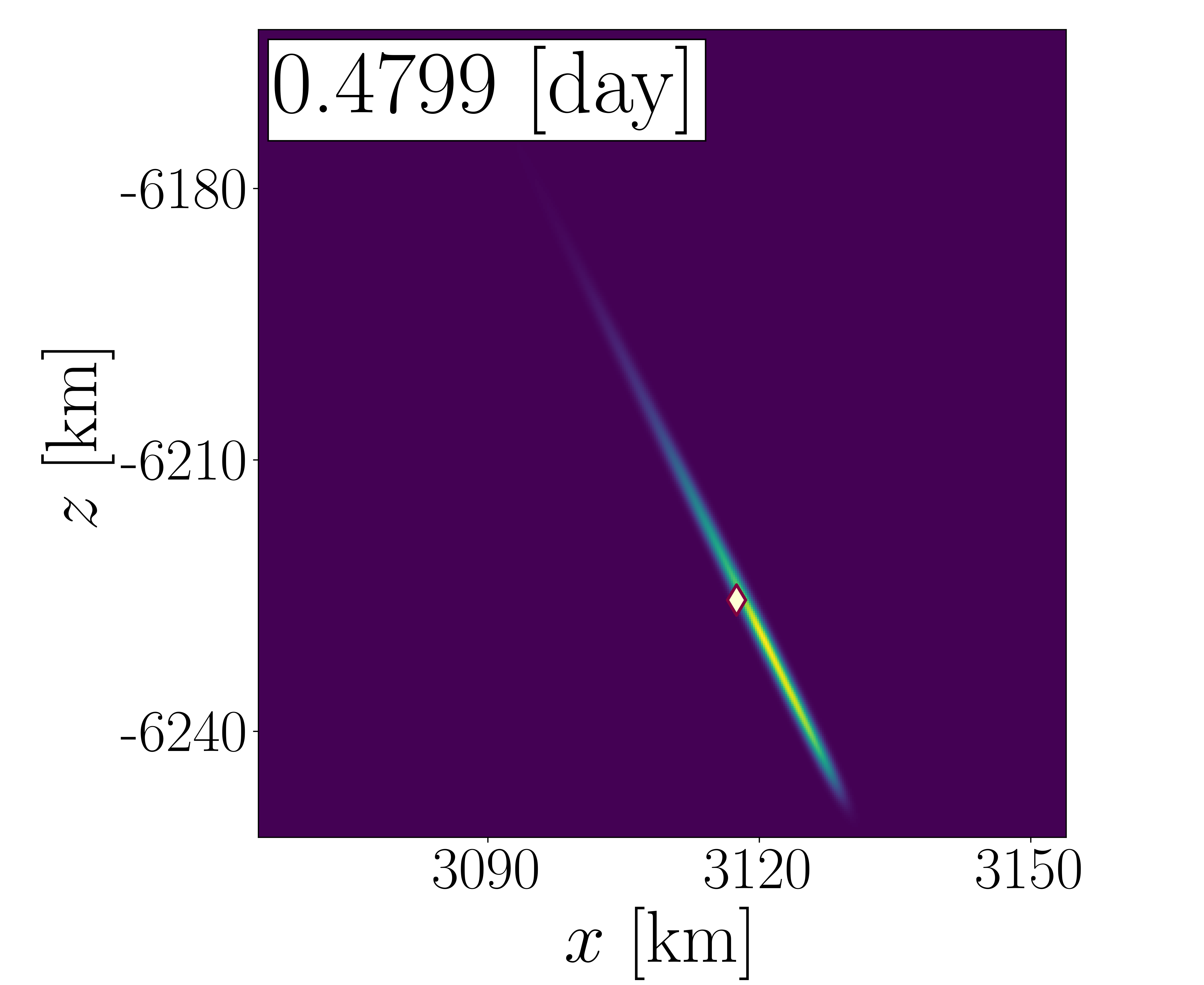}
            \put(8,80){\large (b)}
        \end{overpic}
        \label{fig:MolniyaFilterXZMarginal11-5}
    \end{subfigure} %
    \begin{subfigure}{0.33\figrowwidth}
        \begin{overpic}[width=\linewidth]{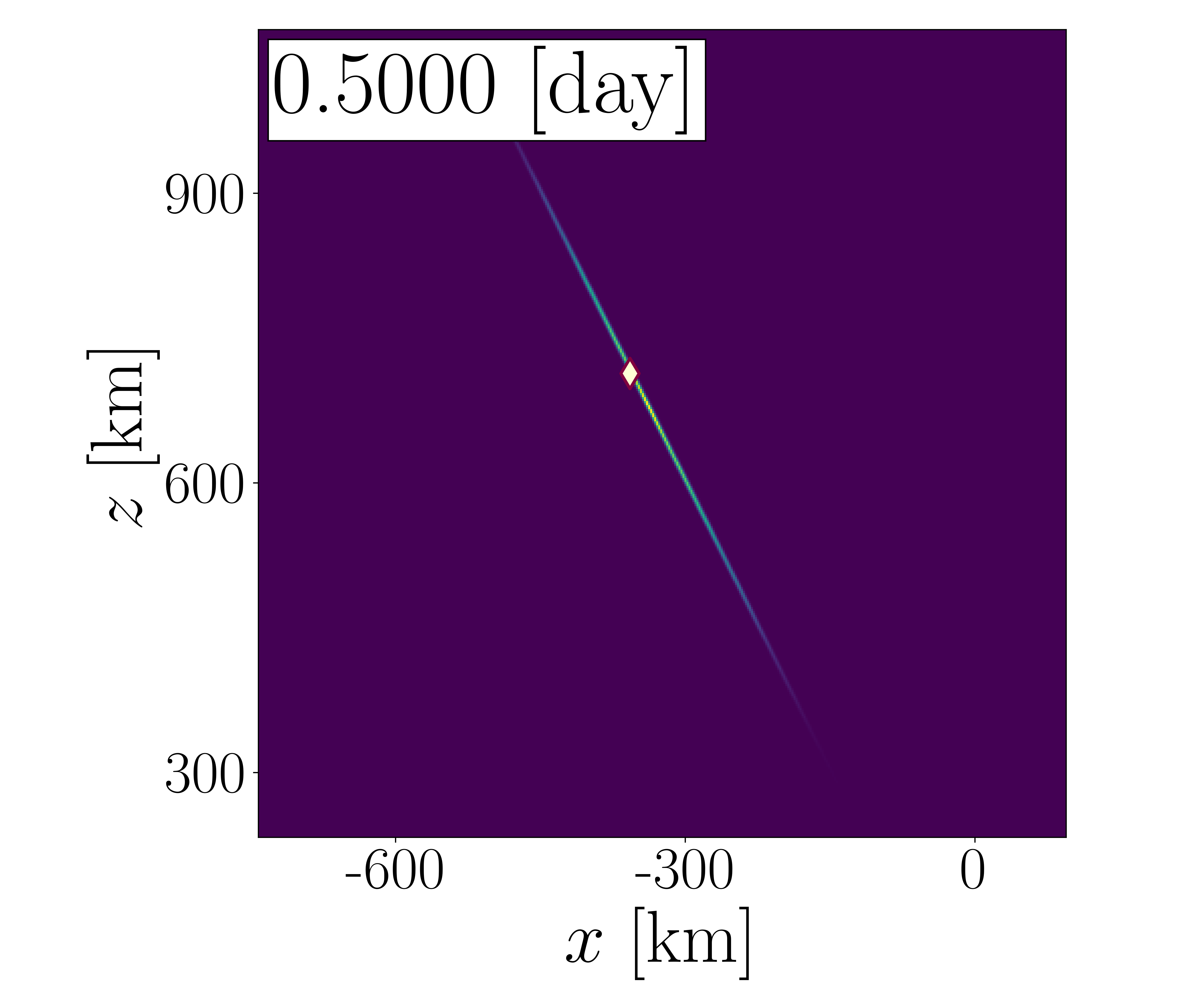}
            \put(8,80){\large (c)}
        \end{overpic}
        \label{fig:MolniyaFilterXZMarginal12}
    \end{subfigure}

    \begin{subfigure}{0.33\figrowwidth}
        \begin{overpic}[width=\linewidth]{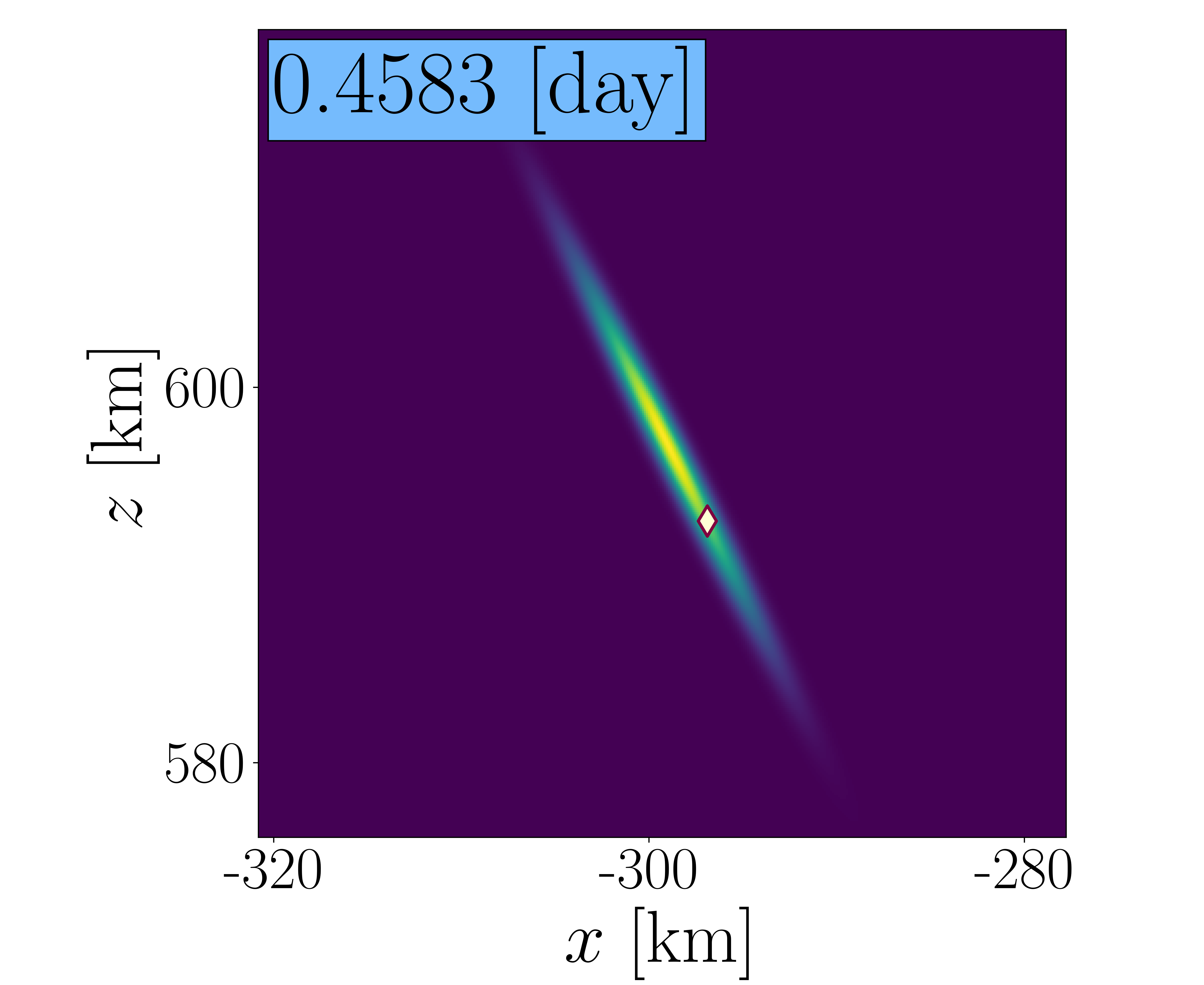}
            \put(8,80){\large (d)}
        \end{overpic}
        \label{fig:MolniyaSmootherXZMarginal11}
    \end{subfigure}%
    \begin{subfigure}{0.33\figrowwidth}
        \begin{overpic}[width=\linewidth]{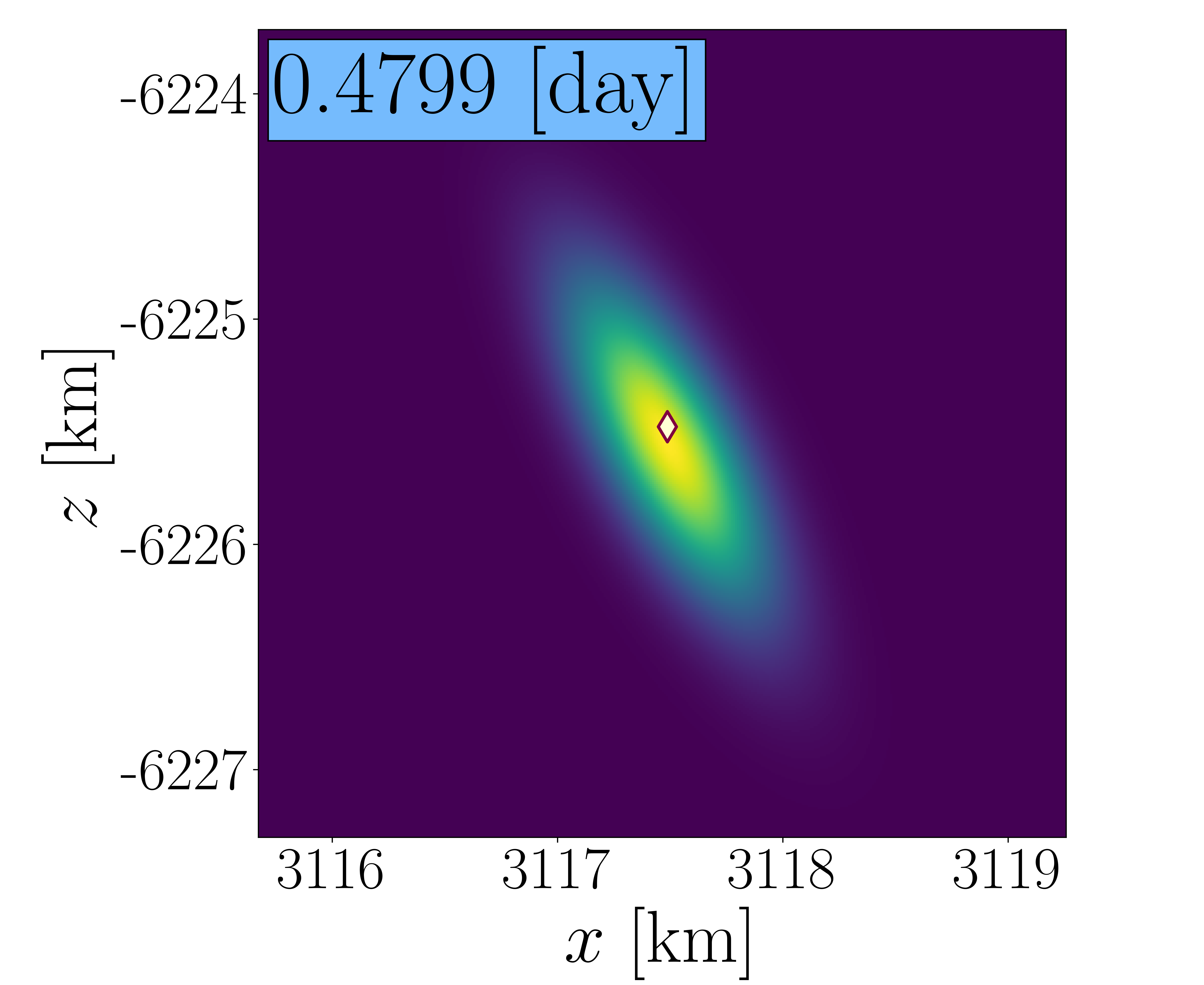}
            \put(8,80){\large (e)}
        \end{overpic}
        \label{fig:MolniyaSmootherXZMarginal11-5}
    \end{subfigure}%
    \begin{subfigure}{0.33\figrowwidth}
        \begin{overpic}[width=\linewidth]{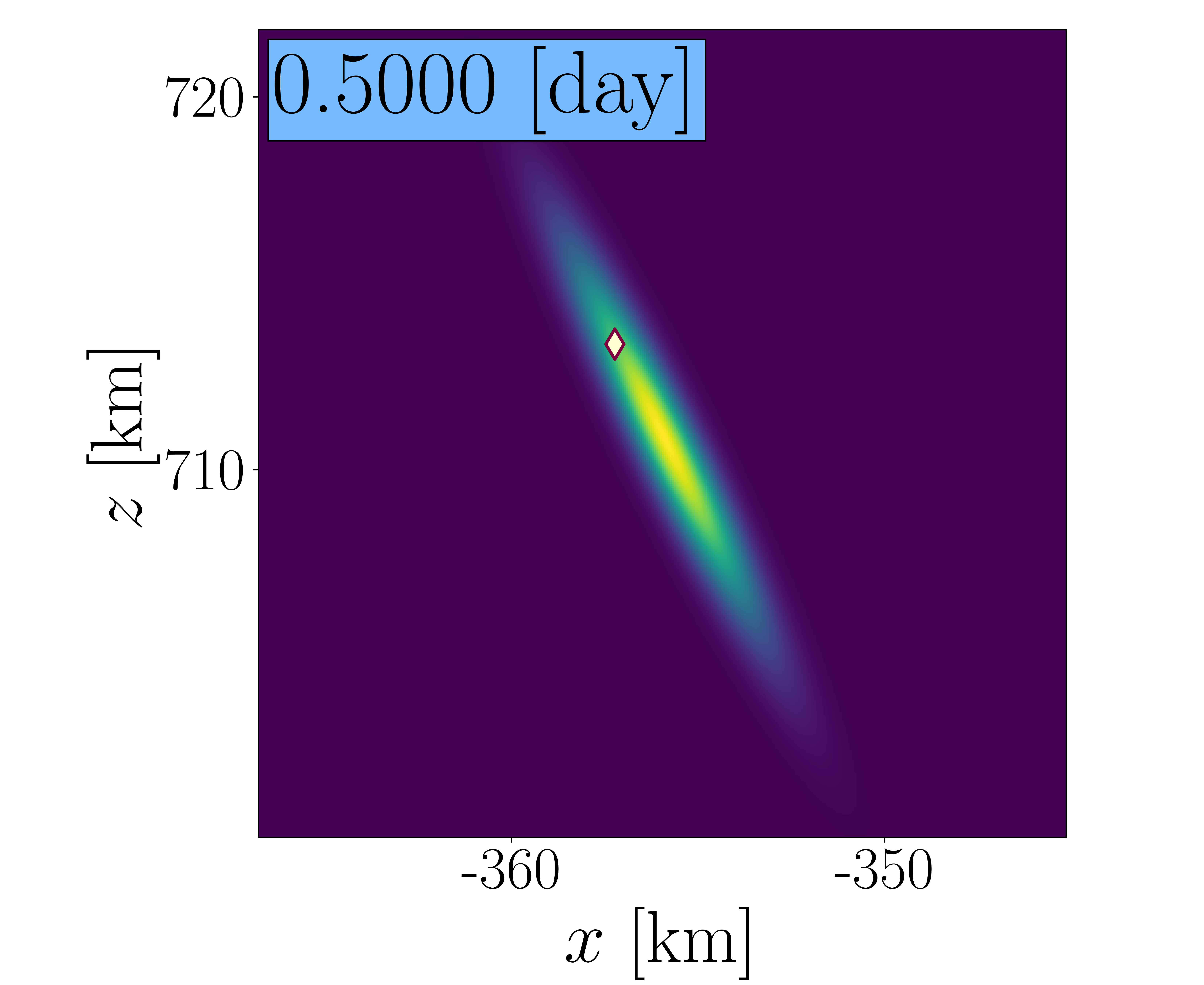}
            \put(8,80){\large (f)}
        \end{overpic}
        \label{fig:MolniyaSmootherXZMarginal12}
    \end{subfigure}%
    \caption{Filter and smoother $x$-$z$ marginal densities during perigee for Molniya orbit scenario.}
    \label{fig:MolniyaXZMarginalComparison}
\end{figure}

\begin{figure}[bht!]
    \centering
    \includegraphics[width=0.5\linewidth]{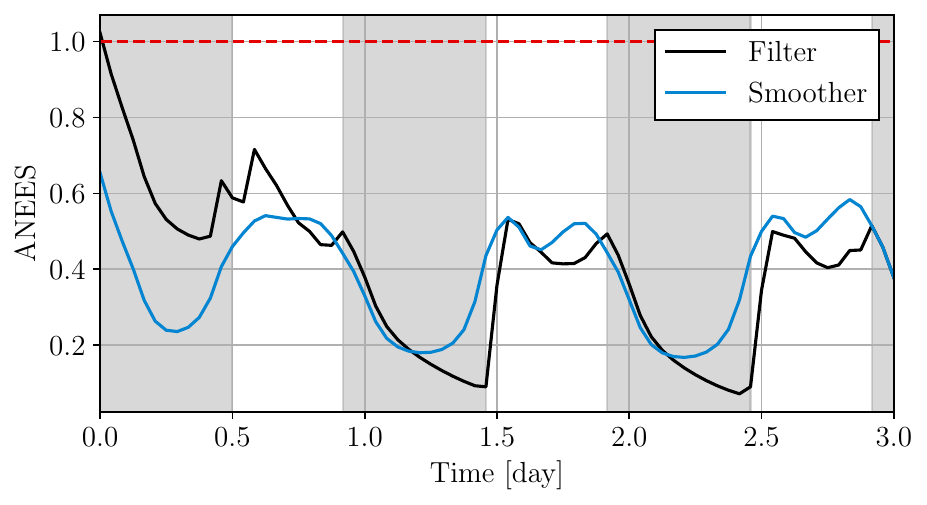}
    \caption{ANEES over 250 Monte Carlo trials for Molniya orbit scenario.}
    \label{fig:MolniyaANEES}
\end{figure}

The resulting time series of \ac{ANEES} for the filter and smoother is shown in Fig. \ref{fig:MolniyaANEES}. The dashed red line denotes the credible \ac{ANEES} value of 1. The filter and smoother are both conservative in their estimates, which is a result of process noise being included in the dynamics model despite the true dynamics being deterministic. This process noise is included to counteract errors stemming from linearization of the true nonlinear dynamics, which could otherwise diverge the filter.
The smoother is also generally more credible than the filter near the end of measurement gaps, as the backward filter has access to more recent measurement information and is therefore more credible during these periods.

\subsection{Southern $\text{L}_2$ Halo Orbit}
For modeling cislunar dynamics, the assumptions of the \ac{CR3BP} are adopted for simplicity \cite{koon2017dynamical}.
Within this dynamical framework, the tracking scenario is defined by a space-based observer architecture.
The measurements consist of azimuth and elevation angles taken by the observer deployed in a \ac{DRO}.
The observer has a body frame defined by
\begin{equation}
    \mathcal{B} = \left(\mathbf{r}_{\text{obs}},\hat{\mathbf{b}}_1,\hat{\mathbf{b}}_2,\hat{\mathbf{b}}_3\right),
\end{equation}
where $\mathbf{r}_{\text{obs}}$ is the position of the observer  in the synodic frame and $\hat{\mathbf{b}}_1$, $\hat{\mathbf{b}}_2$, and $\hat{\mathbf{b}}_3$ form an orthonormal basis. The observer's camera boresight is assumed to be aligned with $\hat{\mathbf{b}}_2$, corresponding to the time-varying direction of the observer-state relative position. The measurement function is given by
\begin{equation}
    \mathbf{h}_k(\mathbf{x}) = \begin{bmatrix}
        \arctan\left(\displaystyle\frac{\hat{\mathbf{u}}\cdot\hat{\mathbf{b}}_1}{\hat{\mathbf{u}}\cdot\hat{\mathbf{b}}_2}\right) \\
        \arcsin\left(\hat{\mathbf{u}}\cdot\hat{\mathbf{b}}_3\right)
    \end{bmatrix}, \quad \hat{\mathbf{u}} = \displaystyle\frac{\mathbf{r}-\mathbf{r}_{\text{obs}}}{||\mathbf{r}-\mathbf{r}_{\text{obs}}||},
\end{equation}
and the measurements are assumed to be corrupted by white Gaussian noise with a standard deviation of 10 [arcsec].

\begin{figure}[bht!]
    \centering
    \includegraphics[width=0.5\linewidth]{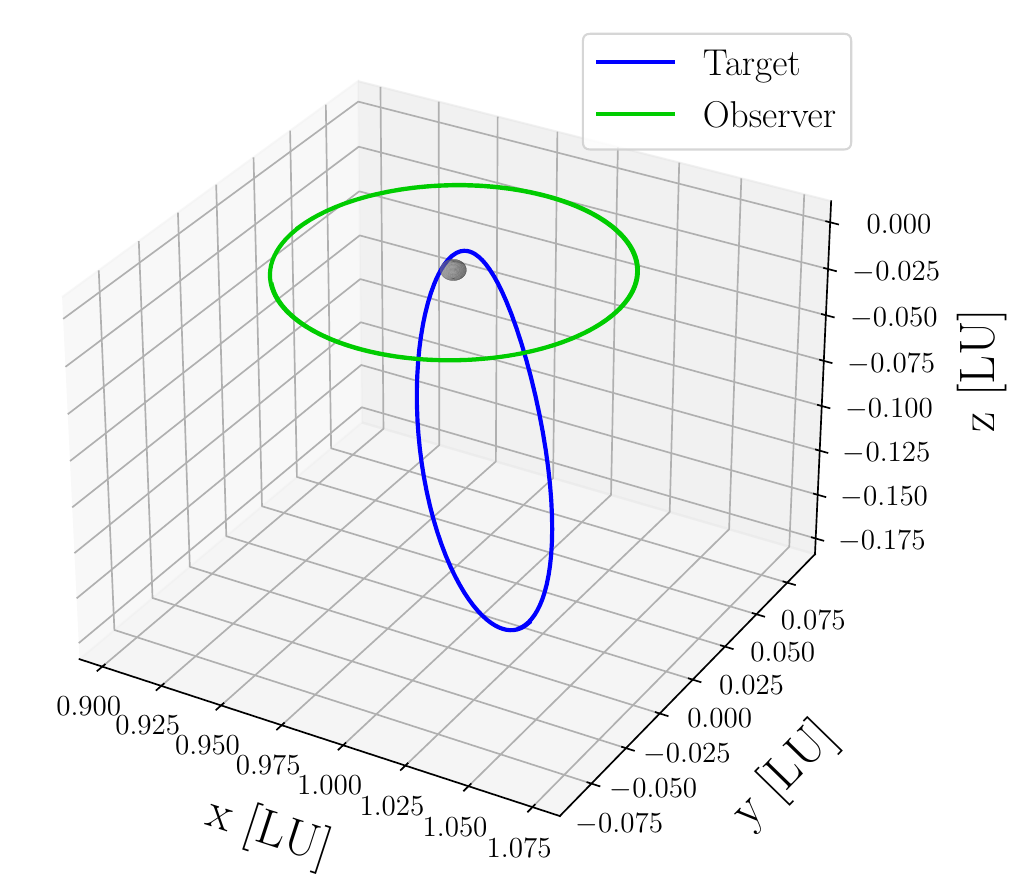}
    \caption{Orbits for cislunar scenario.}
    \label{fig:CislunarOrbits}
\end{figure}

Table~\ref{tab:CislunarScenario} summarizes the parameters for the cislunar scenario, including the continuous-time process noise, the initial target state uncertainty, and the initial states of both vehicles. The tracking setup considers a target in a \ac{NRHO} with a 6.5-day orbital period. Over a 28-day simulation span, the observer takes one measurement per hour when the target satisfies the assumed visibility conditions, which amounts to 86 observations per target orbit on average. To satisfy the visibility conditions, the target must be illuminated, non-occluded, and outside of the Moon exclusion cone \cite{Iannamorelli_AGMIMM_2025,Klonowski_CislunarCoopArchitecture_2024}. The orbits of the target and observer are shown in Fig.~\ref{fig:CislunarOrbits}, where the target is in blue and the observer is in green.

\begin{table}[hbt!]
    \centering
    \caption{Parameters for cislunar scenario.}
    \label{tab:CislunarScenario}
    \begin{subtable}{\textwidth}
        \centering
        \caption{Target and Observer Initial Conditions}
        \label{tab:SubCislunarStates}
        \vspace{2pt}
        \begin{tabular}{lcc}
            \toprule\toprule
            \textbf{State Component} & \textbf{Symbol} & \textbf{Value} \\
            \midrule
            Target Position & $\mathbf{r}_0$ & $\begin{bmatrix} 1.0193276 & 0 & -0.1801721 \end{bmatrix}^\top\,[\text{LU}]$ \\
            Target Velocity & $\mathbf{v}_0$ & $\begin{bmatrix} 0 & -0.09731243 & 0 \end{bmatrix}^\top\,[\text{LU/TU}]$ \\
            \addlinespace[0.5em]
            Observer Position & $\mathbf{r}_{\text{obs},0}$ & $\begin{bmatrix} 0.9243621 & 0 & 0 \end{bmatrix}^\top\,[\text{LU}]$ \\
            Observer Velocity & $\mathbf{v}_{\text{obs},0}$ & $\begin{bmatrix} 0 & 0.5095987 & 0 \end{bmatrix}^\top\,[\text{LU/TU}]$ \\
            \bottomrule\bottomrule
        \end{tabular}
    \end{subtable}
    \vspace{15pt} %

    \begin{subtable}[t]{0.48\textwidth}
        \centering
        \caption{Filter Parameters \& Noise}
        \label{tab:SubCislunarFilter}
        \vspace{2pt}
        \begin{tabular}{lcc}
            \toprule\toprule
            \textbf{Parameter} & \textbf{Symbol} & \textbf{Value} \\
            \midrule
            Pos. uncertainty & $\sigma_r$ & $10\,[\text{km}]$ \\
            Vel. uncertainty & $\sigma_v$ & $1\,[\text{m/s}]$ \\
            Process noise & $q_{w_{x,y,z}}$ & $10^{-3}\,[\text{LU}/\text{TU}^{5/2}]$ \\
            Sensor noise & $v$ & $10\,[\text{arcsec}]$ \\
            \bottomrule\bottomrule
        \end{tabular}
    \end{subtable}
    \hfill
\end{table}

\begin{figure}[bht!]
    \centering
    \includegraphics[width=0.5\linewidth]{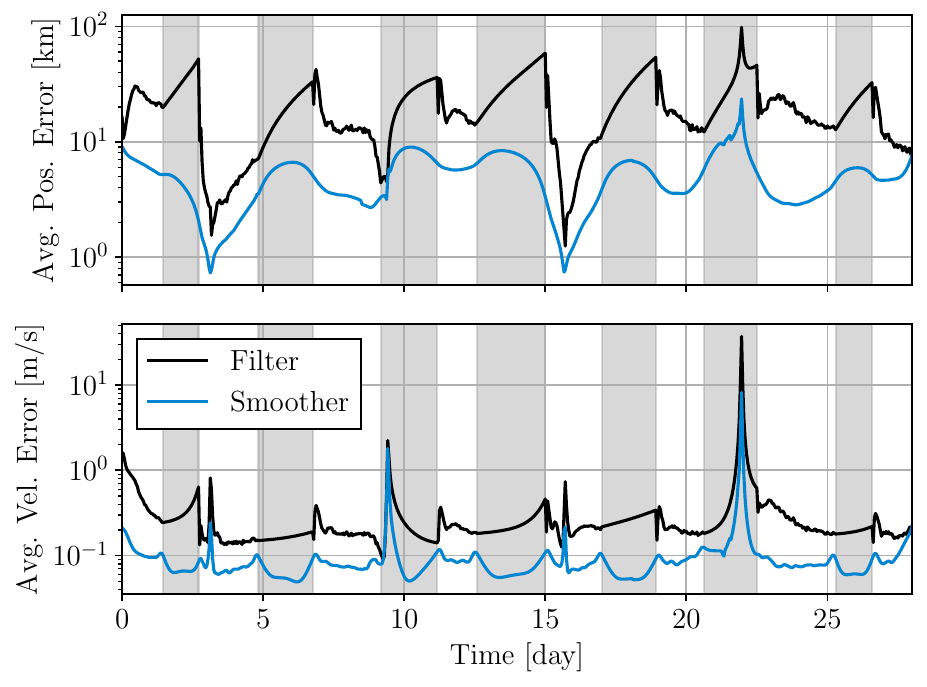}
    \caption{Average position and velocity error over 250 Monte Carlo trials for cislunar scenario.}
    \label{fig:CislunarError}
\end{figure}

Figure~\ref{fig:CislunarError} shows the average position and velocity error for the filter and smoother across the Monte Carlo trials. The smoother, as for the Molniya case, refines the filter estimates at all time steps, and achieves an order of magnitude improvement in position error for time steps near the end of measurement gaps. The improvements of smoothing can be limited when the future measurement information has become stale after being propagated through highly chaotic dynamics. This can be seen by the spike in smoothing error near 9.5~[day], which occurs during a target perilune passage within a measurement gap. However, these spikes in error are much less prevalent for the smoother than the filter, which highlights its utility in refining state estimates for highly nonlinear systems with intermittent measurement availability.

\begin{figure}[bht!]
    \par\medskip
    \centering
    \begin{subfigure}{0.33\figrowwidth}
        \begin{overpic}[width=\linewidth]{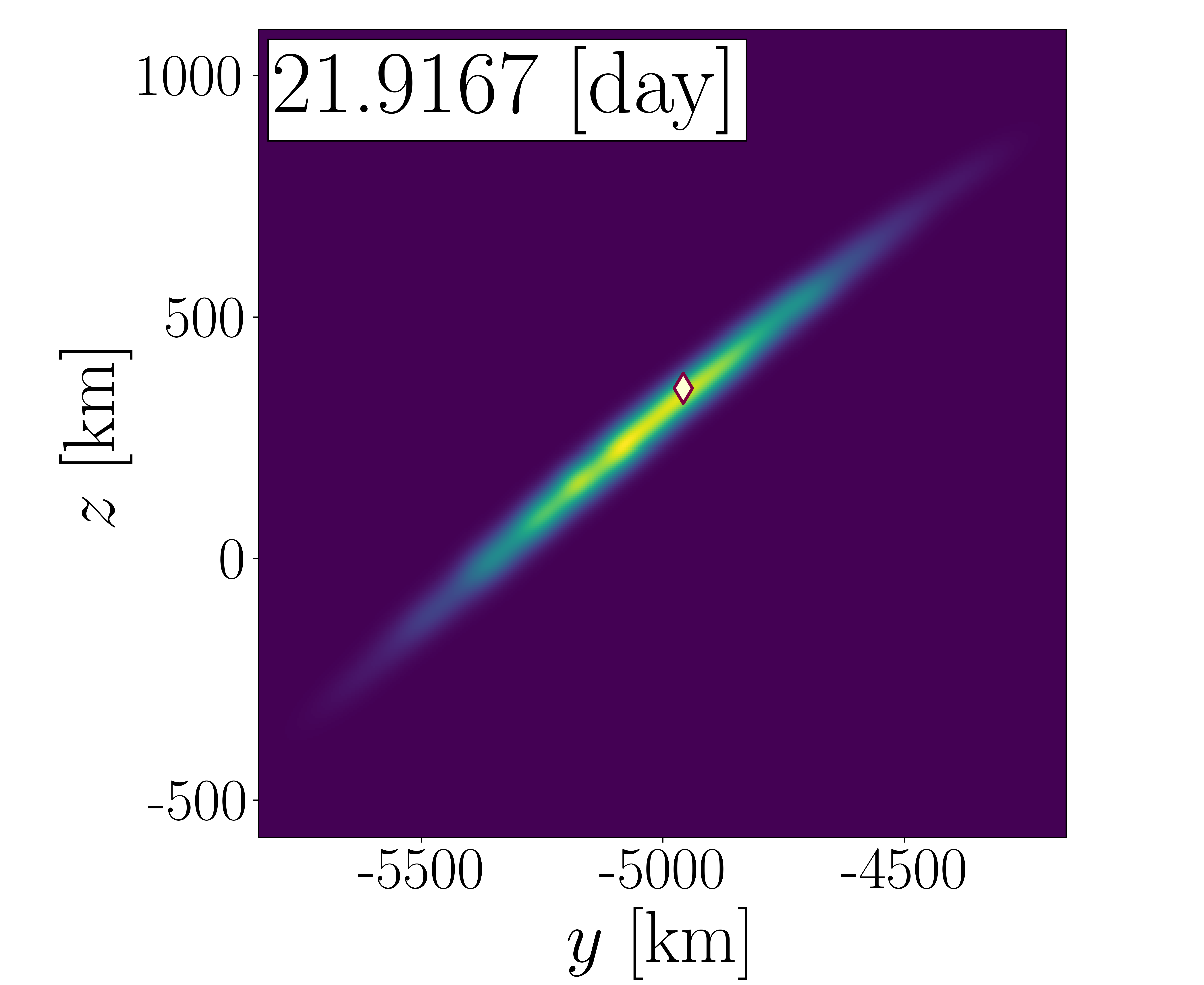}
            \put(8,80){\large (a)}
        \end{overpic}
        \label{fig:CislunarFilterYZMarginal526}
    \end{subfigure}
    \begin{subfigure}{0.33\figrowwidth}
        \begin{overpic}[width=\linewidth]{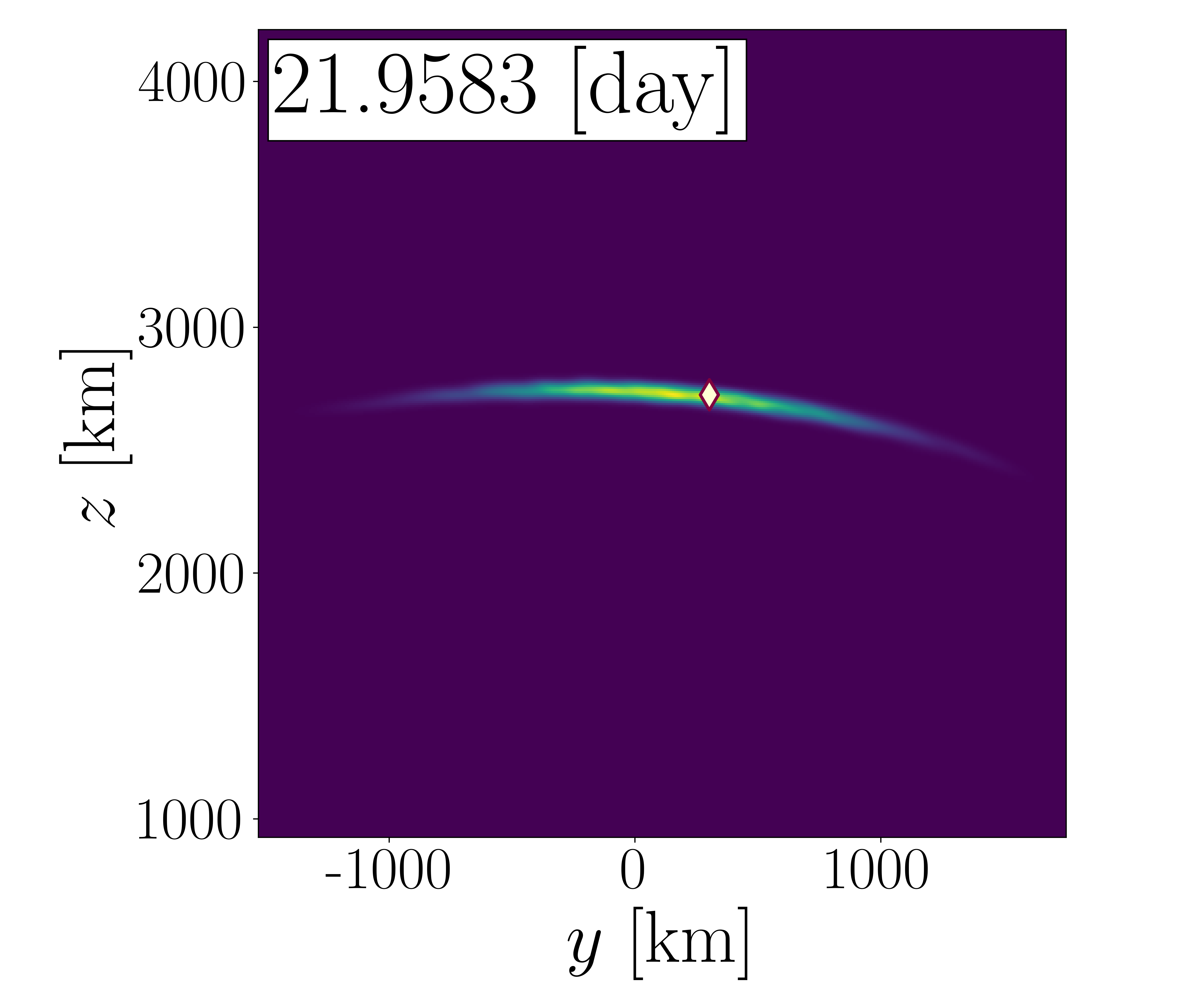}
            \put(8,80){\large (b)}
        \end{overpic}
        \label{fig:CislunarFilterYZMarginal527}
    \end{subfigure}
    \begin{subfigure}{0.33\figrowwidth}
        \begin{overpic}[width=\linewidth]{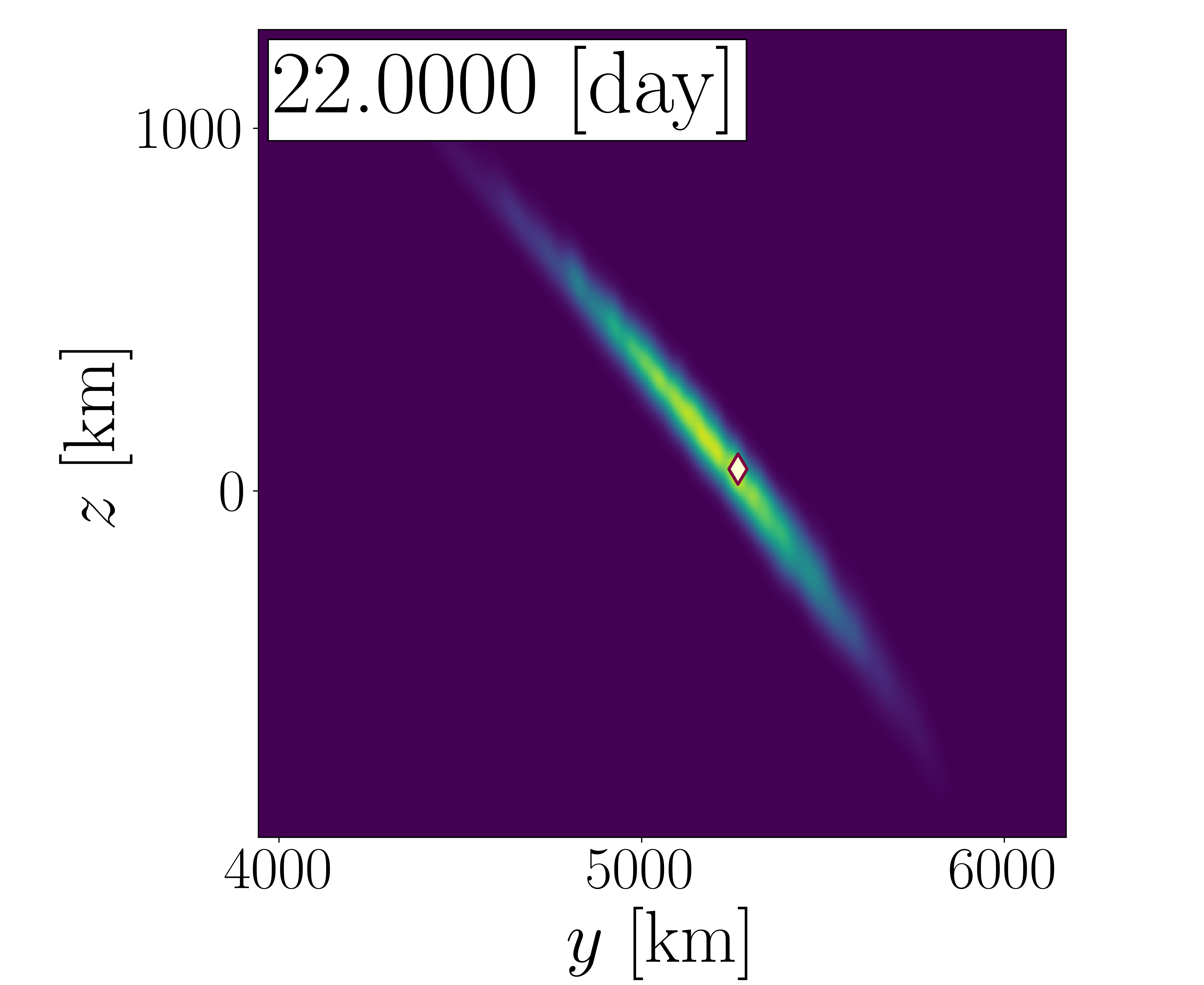}
            \put(8,80){\large (c)}
        \end{overpic}
        \label{fig:CislunarFilterYZMarginal528}
    \end{subfigure}

    \begin{subfigure}{0.33\figrowwidth}
        \begin{overpic}[width=\linewidth]{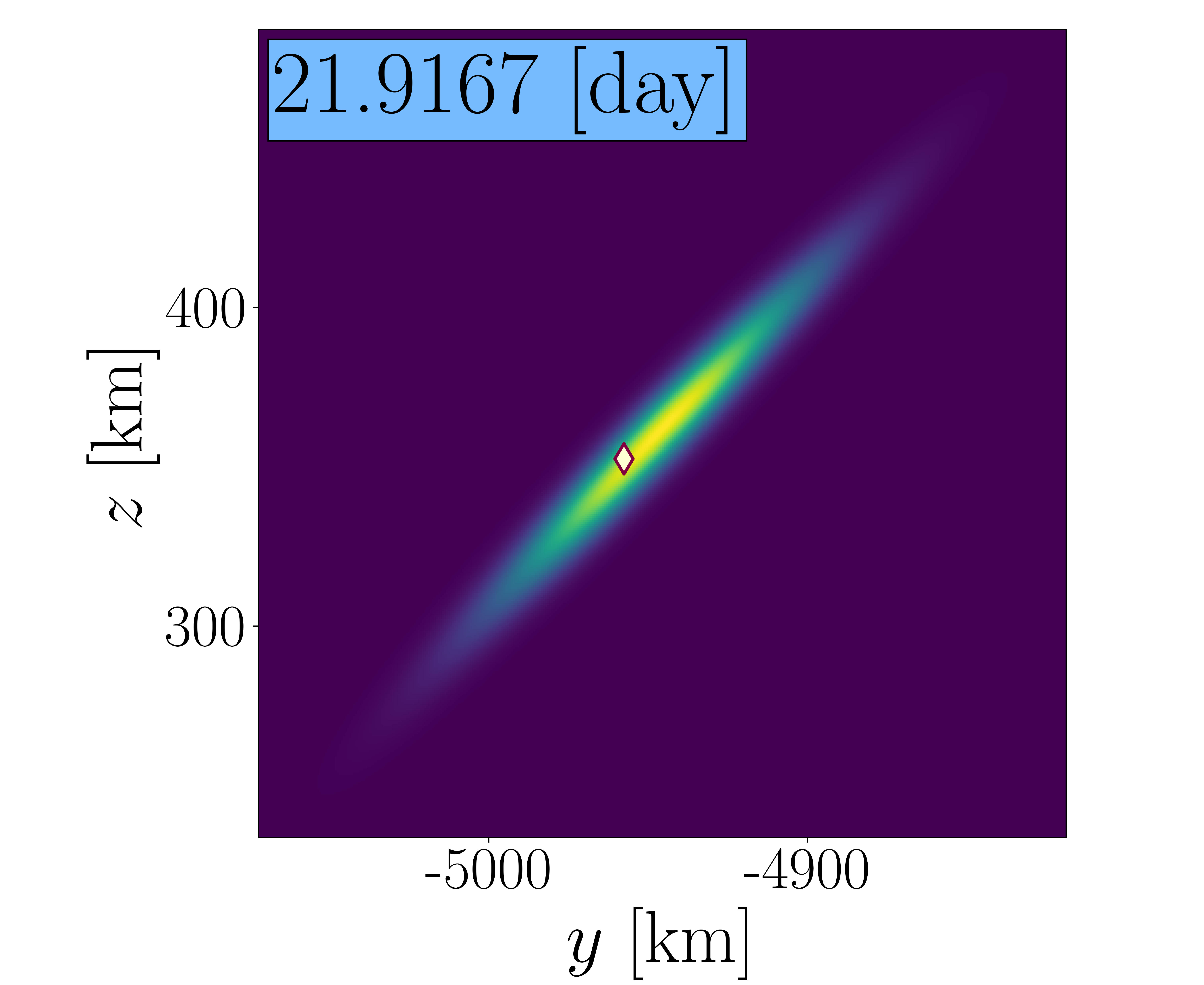}
            \put(8,80){\large (d)}
        \end{overpic}
        \label{fig:CislunarSmootherYZMarginal526}
    \end{subfigure}%
    \begin{subfigure}{0.33\figrowwidth}
        \begin{overpic}[width=\linewidth]{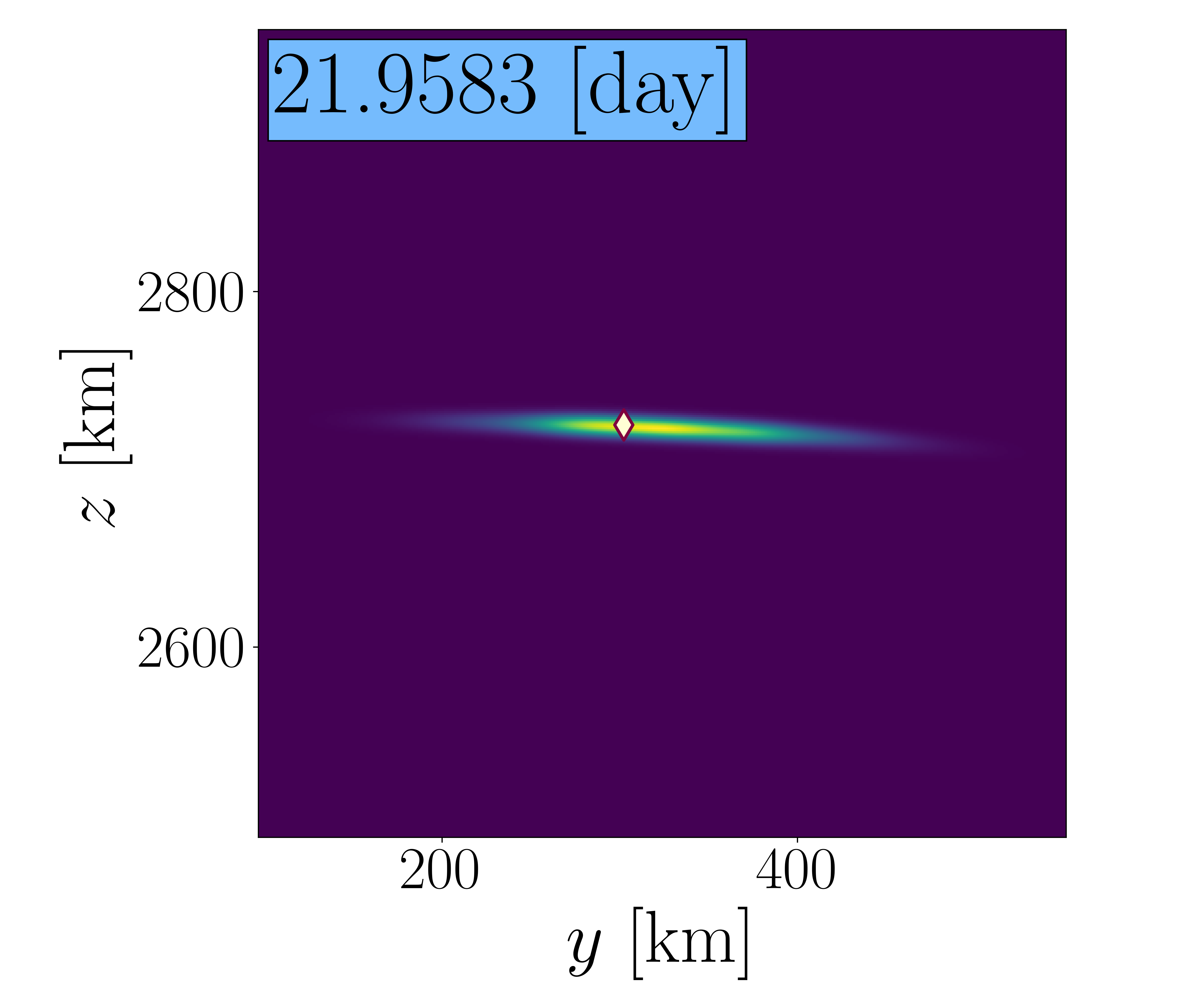}
            \put(8,80){\large (e)}
        \end{overpic}
        \label{fig:CislunarSmootherYZMarginal527}
    \end{subfigure}%
    \begin{subfigure}{0.33\figrowwidth}
        \begin{overpic}[width=\linewidth]{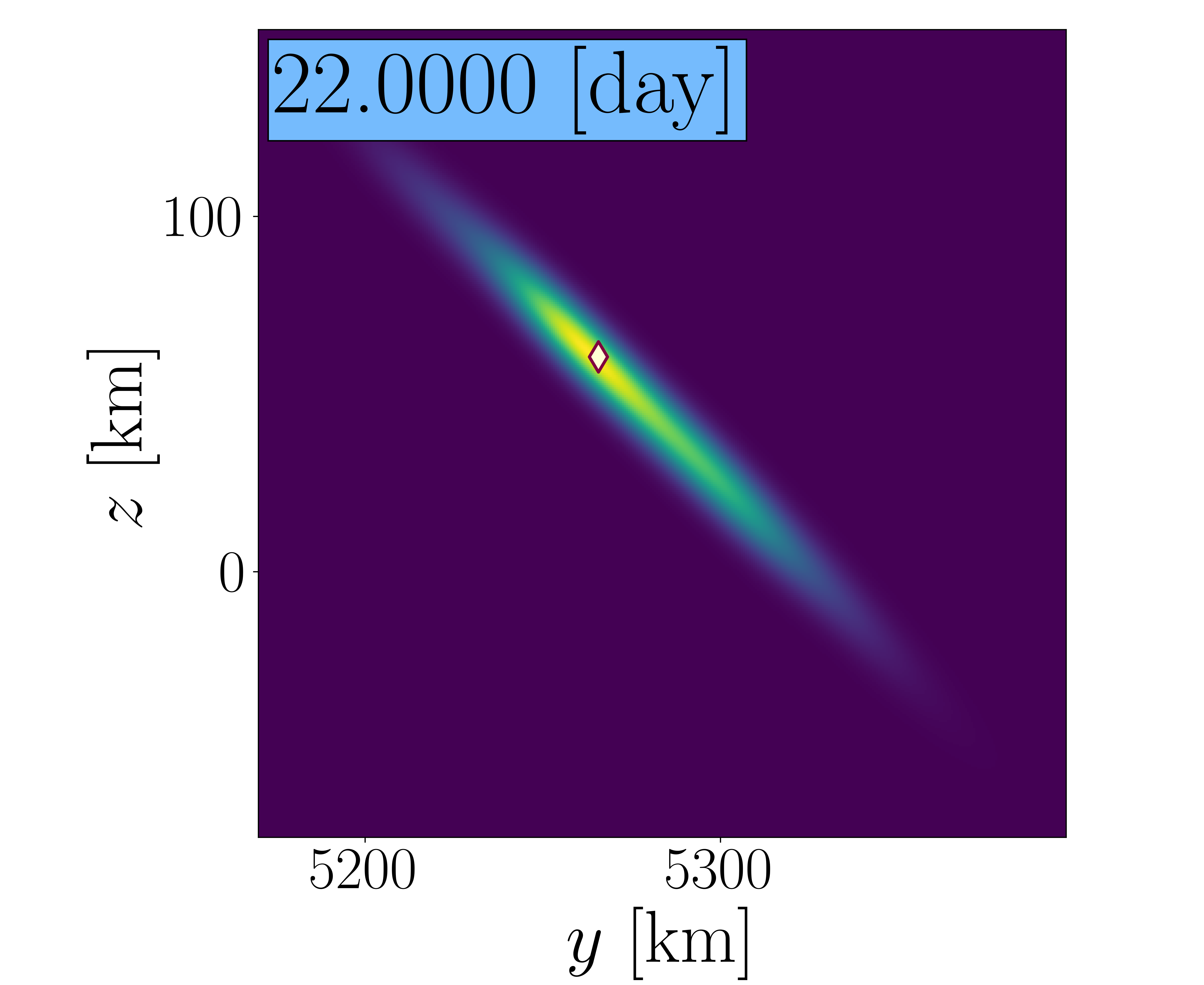}
            \put(8,80){\large (f)}
        \end{overpic}
        \label{fig:CislunarSmootherYZMarginal528}
    \end{subfigure}%
    \caption{Filter and smoother $y$-$z$ marginal densities during perilune of cislunar scenario.}
    \label{fig:CislunarYZMarginalComparison}
\end{figure}

Figure~\ref{fig:CislunarYZMarginalComparison} shows the~$y$-$z$ marginals for the filter and smoother during the measurement gap centered around a target perilune passage at 22.75~[day]. The time since epoch is displayed in a box in the top left of each plot. The color of the box indicates whether the marginal corresponds to the filter (white) or smoother (blue). The red and white diamond denotes the true target state. These time steps correspond to the highly nonlinear dynamics of perilune, and thus represent some of the most challenging steps for the forward and backward filters. Despite this, the smoothed marginals still significantly reduce the uncertainty of the filter, with the standard deviation in the $y$-direction at 21.9583~[day] being reduced from 488.8~[km] to 61.8~[km].

\begin{figure}[bht!]
    \centering
    \includegraphics[width=0.5\linewidth]{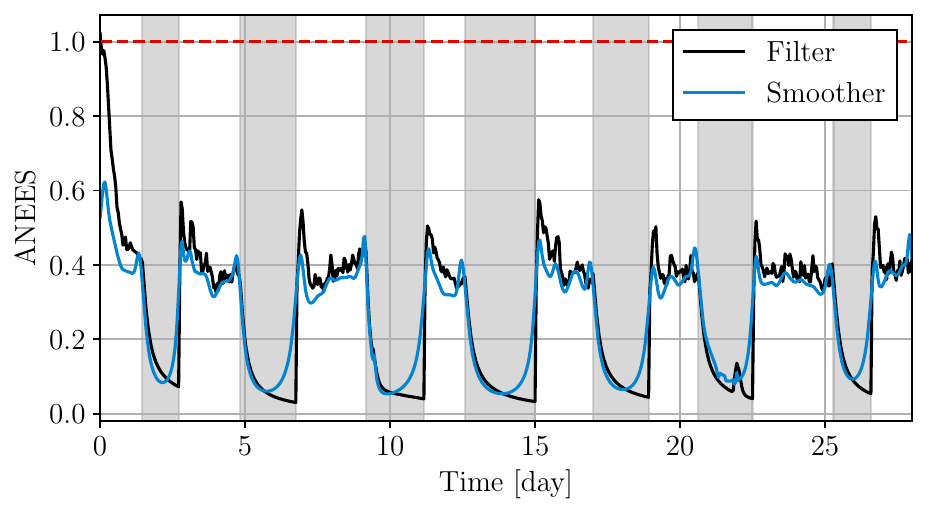}
    \caption{ANEES over 250 Monte Carlo trials for cislunar scenario.}
    \label{fig:CislunarANEES}
\end{figure}

Figure~\ref{fig:CislunarANEES} shows the \ac{ANEES} for the filter and smoother over time.
Again, the filter and smoother are both conservative due to the process noise, and the smoother is more credible than the filter toward the end of measurement gaps due to the better credibility of the backward filter at these times.

\section{Conclusion}\label{sec:conclusion}
This paper presents a novel adaptive Gaussian mixture smoothing algorithm for Bayesian nonlinear state estimation. The backward filter likelihood is approximated by a mixture of exponential quadratic likelihood components in the state variable, and is recursively updated and propagated through the nonlinear system using linearization approximations in a similar manner to a forward \acf{EKF}. To maintain the tractability and accuracy of this approximation under highly nonlinear dynamics and measurements, methodologies for splitting and merging likelihood components are proposed that increase fidelity in regions of high nonlinearity and improve efficiency by combining similar components. The smoothing algorithm performance is analyzed on two space object tracking problems with targets in a Molniya orbit and a \acf{NRHO}, and in both scenarios the smoother greatly reduced uncertainty and non-Gaussianity compared to the forward filter alone while maintaining credibility. The algorithm is especially well-suited for applications with limited or sporadic measurement availability, as smoothing can reduce estimation error and uncertainty by orders of magnitude for estimates based on outdated measurement information.

\section*{Appendix}\label{sec:appendix}
\appendix
\counterwithin{equation}{subsection}
\renewcommand{\thesubsection}{\Alph{subsection}}
\renewcommand{\theequation}{\thesubsection\arabic{equation}}
\subsection{Adaptive Gaussian Mixture Extended Kalman Filter}\label{sec:appendix_GMfilter}
The initial distribution is assumed to be given in the form of a \ac{GM} as
\begin{align}
    p(\mathbf{x}_0) = \sum_{n=1}^{N_0} w_0^{(n)} \mathcal{N}\!\left(\mathbf{x}_0;\mathbf{m}_0^{(n)},\mathbf{P}_0^{(n)}\right)
\end{align}

By applying the assumed forms of the posterior and transition densities to \eqref{eqn:bayes_pred_k1}, the prior density can be computed by
\begin{align}
    p(\mathbf{x}_k|\mathbf{z}_{1:k-1}) = \sum_{n=1}^{N_{k-1}} w_{k-1}^{(n)} \int \mathcal{N}\!\left(\mathbf{x}_{k-1};\mathbf{m}_{k-1}^{(n)},\mathbf{P}_{k-1}^{(n)}\right) \mathcal{N}\!(\mathbf{x}_k; \bm\varphi_{k-1}^k(\mathbf{x}_{k-1}),\left.\mathbf{Q}_{k-1}\right|_{\mathbf{x}_{k-1}}) \textrm{d}\mathbf{x}_{k-1}
\end{align}

If a linear approximation of the dynamics function is made about each mixand mean, then the prior can be computed as a \ac{GM} in closed-form as
\begin{align}
    p(\mathbf{x}_k|\mathbf{z}_{1:k-1}) \approx \sum_{n=1}^{N_{k|k-1}} w_{k|k-1}^{(n)} \mathcal{N}\!\left(\mathbf{x}_k;\mathbf{m}_{k|k-1}^{(n)},\mathbf{P}_{k|k-1}^{(n)}\right)
\end{align}
where
\begin{subequations}
    \begin{align}
        N_{k|k-1} &= N_{k-1} \\
        w_{k|k-1}^{(n)} &= w_{k-1}^{(n)} \\
        \mathbf{m}_{k|k-1}^{(n)} &= \bm\varphi_{k-1}^{k}\left(\mathbf{m}_{k-1}^{(n)}\right) \\
        \mathbf{P}_{k|k-1}^{(n)} &= \left.\bm\Phi_{k-1}^{k} \right|_{\mathbf{m}_{k-1}^{(n)}}\mathbf{P}_{k-1}^{(n)} \left(\left.\bm\Phi_{k-1}^{k} \right|_{\mathbf{m}_{k-1}^{(n)}}\right)^\top + \left.\mathbf{Q}_k\right|_{\mathbf{m}_{k-1}^{(n)}}
    \end{align}
\end{subequations}

The posterior density can be found by applying the assumed forms of the prior density and measurement likelihood to~\eqref{eqn:bayes_upd_k1}, yielding
\begin{align}
    p(\mathbf{x}_k|\mathbf{z}_{1:k}) \propto \sum_{n=1}^{N_{k|k-1}} w_{k|k-1}^{(n)} \mathcal{N}\!\left(\mathbf{x}_k;\mathbf{m}_{k|k-1}^{(n)},\mathbf{P}_{k|k-1}^{(n)}\right)\mathcal{N}\!\left(\mathbf{z}_k;\mathbf{h}_k(\mathbf{x}_k),\mathbf{R}_k\right)
\end{align}

Again, if a linear approximation of the measurement function is made about each mixand mean, then the posterior can be computed in closed form as
\begin{align}
    p(\mathbf{x}_k|\mathbf{z}_{1:k}) \approx \sum_{n=1}^{N_{k}} w_{k}^{(n)} \mathcal{N}\!\left(\mathbf{x}_k;\mathbf{m}_{k}^{(n)},\mathbf{P}_{k}^{(n)}\right)
\end{align}
where
\begin{subequations}
    \begin{align}
        N_k &= N_{k|k-1} \\
        w_{k}^{(n)} &= \displaystyle\frac{w_{k|k-1}^{(n)}q_k^{(n)}}{\sum_{n=1}^{N_{k|k-1}}w_{k|k-1}^{(n)}q_k^{(n)}} \\
        q_k^{(n)} &= \mathcal{N}\!\left(\mathbf{z}_k;\mathbf{h}_k\left(\mathbf{m}_{k|k-1}^{(n)}\right),\mathbf{H}_{k}|_{\mathbf{m}_{k|k-1}^{(n)}}\mathbf{P}_{k|k-1}^{(n)}\left(\mathbf{H}_{k}|_{\mathbf{m}_{k|k-1}^{(n)}}\right)^\top + \mathbf{R}_k\right) \\
        \mathbf{m}_k^{(n)} &= \mathbf{m}_{k|k-1}^{(n)} + \mathbf{K}_k^{(n)}\left(\mathbf{z}_k - \mathbf{h}_k\left(\mathbf{m}_{k|k-1}^{(n)}\right)\right)\\
        \mathbf{P}_k^{(n)} &= \left(\mathbf{I} - \mathbf{K}_k^{(n)}\mathbf{H}_k^{(n)}|_{\mathbf{m}_{k|k-1}^{(n)}}\right)\mathbf{P}_{k|k-1}^{(n)}\left(\mathbf{I} - \mathbf{K}_k^{(n)}\mathbf{H}_k^{(n)}|_{\mathbf{m}_{k|k-1}^{(n)}}\right)^\top + \mathbf{K}_k^{(n)}\mathbf{R}_k\left(\mathbf{K}_k^{(n)}\right)^\top \\
        \mathbf{K}_k^{(n)} &= \mathbf{P}_{k|k-1}^{(n)}\left(\mathbf{H}_k^{(n)}|_{\mathbf{m}_{k|k-1}^{(n)}}\right)^\top \left(\mathbf{H}_{k}|_{\mathbf{m}_{k|k-1}^{(n)}}\mathbf{P}_{k|k-1}^{(n)}\left(\mathbf{H}_{k}|_{\mathbf{m}_{k|k-1}^{(n)}}\right)^\top + \mathbf{R}_k\right)^{-1}
    \end{align}
\end{subequations}
where~$q_k^{(n)}$ is the measurement likelihood agreement and~$\mathbf{K}_k^{(n)}$ is the Kalman gain.

Before the prediction and update steps, the input densities are split according to the methodology of Section~\ref{sec:background_split} and the nonlinear functions~$\bm\varphi_{k-1}^k(\mathbf{x}_{k-1})$ and~$\mathbf{h}_k(\mathbf{x}_k)$ respectively. Following each step, the output densities are merged according to the methodology of Section~\ref{sec:background_merge}.

\subsection{Proof of Backward Information Filter Update}\label{sec:appendix_upd}
For the sake of clarity, the relationship of~\eqref{eqn:BIF_update}-\eqref{eqn:BIF_update_params} will be shown for a single component with the time index, component index, and evaluation points omitted. Substituting the definitions for likelihood and Gaussian components into~\eqref{eqn:BIF_update} yields
\begin{align}
    &\mathcal{L}\!(\mathbf{x};\bm\mu,\bm\Lambda)\mathcal{N}\!(\mathbf{z};\mathbf{h} + \mathbf{H}(\mathbf{x} - \bm\mu),\mathbf{R}) \\&= \exp\left\{-\displaystyle\frac{1}{2}(\mathbf{x}-\bm\mu)^\top \bm\Lambda (\mathbf{x}-\bm\mu)\right\} \cdot|2\pi\mathbf{R}|^{-1/2}\exp\left\{-\displaystyle\frac{1}{2}(\mathbf{z} - \mathbf{h} - \mathbf{H}(\mathbf{x} - \bm\mu))^\top\mathbf{R}^{-1}(\mathbf{z} - \mathbf{h} - \mathbf{H}(\mathbf{x} - \bm\mu))\right\} \\
    &= \mathcal{N}\!(\mathbf{z};\mathbf{h},\mathbf{R})\exp\left\{-\displaystyle\frac{1}{2}\left[(\mathbf{x}-\bm\mu)^\top(\bm\Lambda + \mathbf{H}^\top\mathbf{R}^{-1}\mathbf{H})(\mathbf{x}-\bm\mu) - 2(\mathbf{x}-\bm\mu)^\top \mathbf{H}^\top \mathbf{R}^{-1}(\mathbf{z}-\mathbf{h})\right]\right\} \label{eqn:update_paused_derivation}
\end{align}

Examining the second term of~\eqref{eqn:update_paused_derivation}, further simplification requires a quadratic of the form
\begin{equation}
    (\mathbf{x}-(\bm\mu + \bm\Delta\bm\mu))^\top (\bm\Lambda + \mathbf{H}^\top \mathbf{R}^{-1}\mathbf{H})(\mathbf{x}-(\bm\mu + \bm\Delta\bm\mu))
\end{equation}
where~$\bm\Delta\bm\mu$ must satisfy
\begin{align}
    -2(\mathbf{x}-\bm\mu)^\top \mathbf{H}^\top \mathbf{R}^{-1}(\mathbf{z}-\mathbf{h}) = -2(\mathbf{x}-\bm\mu)^\top(\bm\Lambda + \mathbf{H}^\top\mathbf{R}^{-1}\mathbf{H})\bm\Delta\bm\mu \label{eqn:update_completing_square}
\end{align}
If~$\bm\Lambda + \mathbf{H}^\top\mathbf{R}^{-1}\mathbf{H}$ is nonsingular, then~$\bm\Delta \bm\mu$ can be found through a standard linear system solve. If~$\bm\Lambda + \mathbf{H}^\top\mathbf{R}^{-1}\mathbf{H}$ is singular, then more care must be taken.

Let~$\mathcal{R}(\mathbf{A})$ and~$\textrm{Null}(\mathbf{A})$ denote the range and nullspace of a matrix~$\mathbf{A}$ respectively. By definition, it is true that~$\mathbf{H}^\top \mathbf{R}^{-1}(\mathbf{z}-\mathbf{h}) \in \mathcal{R}(\mathbf{H}^\top)$. It is also true that~$\mathcal{R}(\mathbf{H}^\top) \subseteq \mathcal{R}(\bm\Lambda + \mathbf{H}^\top \mathbf{R}^{-1} \mathbf{H})$, which will be shown subsequently. Let~$\mathbf{y} \in \textrm{Null}(\bm\Lambda + \mathbf{H}^\top \mathbf{R}^{-1} \mathbf{H})$. Then,
\begin{align}
    &(\bm\Lambda + \mathbf{H}^\top \mathbf{R}^{-1}\mathbf{H})\mathbf{y} = \mathbf{0} \\
    \implies & \mathbf{y}^\top\bm\Lambda\mathbf{y} + (\mathbf{H}\mathbf{y})^\top \mathbf{R}^{-1} (\mathbf{H}\mathbf{y}) = 0 \\
    \implies & \mathbf{H}\mathbf{y} = \mathbf{0}
\end{align}
where the last step is due to the positive definiteness of~$\mathbf{R}$ and positive semi-definiteness of~$\bm\Lambda$. This means that~$\textrm{Null}(\bm\Lambda + \mathbf{H}^\top \mathbf{R}^{-1}\mathbf{H}) \subseteq \textrm{Null}(\mathbf{H})$, which is an equivalent assertion to~$\mathcal{R}(\mathbf{H}^\top) \subseteq \mathcal{R}(\bm\Lambda + \mathbf{H}^\top \mathbf{R}^{-1}\mathbf{H})$.

So, since~$\mathbf{H}^\top \mathbf{R}^{-1}(\mathbf{z}-\mathbf{h}) \in \mathcal{R}(\bm\Lambda + \mathbf{H}^\top \mathbf{R}^{-1}\mathbf{H})$, a Moore-Penrose inverse can be used to solve~\eqref{eqn:update_completing_square} for~$\bm\Delta\bm\mu$, defined via an \ac{SVD} as
\begin{align}
    \bm\Lambda + \mathbf{H}^\top \mathbf{R}^{-1}\mathbf{H} &= \mathbf{U}\bm\Sigma \mathbf{U}^\top \\
    (\bm\Lambda + \mathbf{H}^\top \mathbf{R}^{-1}\mathbf{H})^\dagger &= \mathbf{U}\bm\Sigma^\dagger \mathbf{U}^\top \\
    \bm\Sigma^\dagger_{ii} = \begin{cases}
        1/\bm\Sigma_{ii} & \bm\Sigma_{ii} \neq 0 \\
        0 & \bm\Sigma_{ii} = 0
    \end{cases}
\end{align}
In practice, this Moore-Penrose inverse does not need to be explicitly computed, and~\eqref{eqn:update_completing_square} can instead be solved for~$\bm\Delta\bm\mu$ via a least-squares solve.

For clarity, define
\begin{align}
    \bm\Lambda^+ = \bm\Lambda + \mathbf{H}^\top \mathbf{R}^{-1}\mathbf{H},\qquad\bm\mu^+ = \bm\mu + \bm\Delta\bm\mu = \bm\mu + (\bm\Lambda^+)^\dagger\mathbf{H}^\top\mathbf{R}^{-1}(\mathbf{z}-\mathbf{h})
\end{align}
Then,~\eqref{eqn:update_paused_derivation} can be further rearranged as
\begin{align}
    \mathcal{L}\!(\mathbf{x};\bm\mu,\bm\Lambda)\mathcal{N}\!(\mathbf{z};\mathbf{h} + \mathbf{H}(\mathbf{x} - \bm\mu),\mathbf{R}) &= \mathcal{N}\!(\mathbf{z};\mathbf{h},\mathbf{R}) \exp\left\{-\displaystyle\frac{1}{2}\left[(\mathbf{x}-\bm\mu^+)^\top \bm\Lambda^+ (\mathbf{x}-\bm\mu^+) - \bm\Delta\bm\mu^\top \bm\Lambda^+ \bm\Delta\bm\mu\right]\right\} \\
    &= \mathcal{N}\!(\mathbf{z};\mathbf{h},\mathbf{R})\exp\left\{-\displaystyle\frac{1}{2}(\mathbf{z}-\mathbf{h})^\top\mathbf{R}^{-1}\mathbf{H}(\bm\mu - \bm\mu^+)\right\} \mathcal{L}\!(\mathbf{x};\bm\mu^+,\bm\Lambda^+)
\end{align}
which, when applied to each component, is equivalent to~\eqref{eqn:BIF_update}-\eqref{eqn:BIF_update_params}.

\subsection{Proof of Backward Information Filter Prediction}\label{sec:appendix_prop}
The relationship in \eqref{eqn:BIF_prop}--\eqref{eqn:BIF_prop_params} is verified here for a single likelihood component, with the component index and expansion points omitted for notation clarity. Substituting the definitions of the likelihood and the Gaussian components into \eqref{eqn:BIF_prop}, assuming $\mathbf{Q}_{k-1} \succ \mathbf{0}$ yields:
\begin{align}
\label{eq:C1}
    &\int \mathcal{L}\!(\mathbf{x}_k;\bm\mu_k,\bm\Lambda_k)\mathcal{N}\!(\mathbf{x}_k;\bm\mu_k + \bm\Phi_{k-1}^k(\mathbf{x}_{k-1} - \bm\varphi_{k}^{k-1}), \mathbf{Q}_{k-1}\!) \textrm{d}\mathbf{x}_k \\
    &= \int \exp\left\{-\frac{1}{2}(\mathbf{x}_k-\bm\mu_k)^\top\bm\Lambda_k(\mathbf{x}_k-\bm\mu_k)\right\} \nonumber \\
    &\quad\quad \cdot |2\pi\mathbf{Q}_{k-1}|^{-1/2}\exp\left\{-\frac{1}{2}(\mathbf{x}_k-\bm\mu_k - \bm\Phi_{k-1}^k(\mathbf{x}_{k-1}-\bm\varphi_{k}^{k-1}))^\top\ (\mathbf{Q}_{k-1})^{-1}(\mathbf{x}_k-\bm\mu_k - \bm\Phi_{k-1}^k(\mathbf{x}_{k-1}-\bm\varphi_{k}^{k-1}))\right\}\textrm{d}\mathbf{x}_k \label{eqn:expanded_prop}
\end{align}

Define~$\bm\xi_k = \bm\mu_k + \bm\Phi_{k-1}^k(\mathbf{x}_{k-1} - \bm\varphi_{k}^{k-1})$,
$\mathbf{M}_k = \bm\Lambda_k + (\mathbf{Q}_{k-1})^{-1}$, and~$\bm\eta_k = (\mathbf{Q}_{k-1})^{-1}\bm\xi_k + \bm\Lambda_k\bm\mu_k$.
Then,~\eqref{eqn:expanded_prop} can be rearranged as
\begin{align}
    &\int \mathcal{L}\!(\mathbf{x}_k;\bm\mu_k,\bm\Lambda_k)\mathcal{N}\!(\mathbf{x}_k;\bm\mu_k + \bm\Phi_{k-1}^k(\mathbf{x}_{k-1} - \bm\varphi_{k}^{k-1}),\mathbf{Q}_{k-1}) \textrm{d}\mathbf{x}_k \\
    &= |2\pi\mathbf{Q}_{k-1}|^{-1/2} \exp\left\{-\frac{1}{2}\left[\bm\xi_k^\top(\mathbf{Q}_{k-1})^{-1}\bm\xi_k + \bm\mu_k^\top\bm\Lambda_k\bm\mu_k - \bm\eta_k^\top \mathbf{M}_k^{-1}\bm\eta_k\right]\right\} \nonumber \\
    &\quad\quad\cdot\int \exp\left\{-\frac{1}{2}(\mathbf{x}_k-\mathbf{M}_k^{-1}\bm\eta_k)^\top \mathbf{M}_k(\mathbf{x}_k-\mathbf{M}_k^{-1}\bm\eta_k)\right\}\textrm{d}\mathbf{x}_k \\
    &= |2\pi\mathbf{Q}_{k-1}|^{-1/2}|2\pi\mathbf{M}_k^{-1}|^{1/2}\exp\left\{-\frac{1}{2}\left[\bm\xi_k^\top(\mathbf{Q}_{k-1})^{-1}\bm\xi_k + \bm\mu_k^\top\bm\Lambda_k\bm\mu_k - \bm\eta_k^\top \mathbf{M}_k^{-1}\bm\eta_k\right]\right\} \label{eqn:mess_prop}
\end{align}
By expanding the defined variables and simplifying, it can be shown that~\eqref{eqn:mess_prop} is equivalent to
\begin{align}
    &\int \mathcal{L}\!(\mathbf{x}_k;\bm\mu_k,\bm\Lambda_k)\mathcal{N}\!(\mathbf{x}_k;\bm\mu_k + \bm\Phi_{k-1}^k(\mathbf{x}_{k-1} - \bm\varphi_{k}^{k-1}),\mathbf{Q}_{k-1}) \textrm{d}\mathbf{x}_k \\
    &= |\mathbf{I} + \mathbf{Q}_{k-1}\bm\Lambda_k|^{-1/2} \exp\left\{-\frac{1}{2}(\mathbf{x}_{k-1} - \bm\varphi_{k}^{k-1})^\top(\bm\Phi_{k-1}^k)^\top\bm\Lambda_k(\mathbf{I} + \mathbf{Q}_{k-1}\bm\Lambda_k)^{-1}\bm\Phi_{k-1}^k(\mathbf{x}_{k-1} - \bm\varphi_{k}^{k-1})\right\} \\
    &= |\mathbf{I} + \mathbf{Q}_{k-1}\bm\Lambda_k|^{-1/2}\, \mathcal{L}\!(\mathbf{x}_{k-1};\bm\varphi_{k}^{k-1},(\bm\Phi_{k-1}^k)^\top\bm\Lambda_k(\mathbf{I} + \mathbf{Q}_{k-1}\bm\Lambda_k)^{-1}\bm\Phi_{k-1}^k)
    \label{eq:c8}
\end{align}
which, when applied to each component, is equivalent to~\eqref{eqn:BIF_prop}-\eqref{eqn:BIF_prop_params}.

Suppose $\mathbf{Q}_{k-1} \succeq \mathbf{0}$ is positive semi-definite with $\operatorname{rank}(\mathbf{Q}_{k-1}) = r < n_w$, so that the noise is confined to an $r$-dimensional subspace. Consider the nonlinear dynamics
\begin{equation}
    \mathbf{x}_k = \mathbf{f}(\mathbf{x}_{k-1},\, \mathbf{w}_{k-1},\, t_{k-1}), \qquad \mathbf{w}_{k-1} \sim \mathcal{N}(\mathbf{0},\, \mathbf{Q}_{k-1}).
\end{equation}
With the aid of the implicit function theorem, and partitioning $\mathbf{x}_k = [(\mathbf{x}_k^{(1)})^\top\, (\mathbf{x}_k^{(2)})^\top]^\top$ appropriately \cite{Andrew1970}, $f$ always admits the decomposition
\begin{align}
    \mathbf{x}_k^{(1)} &= \mathbf{f}^{(1)}\!\left(\mathbf{x}_{k-1},\, \mathbf{w}_{k-1}^*, t_{k-1}\right), \\
    \mathbf{x}_k^{(2)} &= \mathbf{f}^{(2)}\!\left(\mathbf{x}_{k-1},\, \mathbf{x}_k^{(1)}, t_{k-1}\right),
\end{align}
where $\mathbf{w}_{k-1}^* \sim \mathcal{N}(\mathbf{0},\, \mathbf{Q}_{k-1}^*)$ with $\mathbf{Q}_{k-1}^* \succ \mathbf{0}$ is a reduced $r$-dimensional noise, and $\mathbf{x}_k^{(2)}$ is fully determined by $\mathbf{x}_{k-1}$ and $\mathbf{x}_k^{(1)}$ with no independent stochastic input. The conditional densities of the two components are therefore
\begin{align}
    p\!\left(\mathbf{x}_k^{(1)} \mid \mathbf{x}_{k-1}\right) & = p_{\mathbf{w}_{k-1}}\left((\mathbf{f}^{(1)})^{-1}(\mathbf{x}_{k-1}, \mathbf{x}_k^{(1)}, t_{k-1})\right) \left\| \frac{\partial (\mathbf{f}^{(1)})^{-1}}{\partial \mathbf{x}_k^{(1)}} \right\|.\\
    p\!\left(\mathbf{x}_k^{(2)} \mid \mathbf{x}_k^{(1)},\, \mathbf{x}_{k-1}\right) &= \delta\!\left(\mathbf{x}_k^{(2)} - \mathbf{f}^{(2)}\!\left(\mathbf{x}_{k-1},\, \mathbf{x}_k^{(1)}\right)\right),
\end{align}
so the full transition density factorizes as
\begin{equation}
    p(\mathbf{x}_k \mid \mathbf{x}_{k-1})
    =  p_{\mathbf{w}_{k-1}}\left((\mathbf{f}^{(1)})^{-1}(\mathbf{x}_{k-1}, \mathbf{x}_k^{(1)}, t_{k})\right) \left\| \frac{\partial (\mathbf{f}^{(1)})^{-1}}{\partial \mathbf{x}_k^{(1)}} \right\|
    \cdot\,
    \delta\!\left(\mathbf{x}_k^{(2)} - \mathbf{f}^{(2)}\!\left(\mathbf{x}_{k-1},\, \mathbf{x}_k^{(1)}\right)\right).
\end{equation}

And in the case of Gaussian white noise:
\begin{multline}
    p(\mathbf{x}_k \mid \mathbf{x}_{k-1})
    =  \left[ (2\pi)^{n/2} |\mathbf{Q}^*_{k-1}|^{1/2} \right]^{-1} \cdot \exp\left\{ -\frac{1}{2} \left((\mathbf{f}^{(1)})^{-1}(\mathbf{x}_{k-1}, \mathbf{x}_k^{(1)}, t_k)\right)^{\top} (\mathbf{Q}_{k+1}^*)^{-1} \left((\mathbf{f}^{(1)})^{-1}(\mathbf{x}_{k-1},\mathbf{x}_k^{(1)}, t_k)\right) \right\}\\
    \cdot
    \delta\!\left(\mathbf{x}_k^{(2)} - \mathbf{f}^{(2)}\!\left(\mathbf{x}_{k-1},\, \mathbf{x}_k^{(1)}\right)\right).
\end{multline}

 Then,~\eqref{eq:C1} becomes:

\begin{multline}
    \int \mathcal{L}\!(\mathbf{x}_k;\bm\mu_k,\bm\Lambda_k)\left[ (2\pi)^{n/2} |\mathbf{Q}^*_{k-1}|^{1/2} \right]^{-1} \cdot \exp\left\{ -\frac{1}{2} \left((\mathbf{f}^{(1)})^{-1}(\mathbf{x}_{k-1}, \mathbf{x}_k^{(1)}, t_k)\right)^{\top} (\mathbf{Q}_{k-1}^*)^{-1} \left((\mathbf{f}^{(1)})^{-1}(\mathbf{x}_{k-1},\mathbf{x}_k^{(1)}, t_k)\right) \right\} \\
    \cdot \delta\!\left(\mathbf{x}_k^{(2)} - \mathbf{f}^{(2)}\!\left(\mathbf{x}_{k-1},\, \mathbf{x}_k^{(1)}\right)\right) \textrm{d}\mathbf{x}_k
\end{multline}

By linearizing the dynamics, it is possible to retrieve a final formulation that mirrors~\eqref{eq:c8}.

\subsection{Proof of Backward Information Filter Smoothing Operation}\label{sec:appendix_smooth}
The relationship of~\eqref{eqn:smoothed_dist}-\eqref{eqn:smoothed_params} will be shown for a single component pair with component and time indices omitted. Substituting the definitions for likelihood and Gaussian components into~\eqref{eqn:smoothed_dist} yields
\begin{align}
    w\mathcal{N}\!(\mathbf{x};\mathbf{m},\mathbf{P}) \cdot \nu\mathcal{L}\!(\mathbf{x};\bm\mu,\bm\Lambda) &= w |2\pi\mathbf{P}|^{-1/2} \exp\left\{-\displaystyle\frac{1}{2}(\mathbf{x}-\mathbf{m})^\top \mathbf{P}^{-1}(\mathbf{x}-\mathbf{m})\right\} \cdot \nu\exp\left\{ -\displaystyle\frac{1}{2}(\mathbf{x}-\bm\mu)^\top \bm\Lambda (\mathbf{x}-\bm\mu)\right\} \\
    &= w\nu |2\pi\mathbf{P}|^{-1/2}\exp\left\{-\displaystyle\frac{1}{2}\left[\mathbf{x}^\top (\mathbf{P}^{-1} + \bm\Lambda)\mathbf{x} - 2\mathbf{x}^\top (\mathbf{P}^{-1}\mathbf{m} + \bm\Lambda\bm\mu) + \mathbf{m}^\top \mathbf{P}^{-1}\mathbf{m} + \bm\mu^\top \bm\Lambda\bm\mu\right]\right\} \label{eqn:expanded_smooth}
\end{align}

Define~$\mathbf{P}^* = (\mathbf{P}^{-1} + \bm\Lambda)^{-1}$ and~$\mathbf{m}^* = \mathbf{P}^*(\mathbf{P}^{-1}\mathbf{m} + \bm\Lambda\bm\mu)$. Then,~\eqref{eqn:expanded_smooth} continues as
\begin{align}
    &w\mathcal{N}\!(\mathbf{x};\mathbf{m},\mathbf{P}) \cdot \nu\mathcal{L}\!(\mathbf{x};\bm\mu,\bm\Lambda) \\
    &= w\nu |2\pi\mathbf{P}|^{-1/2} \exp\left\{-\displaystyle\frac{1}{2}\left[(\mathbf{x}-\mathbf{m}^*)^\top (\mathbf{P}^*)^{-1}(\mathbf{x}-\mathbf{m}^*) - (\mathbf{m}^*)^\top(\mathbf{P}^*)^{-1}\mathbf{m}^* + \mathbf{m}^\top \mathbf{P}^{-1}\mathbf{m} + \bm\mu^\top \bm\Lambda\bm\mu\right]\right\} \\
    &= w\nu |2\pi\mathbf{P}|^{-1/2}|2\pi\mathbf{P}^*|^{1/2} \exp\left\{\displaystyle\frac{1}{2}\left[(\mathbf{m}^*)^\top(\mathbf{P}^*)^{-1}\mathbf{m}^* - \mathbf{m}^\top \mathbf{P}^{-1}\mathbf{m} - \bm\mu^\top \bm\Lambda\bm\mu\right]\right\} \mathcal{N}\!(\mathbf{x};\mathbf{m}^*,\mathbf{P}^*) \\
    &= w\nu |\mathbf{I} + \mathbf{P}\bm\Lambda|^{-1/2}\exp\left\{\displaystyle\frac{1}{2}\left[(\mathbf{m}^*)^\top(\mathbf{P}^*)^{-1}\mathbf{m}^* - \mathbf{m}^\top \mathbf{P}^{-1}\mathbf{m} - \bm\mu^\top \bm\Lambda\bm\mu\right]\right\} \mathcal{N}\!(\mathbf{x};\mathbf{m}^*,\mathbf{P}^*)
\end{align}
which, when applied to each component pair, is equivalent to~\eqref{eqn:smoothed_dist}-\eqref{eqn:smoothed_params}.

\section*{Acknowledgments}\label{sec:acknowledgements}
Part of this research was sponsored by the United States Air Force Research Laboratory and the United States AFRL Regional Hub and was accomplished under Cooperative Agreement Number FA8750-22- 2-0501.
The views and conclusions contained in this document are those of the authors and should not be interpreted as representing the official policies, either expressed or implied, of the United States Air Force or the U.S. Government.
The U.S. Government is authorized to reproduce and distribute reprints for Government purposes notwithstanding any copyright notation herein.

\bibliography{references}

\end{document}